\documentclass[aps, pra, floatfix, showpacs, 10pt]{revtex4-1}
\usepackage{eurosym}
\usepackage{dcolumn}
\usepackage{bm}
\usepackage{amssymb, amsmath}
\usepackage{graphicx}
\usepackage{hyperref}
\usepackage{epstopdf}
\usepackage{subfigure}
\usepackage{color}
\usepackage[normalem]{ulem}
\usepackage[utf8]{inputenc}
\usepackage[sort&compress]{natbib}

\pdfoutput=1

\begin{document}

\title{Single and double linear and nonlinear flatband chains: spectra and
modes}
\author{Krzysztof Zegadlo$^{1,2}$, Nir Dror$^3$, Nguyen Viet Hung$^4$, Marek
Trippenbach$^1$, and Boris A. Malomed $^{3,5}$}
\address{$^1$Faculty of Physics, University of Warsaw, ul. Pasteura 5,
02-093 Warszawa, Poland\\
$^2$Faculty of Physics, Astronomy and Applied Computer Science, Jagiellonian University, ul. Łojasiewicza  11, 30-348 Kraków, Poland\\
$^3$Department of Physical Electronics, School of Electrical Engineering,
Tel Aviv University, Tel Aviv 69978, Israel\\
$^4$Advanced Institute for Science and Technology, Hanoi University of
Science and Technology, Hanoi, Vietnam\\
$^5$Laboratory of Nonlinear-Optical Informatics, ITMO University, St.
Petersburg 197101, Russia}

\begin{abstract}
We report results of systematic analysis of various modes in the flatband
lattice, based on the diamond-chain model with the on-site cubic
nonlinearity, and its double version with the linear on-site mixing between
the two lattice fields. In the single-chain system, a full analysis is
presented, first, for the single nonlinear cell, making it possible to find
all stationary states, \textit{viz}., antisymmetric, symmetric, and
asymmetric ones, including an exactly investigated symmetry-breaking
bifurcation of the subcritical type. In the nonlinear infinite
single-component chain, compact localized states (CLSs) are found in an
exact form too, as an extension of known compact eigenstates of the linear
diamond chain. Their stability is studied by means of analytical and
numerical methods, revealing a nontrivial stability boundary. In addition to
the CLSs, various species of extended states and exponentially localized
lattice solitons of symmetric and asymmetric types are studied too, by means
of numerical calculations and variational approximation. As a result,
existence and stability areas are identified for these modes. Finally, the
linear version of the double diamond chain is solved in an exact form,
producing two split flatbands in the system's spectrum.
\end{abstract}

\pacs{03.75.Lm, 05.45.Yv, 42.82.Et, 63.22.+m}
\maketitle





\section{Introduction}

The behavior of dynamical lattices, which model a vast variety of physical
systems, is determined by the interplay of their linear-excitation spectra
and local nonlinearity \cite{Moti}-\cite{Jena}. An essential peculiarity of
many lattices is the presence of at least one \textit{flatband} (FB), i.e.,
a dispersionless branch in the spectrum \cite{richter}. Interest to FBs was
drawn by their discovery in the Hubbard \cite{Tasaki} and
Su-Schrieffer-Heeger models, where they underlie the existence of a stable
ferromagnetic phase. It was also demonstrated that the FB may cause
insulator-metal transitions in the underlying lattices \cite{transition}.
Further, quantum-Hall and topological-insulator states were predicted in FB
systems \cite{Hall}. Lattices can support FBs in diverse physical settings,
including arrayed optical waveguides \cite{diamond,guzman-silva14},
exciton-polariton condensates in semiconductor microcavities with lattice
patterns etched into them \cite{masumoto12,cavities}, and atomic
Bose-Einstein condensates (BECs) loaded into an appropriately designed
optical lattice \cite{ZhangZhang,taie15}.

A remarkable property of linear dynamical lattices featuring the FB spectrum
is that, in addition to the usual dispersive \textquotedblleft phonon"
excitations, they support exact eigenmodes in the form of compact localized
states (CLSs), which include a finite number of lattice sites with nonzero
amplitudes \cite{richter,flach14,add1}. In particular, these states may be
robust against the presence of disorder in the system \cite{disorder}. A
local resonant mechanism, which is similar to the Fano resonance \cite%
{miroshnichenko10}, may hybridize the CLS with the phonon modes, thus giving
rise to new varieties of lattice excitations \cite{flach14}. Experimentally,
the existence of CLSs has been demonstrated in the above-mentioned settings
which admit the realization of the FB spectra, \textit{viz}., optical
waveguiding arrays \cite{vicencio15,add2,add3,mukherjee15,weimann16},
exciton-polariton condensates \cite{cavities}, and atomic BECs \cite{taie15}.

Unlike the self-trapped discrete solitons in nonlinear lattices \cite{Moti}-%
\cite{Jena}, the localization of CLSs does not require the presence of
nonlinearity. On the other hand, nonlinearities, represented by the Kerr
term in optics and collisional term in BEC, are naturally present in
physical settings which admit the realization of FBs and CLSs. This fact
suggests to explore the existence and dynamics of CLSs in nonlinear
lattices, and their possible relations to exponentially localized (but not
compact) discrete solitons, which are generic modes in nonlinear lattices.
In recent works, it was demonstrated that the nonlinearity may produce
various effects in FB systems, such as stabilization or destabilization of
the compact states, detuning of their frequencies, interactions between
CLSs, and coexistence and interactions between CLSs and lattice solitons
\cite{vicencio13,add4,flach14,johannson15,lopez-gonzalez16,Sandra,Maimistov}.

The purpose of this paper is two-fold. First, we aim to develop a nonlinear
version of the FB system based on the known configuration in the form of the
\textquotedblleft diamond chain" \cite{flach14} (see Fig. \ref{fig:diamond}%
), by adding the on-site cubic nonlinearity to it. Working in this
direction, we aim to construct CLSs and usual (exponentially localized, but
not compact) discrete solitons in this chain, and analyze their stability.
Results obtained for the CLSs demonstrate specific extension of this type of
modes into the realm of the nonlinearity: they keep to exists as compact
states, featuring a nontrivial stability boundary inside their family, which
is a manifestation of the nonlinearity. Stability boundaries for families of
exponentially localized discrete solitons are found too. Second, we
introduce a double diamond chain, with on-site linear coupling between two
components. It can also be readily implemented in optics, considering arrays
of double-core waveguides \cite{Christo}, and in BEC, as a binary condensate
whose components are Rabi-coupled by a resonant electromagnetic field \cite%
{binary}.

Analytical results for the nonlinear one-chain model are reported in Section
II. They present a full exact solution for the single nonlinear diamond cell
(rather than in the chain), including both symmetric and asymmetric states,
and an exact solution for the CLS in the infinite nonlinear chain, including
a partial analysis of its stability. Numerical findings for the single
nonlinear chain are presented in Section III. These include extended modes,
which may be considered as fragments of continuous-wave (CW) states,
symmetric and asymmetric exponentially localized lattice solitons, and the
full analysis of the stability of the nonlinear CLS. Further, in Section V
we present analytical results for the lattice solitons in the nonlinear
chain, obtained by means of the variational approximation (VA), which
compare well to their numerical counterparts. Finally, section VI concludes
the paper.

\section{The single-component model: analytical results}

\subsection{The formulation}

\begin{figure}[h]
\centering
\subfigure{\includegraphics[width=0.45\columnwidth]{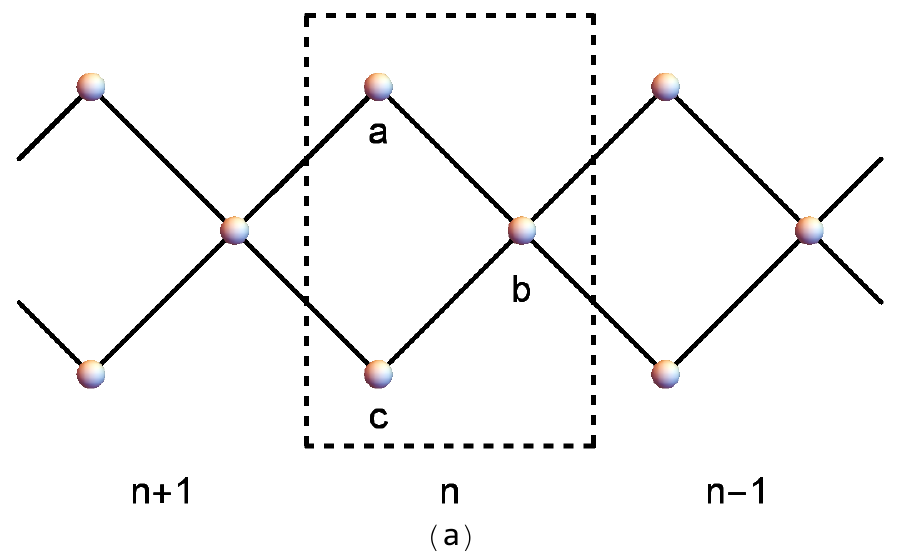}} %
\subfigure{\includegraphics[width=0.45\columnwidth]{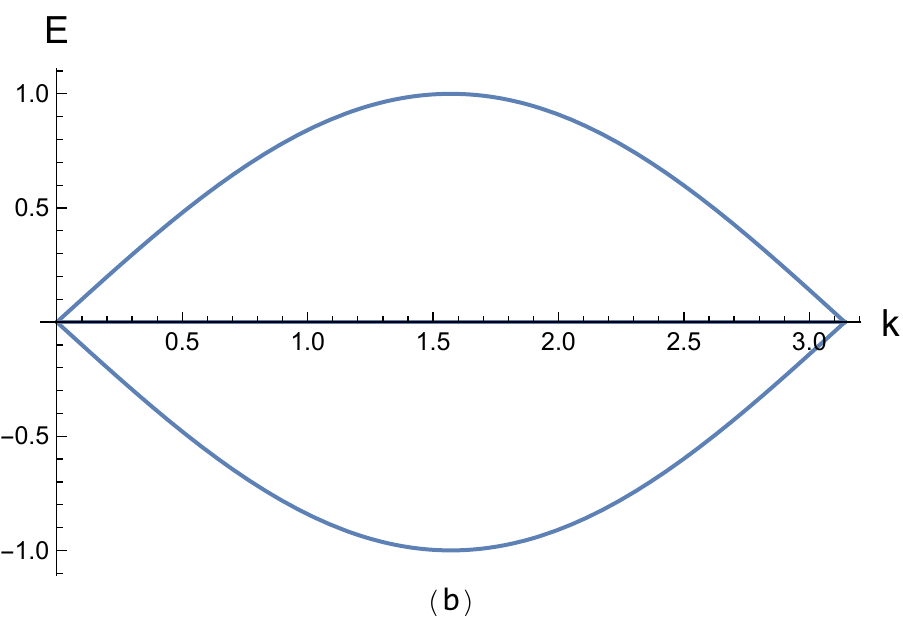}}
\caption{(a) The single-component diamond chain. (b) The respective
dispersion relation.}
\label{fig:diamond}
\end{figure}

We consider the quasi-1D lattice in the form of the \textquotedblleft
diamond chain" with the on-site cubic nonlinearity, which is shown in Fig. %
\ref{fig:diamond}(a). The band structure of the chain's linear version,
plotted in Fig. \ref{fig:diamond}(b), is produced by Eq. (\ref{spectrum}),
see below. The spectrum contains two dispersive bands which intersect with
the third, flat band, at edges of the Brillouin zone. The intersections
create Dirac points with conical dispersion around them.

The dynamics of the diamond lattice is governed by equations for complex
amplitudes at sites labeled $a$, $b$, and $c$ in Fig. \ref{fig:diamond}(a):
\begin{align}
& i\frac{da_{n}}{dz}+(b_{n}+b_{n+1})+\beta |a_{n}|^{2}a_{n}=0,  \notag \\
& i\frac{db_{n}}{dz}+(a_{n}+a_{n-1}+c_{n}+c_{n-1})+\beta |b_{n}|^{2}b_{n}=0,
\label{abc1} \\
& i\frac{dc_{n}}{dz}+(b_{n}+b_{n+1})+\beta |c_{n}|^{2}c_{n}=0.  \notag
\end{align}%
Here evolution variable $z$ is the propagation distance, if the lattice is
realized as an array of optical waveguides \cite%
{diamond,guzman-silva14,vicencio15,add2,add3,mukherjee15,weimann16}, $n$ is
the cell's number in the chain, the inter-site coupling constant is scaled
to be $1$, and $\beta $ is the strength of the on-site nonlinearity. The
propagation equations can be derived from the corresponding Hamiltonian,
\begin{gather}
\mathcal{H}=\sum_{n}\big(\lbrack (a_{n}^{\ast }+c_{n}^{\ast
})(b_{n}+b_{n+1})+\mathrm{c.c.}]  \notag \\
+\frac{\beta }{2}(|a_{n}|^{4}+|b_{n}|^{4}+|c_{n}|^{4})]\big),  \label{Ham}
\end{gather}%
where both the asterisk and $\mathrm{c.c.}$ stand for the complex-conjugate
expression. The Hamiltonian is a dynamical invariant of system (\ref{abc1}),
along with the total norm,%
\begin{equation}
N=\sum_{n}(|a_{n}|^{2}+|b_{n}|^{2}+|c_{n}|^{2}).  \label{N=1}
\end{equation}%
Here we assume that $\beta $ is positive. Then, using the scaling invariance
of the system, one can fix $\beta =1$, while $N$ will play the role of a
parameter of families of stationary states (in the analytical part of the
work, we do not fix $\beta =1$, since keeping $\beta >0$ as a free
coefficient does not make analytical results cumbersome). In Eqs. (\ref{abc}%
) with $\beta <0$, the sign of the nonlinearity coefficient can be reversed
by changing $b_{n}\rightarrow -b_{n}$ and replacing the equations by their
complex-conjugate version.

Eigenmodes of {system (\ref{abc1})} with real propagation constant $E$ are
looked for as
\begin{equation}
\{{a}_{n}(z),{b}_{n}(z),{c}_{n}(z)\}=\left\{ A_{n},B_{n},C_{n}\right\}
e^{iEz},  \label{aA}
\end{equation}%
with stationary amplitudes $\left\{ A_{n},B_{n},C_{n}\right\} $ satisfying
the following equations:%
\begin{align}
-EA_{n}+(B_{n}+B_{n+1})+\beta |A_{n}|^{2}A_{n}& =0,  \label{A_n} \\
-EB_{n}+(A_{n}+A_{n-1}+C_{n}+C_{n-1})+\beta |B_{n}|^{2}B_{n}& =0,
\label{B_n} \\
-EC_{n}+(B_{n}+B_{n+1})+\beta |C_{n}|^{2}C_{n}& =0.  \label{C_n}
\end{align}%
The spectrum of the linearized version of (\ref{A_n})-(\ref{C_n}) contains
three branches,
\begin{equation}
E(k)=0,~\pm 2\sqrt{2}\cos (k/2),  \label{spectrum}
\end{equation}%
which are shown above in Fig.~\ref{fig:diamond}(b). Obviously, $E(k)=0$
represents the FB. The branches were derived by substituting, in the
linearized equations, the continuous-wave (CW) ansatz, $\left\{
A_{n},B_{n},C_{n}\right\} =\left\{ A,B,C\right\} e^{ikn}$, with real
wavenumber $k$ which takes values in the first Brillouin zone, $-\pi \leq
k\leq +\pi $. The eigenmodes corresponding to the flat (``0") and dispersive
(``$\pm$") branches (\ref{spectrum}) amount, respectively, to the following
sets of constant amplitudes:
\begin{equation}
\left\{ A,B,C\right\}_{0}=\frac{1}{\sqrt{2}}(1,0,-1),\,\left\{ A,B,C\right\}
_{\pm }=\frac{1}{2}(1,\pm \sqrt{2}e^{-ik/2},1).  \label{branches}
\end{equation}

The CLSs are produced by the exact solution in the form of Eq. (\ref{aA})
with $E=0$ (and zero group velocity), which is completely localized in a
single unit cell, centered at an arbitrary position, $n=N_{0}$:
\begin{equation}
\left\{ A_{n},B_{n},C_{n}\right\} =\frac{1}{\sqrt{2}}\delta
_{n,N_{0}}(1,0,-1),  \label{eqn:flatstate}
\end{equation}%
where $\delta _{m,n}$ is the Kronecker's symbol \cite{richter,flach14}. This
solution exists due to the destructive interference (cancellation of the
field) at site $b$ [see Fig. \ref{fig:diamond}(a)], which is provided by the
$\pi $ phase difference between the $a$ and $c$ sites. In contrast,
eigenmodes (\ref{branches})\ corresponding to the dispersive bands are
completely delocalized plane waves. Because CLSs are degenerate modes in the
linear system, with respect to their placement in the lattice, their
arbitrary combinations are exact solutions too \cite{vicencio15,add1,add2}.
In this way, one can easily construct extended states (fragments of CWs)
localized on several adjacent cells of the lattice.

As mentioned above, experimental realization of flat bands and CLSs was
reported in Refs. \cite{vicencio15,mukherjee15,weimann16}, while
counterparts of CLSs in nonlinear lattices were discussed in Refs. \cite%
{vicencio13,flach14,johannson15,lopez-gonzalez16,Sandra,add4}. In this work,
we aim to develop the analysis of CLSs and lattice solitons coexisting with
them in the nonlinear diamond-chain system.

\subsection{Reduced problem: one cell of the single nonlinear chain}

\subsubsection{The setting}

\begin{figure}[th]
\centering
\includegraphics[scale=0.2]{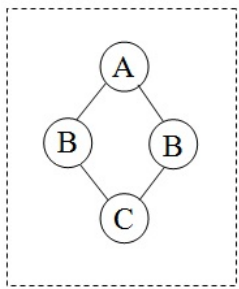}
\caption{The truncated single-cell nonlinear system.}
\label{one cell system}
\end{figure}

We begin the analysis of the nonlinear diamond chain by consideration of the
simplest case of the system truncated to a single cell, which is drawn in
Fig. \ref{one cell system}. It contains single \textquotedblleft A" and
\textquotedblleft C" sites, which communicate to each other via two
\textquotedblleft B" sites (the structure of the system implies that two
fields at the B sited are identical).

The truncated system is based on the following system of equations, in which
the on-site nonlinearity coefficient is scaled to be $\beta \equiv 1$ (if $%
\beta $ is negative, its sign can be inverted by means of a simple
transformation, $b\rightarrow -b,$ $z\rightarrow -z$):%
\begin{gather}
i\frac{da}{dz}+2b+|a|^{2}a=0,  \label{a} \\
i\frac{dc}{dz}+2b+|c|^{2}c=0.  \label{c} \\
i\frac{db}{dz}+a+c+|b|^{2}b=0,  \label{b}
\end{gather}%
This system with three degrees of freedom conserves two dynamical
invariants: the Hamiltonian,%
\begin{equation}
H=\frac{1}{2}\left( |a|^{4}+|c|^{4}+2|b|^{4}\right) +2[b(a^{\ast }+c^{\ast
})+b^{\ast }(a+c)]  \label{H}
\end{equation}%
cf. Eq. (\ref{Ham}), and the norm,
\begin{equation}
N=|a|^{2}+|c|^{2}+2|b|^{2},  \label{N}
\end{equation}%
cf. Eq. (\ref{N=1}). The Hamiltonian representation of Eqs. (\ref{a})-(\ref%
{b}) (i.e., the corresponding Poisson brackets/symplectic structure) is
based on equations%
\begin{equation}
\frac{da}{dz}=i\frac{\partial H}{\partial a^{\ast }},~\frac{dc}{dz}=i\frac{%
\partial H}{\partial c^{\ast }},~\frac{db}{dz}=\frac{i}{2}\frac{\partial H}{%
\partial b^{\ast }}.  \label{sympl}
\end{equation}

Although the present model with three degrees of freedom seems very simple,
to the best of our knowledge it was not explored before. Below, we report a
full analysis of exact stationary solutions of this system and their
stability, which can be easily realized in the experiment.

\subsubsection{Antisymmetric and symmetric stationary solutions}

Stationary solutions to Eqs. (\ref{a})-(\ref{b}) with a real propagation
constant, $E$, are looked for as
\begin{equation}
\{{a}(z),{b}(z),{c}(z)\}=\left\{ A,B,C\right\} e^{iEz},  \label{abc}
\end{equation}%
cf. Eq. (\ref{aA}), with amplitudes $\left\{ A,B,C\right\} $ determined by
the algebraic equations:
\begin{subequations}
\begin{eqnarray}
-EA+2B+A^{3} &=&0,  \label{A} \\
-EC+2B+C^{3} &=&0,  \label{C} \\
-EB+A+C+B^{3} &=&0.  \label{B}
\end{eqnarray}%
These equations have an obvious \emph{antisymmetric} solution,
\end{subequations}
\begin{equation}
B=0,~A=-C=\sqrt{E},  \label{anti}
\end{equation}%
which exists at $E\geq 0$ (all the solutions can be defined as those with $%
A>0$; obviously, there is also a solution with opposite signs in front of
all the amplitudes).

There are also \emph{symmetric} solutions, with $A=C$, $B\neq 0$. Four
different solutions of this type can be identified. The first is%
\begin{equation}
A=C=B=\sqrt{E-2},  \label{E-2}
\end{equation}%
which exists at $E\geq 2$, and there is another solution, with%
\begin{equation}
A=C=-B=\sqrt{E+2},  \label{E+2}
\end{equation}%
which exists at $E\geq -2$. Finally, there are two additional symmetric
solutions, which correspond to signs $\pm $ in Eq. (\ref{pm}), and exist at $%
E\geq 4$:%
\begin{equation}
A=C=\sqrt{\frac{1}{2}\left( E\pm \sqrt{E^{2}-16}\right) },~B=\frac{2}{A}%
\equiv \sqrt{\frac{1}{2}\left( E\mp \sqrt{E^{2}-16}\right) }.  \label{pm}
\end{equation}

\subsubsection{Bifurcation points}

The symmetric solutions undergo a \emph{symmetry-breaking bifurcation}
(SBB), which gives rise to nontrivial \emph{asymmetric solutions}, with $%
A\neq C$. The analysis of the transition from symmetric states to asymmetric
ones is an issue of general interest \cite{book0,book}, including the
present model. Bifurcation points (there are\emph{\ }two such critical
points) can be found analytically. To this end, replacing Eqs. (\ref{A}) and
(\ref{C}) by their sum and difference, and canceling in the latter one a
common factor, $A-C\neq 0$, the system of algebraic equations (\ref{A})-(\ref%
{B}) is replaced by the following one:
\begin{subequations}
\begin{gather}
E=A^{2}+C^{2}+AC,  \label{A-C} \\
-E\left( A+C\right) +4B+A^{3}+C^{3}=0,  \label{A+C} \\
-EB+A+C+B^{3}=0.  \label{BB}
\end{gather}%
Exactly at the SBB point, Eqs. (\ref{A-C})-(\ref{A+C}) must have a solution
with $A=C$. It is easy to find that this is possible at two points (as
mentioned above):
\end{subequations}
\begin{equation}
E=3,~A=C=B=1,  \label{first}
\end{equation}%
and%
\begin{equation}
E=3\sqrt{2},~A=C=2^{1/4},~B=2^{3/4}.  \label{second}
\end{equation}%
Obviously, the bifurcation point (\ref{first}) pertains to symmetric
solution (\ref{E-2}), while point (\ref{second}) pertains to symmetric
solution (\ref{pm}) with sign minus chosen for $\pm $.

The similar analysis for the antisymmetric solution readily demonstrates
that it never undergoes an antisymmetry-breaking bifurcation (formally, the
bifurcation occurs at an unphysical point, with $E^{2}=-2$).

\subsubsection{The analysis of the bifurcations}

To identify the character of the SBB at point (\ref{first}), we consider
solutions of Eqs. (\ref{A-C})-(\ref{BB}) in an infinitesimal vicinity of the
bifurcation, setting%
\begin{eqnarray}
E &=&3-\epsilon ,~0<\epsilon \ll 1,  \label{eps} \\
A &=&1-\alpha \epsilon +\gamma \sqrt{\epsilon },  \label{Aalpha} \\
C &=&1-\alpha \epsilon -\gamma \sqrt{\epsilon },  \label{Calpah} \\
B &=&1-\beta \epsilon .  \label{beta}
\end{eqnarray}%
In Eq. (\ref{eps}), term $\sim \epsilon $ is defined with sign minus in
front of it in the anticipation of the fact that the SBB will be \emph{%
subcritical} \cite{book0} at this point. The substitution of Eqs. (\ref{eps}%
)-(\ref{beta}) into Eqs. (\ref{A-C})-(\ref{BB}) and expanding the result up
to order $\epsilon $ easily yields the following results:%
\begin{eqnarray}
\gamma ^{2} &=&2,  \label{gamma} \\
\alpha &=&1/2,~\beta =7/2  \label{ab}
\end{eqnarray}%
The conclusion is that the SBB at point (\ref{first}) is indeed \emph{%
subcritical}: the asymmetric solutions originally move \emph{backward} (in
the direction of decreasing $E$), as unstable ones, after emerging at the
bifurcation point. Later, they turn forward, passing the respective turning
point, where they undergo stabilization \cite{book0}.

Furthermore, it is relevant to check if Eqs. (\ref{A-C})-(\ref{BB}) would
admit a \emph{regular extension} of the solutions from point (\ref{first})
(i.e., built in terms of $\epsilon $, rather than $\sqrt{\epsilon }$),
instead of the bifurcation. This implies looking for a solution in the form
of the following expansion, instead of Eqs. (\ref{Aalpha})-(\ref{beta}):%
\begin{equation}
A=1-a\epsilon ,~C=1-c\epsilon ,~B=1-b\epsilon .  \label{ABC}
\end{equation}%
Then, the substitution of this into Eq. (\ref{A-C}) yields $a+c=1/3$, while
the substitution into Eq. (\ref{BB}) yields $a+c=1$. This contradiction
implies that regular expansion (\ref{ABC}) cannot produce a solution, the
bifurcation being the only possibility.

Similarly, in a vicinity of bifurcation point (\ref{second}) we seek for a
solution to Eqs. (\ref{A-C})-(\ref{BB}) in the form of%
\begin{eqnarray}
E &=&3\sqrt{2}-\epsilon ,~|\epsilon |\ll 1,  \label{eps2} \\
A &=&2^{1/4}-\alpha \epsilon +\gamma \sqrt{\epsilon },  \label{Aalpha2} \\
C &=&2^{1/4}-\alpha \epsilon -\gamma \sqrt{\epsilon },  \label{Calpha2} \\
B &=&2^{3/4}-\beta \epsilon .  \label{beta2}
\end{eqnarray}%
The same procedure as the one outlined above for the SBB at point (\ref%
{first}) yields $\gamma ^{2}=-1/7,$ $\alpha =(1/7)\cdot 2^{-1/4},$ $\beta
=(4/7)\cdot 2^{-3/4}$. The formal result with $\gamma ^{2}<0$ means that the
bifurcation at point (\ref{second}) actually \emph{does not} take place.
Thus, the actual SBB takes place solely at point (\ref{first}).

\subsubsection{Asymmetric solutions}

The asymmetric solutions produced by the SBB\ can be easily found in the
asymptotic form at $E\rightarrow +\infty $ directly from Eqs. (\ref{A})-(\ref%
{C}):
\begin{equation}
A\approx \sqrt{E},~B\approx A/E\approx 1/\sqrt{E},~C\approx 2B/E\approx
2/E^{3/2}.  \label{approx}
\end{equation}%
The existence of the single solution in the asymptotic form (\ref{approx})
agrees with the conclusion made in the previous subsection, where it was
found that only one bifurcation point, given \ by Eq. (\ref{first}), is a
real one. In the general case, the asymmetry \ of the solutions is defined by%
\begin{equation}
\Theta \equiv \frac{A^{2}-C^{2}}{A^{2}+C^{2}}.  \label{Theta}
\end{equation}

\begin{figure}[th]
\centering
\includegraphics[scale=0.3]{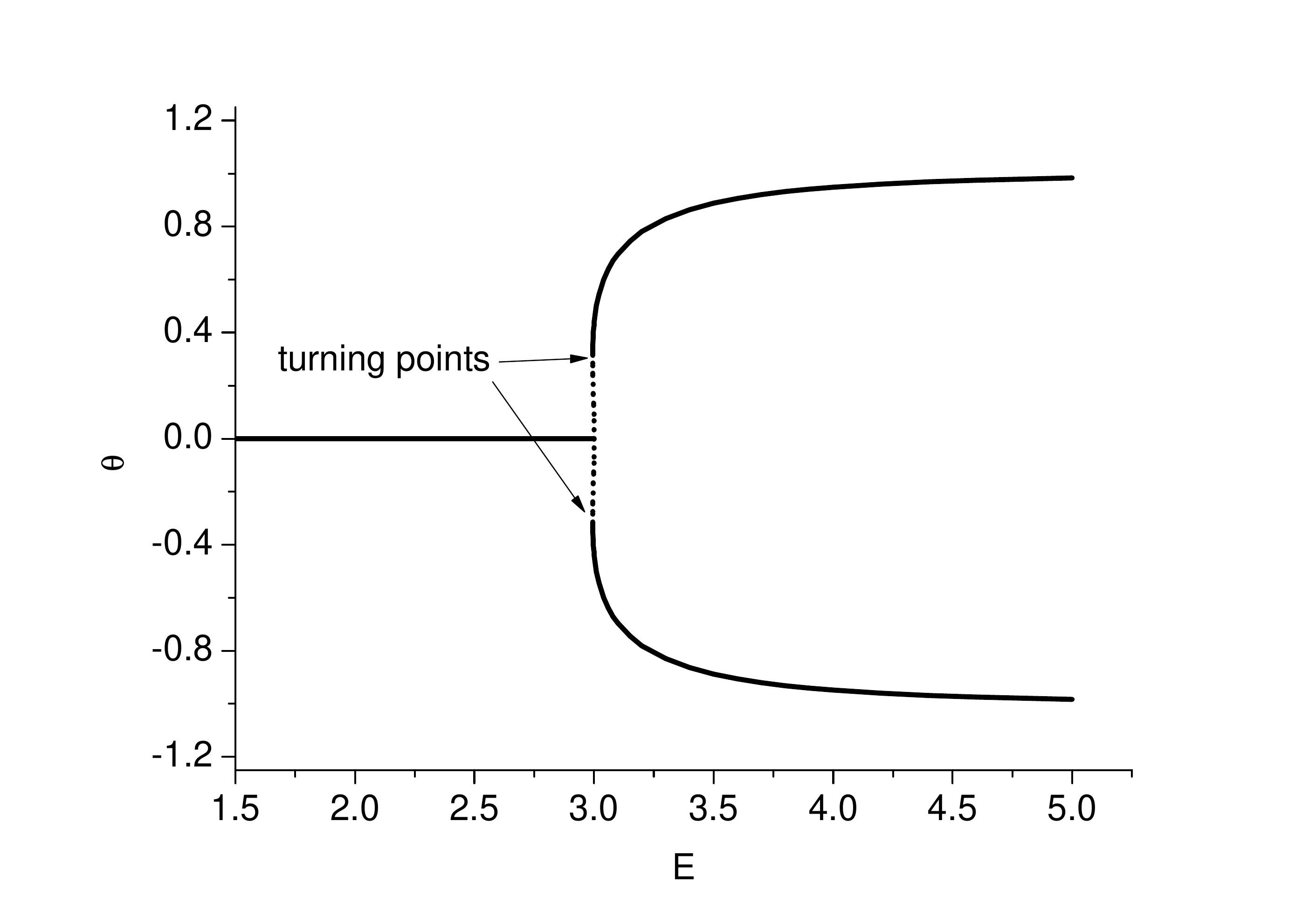}
\caption{The bifurcation diagram for the weakly subcritical transition from
symmetric to asymmetric stationary states in the nonlinear single-cell
system. The asymmetry parameter, defined as per Eq. (\protect\ref{Theta}),
is shown versus propagation constant $E$ of the stationary solutions. Stable
and unstable branches of asymmetric solutions are shown by solid and dashed
curves, respectively.}
\label{one cell bifurcation diagram}
\end{figure}

Next, we aim to address asymmetric solutions emerging from the point of the
subcritical SBB. The system of three coupled cubic algebraic equations (\ref%
{A}), (\ref{B}) and (\ref{C}) has 27 roots, which makes it impossible to
present them in an explicit analytical form. Most roots are complex, hence
they are unphysical. One root remains real at $E>3$, i.e., above the SBB
point (\ref{first}). This solution describes the stable part of the
asymmetric-solution branch where it goes forward, after passing the turning
point (see the continuous lines in Fig. \ref{one cell bifurcation diagram}).
On the other hand, there also exists another solution which remains real in
a very narrow range: $2.994\lesssim E\leq 3$. It represents the initial
short backward-going unstable segment of the asymmetric branch, which is
shown by dashed lines in Fig. \ref{one cell bifurcation diagram}.

\subsection{The single infinite chain: nonlinear compact localized states
(CLSs) and their stability}

Some simple but essential analytical results can be obtained also for the
infinite diamond chain based on nonlinear equations (\ref{abc1}) and (\ref%
{A_n}), (\ref{C_n}), (\ref{B_n}) (here, we keep the nonlinearity coefficient
$\beta $ as a free parameter). It admits an obvious solution for $E>0$,
which is a nonlinear extension of the CLS (\ref{eqn:flatstate}) obtained for
$E=0$ in the linear lattice:%
\begin{equation}
A_{0}=-C_{0}=\sqrt{E/\beta },~\mathrm{all~~other}~A_{n},C_{n},B_{n}=0
\label{CLS}
\end{equation}%
[cf. an equivalent single-cell state given by Eq. (\ref{anti})]. It is
relevant to consider, in an analytical form, the stability of this exact
antisymmetric solution (with $A_{0}=-C_{0}$) against antisymmetry-breaking
perturbations, or, in other words, a possibility of a bifurcation breaking
the antisymmetric form of this state (recall that such a bifurcation does
not occur in the single-cell system, as shown above).

To this end, we consider a small perturbation in an originally empty
semi-infinite lattice, starting from cite $b_{0}$ at $n=0$, which is
followed by sites $a_{0}$ and $c_{0}$ and then by the lattice at sites with $%
n\geq 1$. If the respective field is $B$ at $n=0$, the solution of
linearized equations Eqs. (\ref{A_n}), (\ref{C_n}), (\ref{B_n}) at $n\geq 1$%
, which is localized, exponentially decaying at $n\rightarrow \infty $, can
be easily found:

\begin{eqnarray}
B_{n} &=&Be^{-\lambda n},~A_{n}=C_{n}=\alpha Be^{-\lambda n},  \label{Bn} \\
\alpha &=&\frac{1}{4}\left( E-\sqrt{E^{2}-8}\right) ,  \label{alpha} \\
e^{-\lambda } &=&\frac{4}{E^{2}-4+E\sqrt{E^{2}-8}}<1.  \label{lambda}
\end{eqnarray}%
This solution exists for $E>\sqrt{8}$.

Then, taking into account the coupling of the two sites $b_{0}$ to the
semi-infinite chains originating from them, and to amplitudes $A$ and $C$ at
adjacent sites belonging to the cell at $n=0$, Eqs. (\ref{B_n}) and (\ref%
{A+C}) yield
\begin{eqnarray}
\left( A+C\right) +2\alpha B+B^{3} &=&EB.  \label{EB} \\
\left( A^{2}+C^{2}-AC-E\right) \left( A+C\right) &=&-4B  \label{(A+C)}
\end{eqnarray}%
Finally, the linearization of Eqs. (\ref{EB}) and (\ref{(A+C)}) for small
antisymmetry-breaking perturbations, $\left( A+C\right) $ and $B$, leads to
the condition for the onset of the respective instability:%
\begin{equation}
E^{2}+E\sqrt{E^{2}-8}+4=0.  \label{E}
\end{equation}%
An elementary consideration of Eq. (\ref{E})$\allowbreak $ demonstrates that
its only solution is an unphysical one, $E^{2}=-1$ (recall in the
single-cell model the analysis has predicted the formal
antisymmetry-breaking bifurcation at another unphysical point, $E^{2}=-2$).
Thus, the nonlinear CLS\ (\ref{CLS}) is stable against the
antisymmetry-breaking perturbations. However, in a part of their parameter
space these modes may be destabilized by other perturbations, as shown below.

\section{The nonlinear infinite single chain: numerical results}

\label{sec:NumericalResults} All the stationary solutions described below
were constructed by means of the imaginary-time method, or applying the
Newton-Raphson method for the corresponding nonlinear boundary-value
problem. The stability of the solutions was identified by the analysis of
linearized equations for small perturbations, and using the linear
Cranck-Nicholson scheme for the calculation of the respective eigenvalues.
The so predicted stability/instability was then verified through direct
simulations of the propagation of initially perturbed modes, utilizing the
Crank-Nicholson finite-difference algorithm. In plots presented below,
stable and unstable solutions are indicated by continuous and dashed curves,
respectively. All the results reported below refer to $\beta =1$, fixed by
means of rescaling.

\subsection{Continuous-wave (CW) solutions}

\label{sec:CWSolutions} 
\begin{figure}[th]
\centering%
\subfigure[]{\includegraphics[scale=0.40]{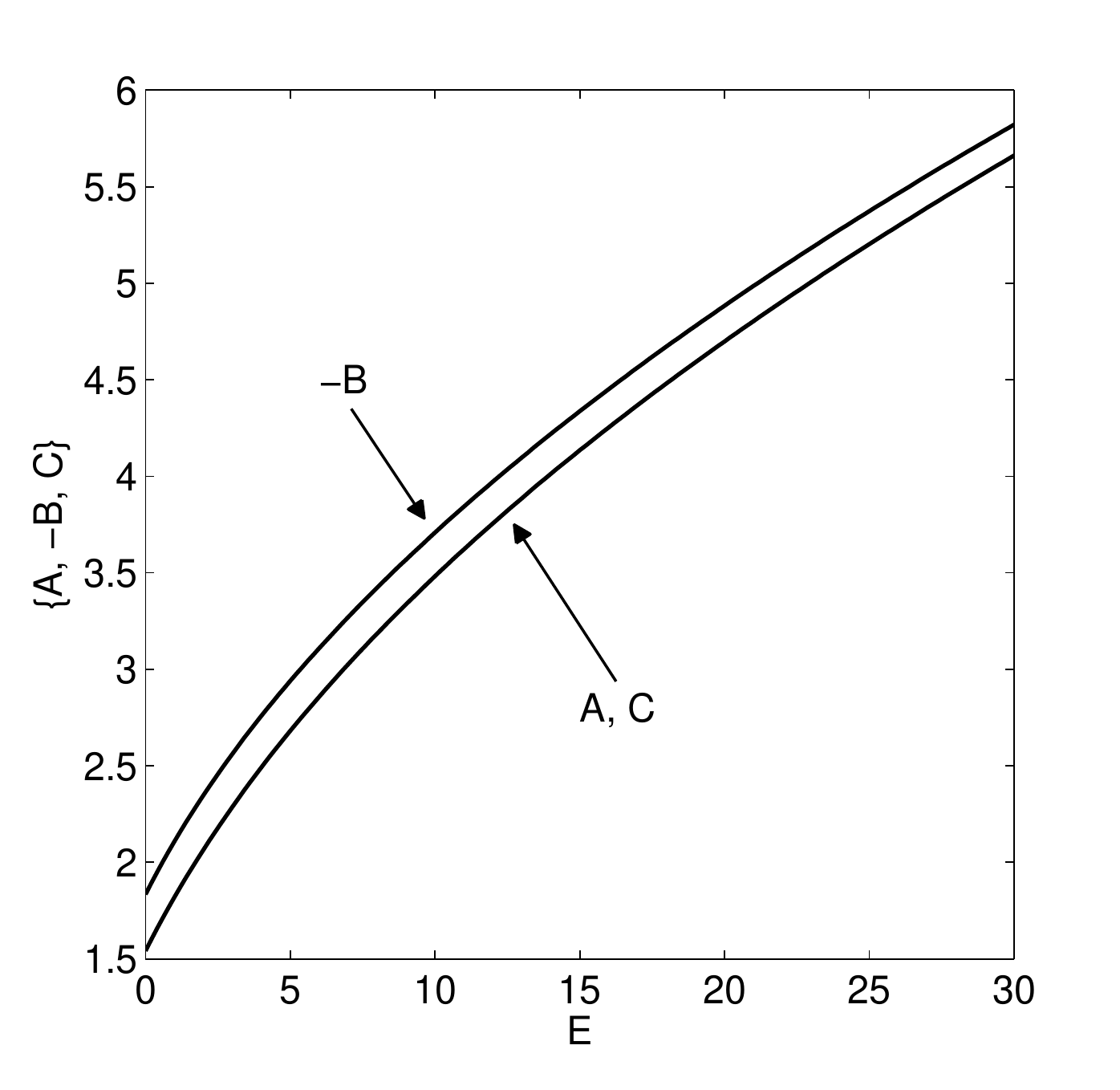}
\label{SymmetricCW1}} \subfigure[]{%
\includegraphics[scale=0.40]{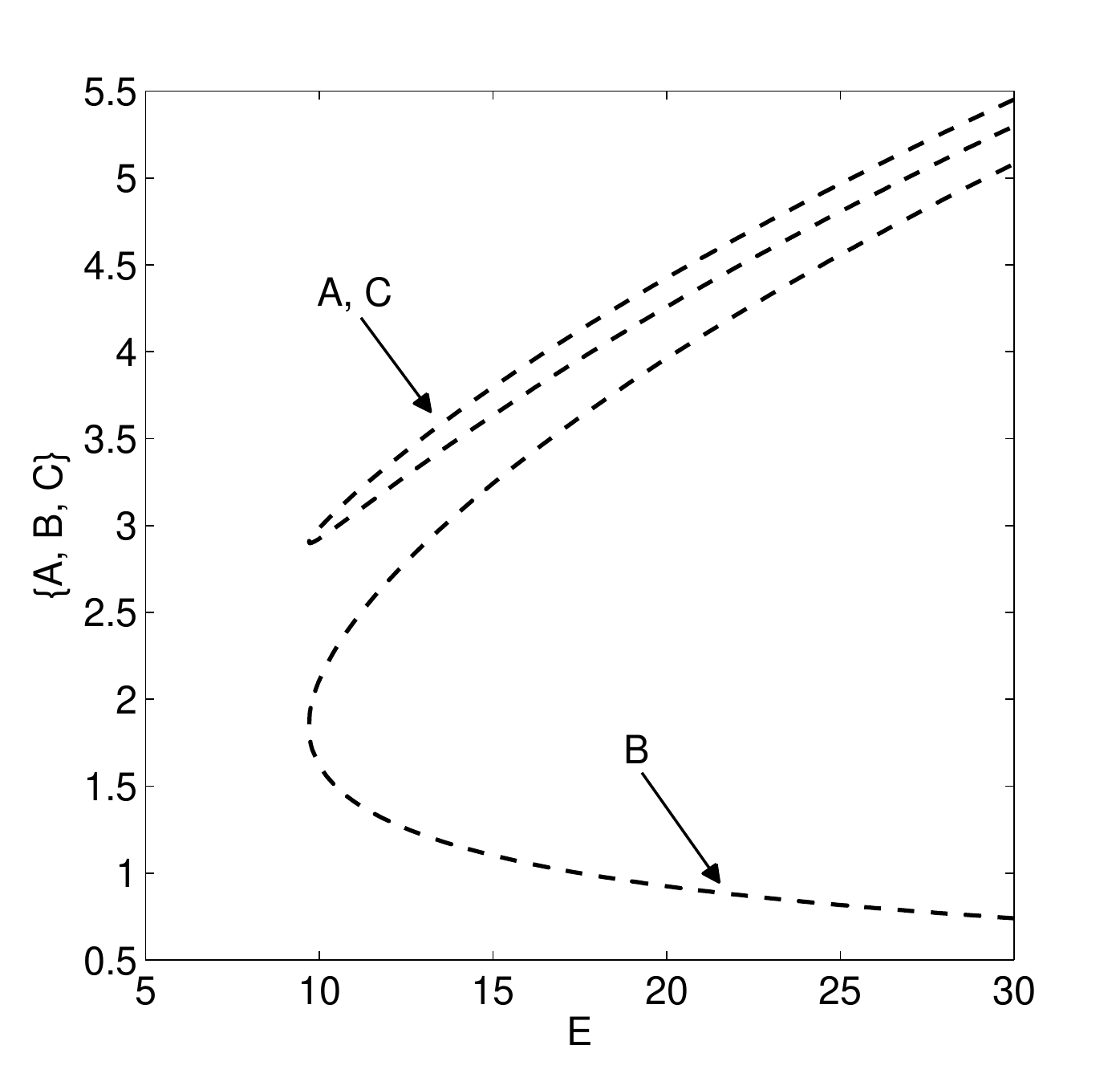}\label{SymmetricCW2}} %
\subfigure[]{\includegraphics[scale=0.40]{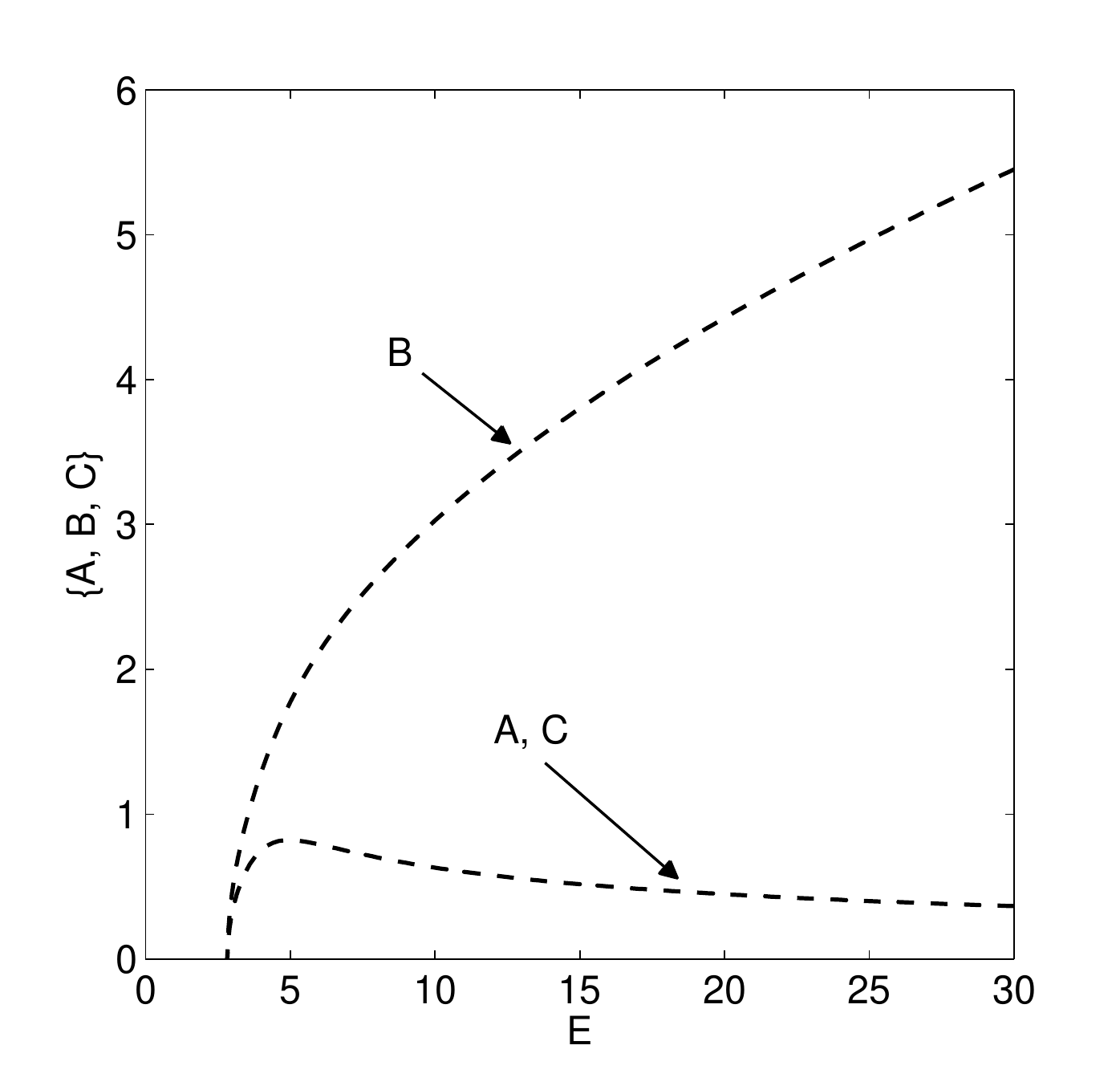}%
\label{SymmetricCW3}}
\caption{Amplitudes of symmetric CW states versus $E$. Panel (a) displays
the stable CW family, whereas two unstable families are shown in panels (b)
and (c). Stable and unstable solutions are shown by continuous and dashed
lines, respectively.}
\label{SymmetricCW}
\end{figure}

\begin{figure}[th]
\centering%
\subfigure[]{\includegraphics[scale=0.50]{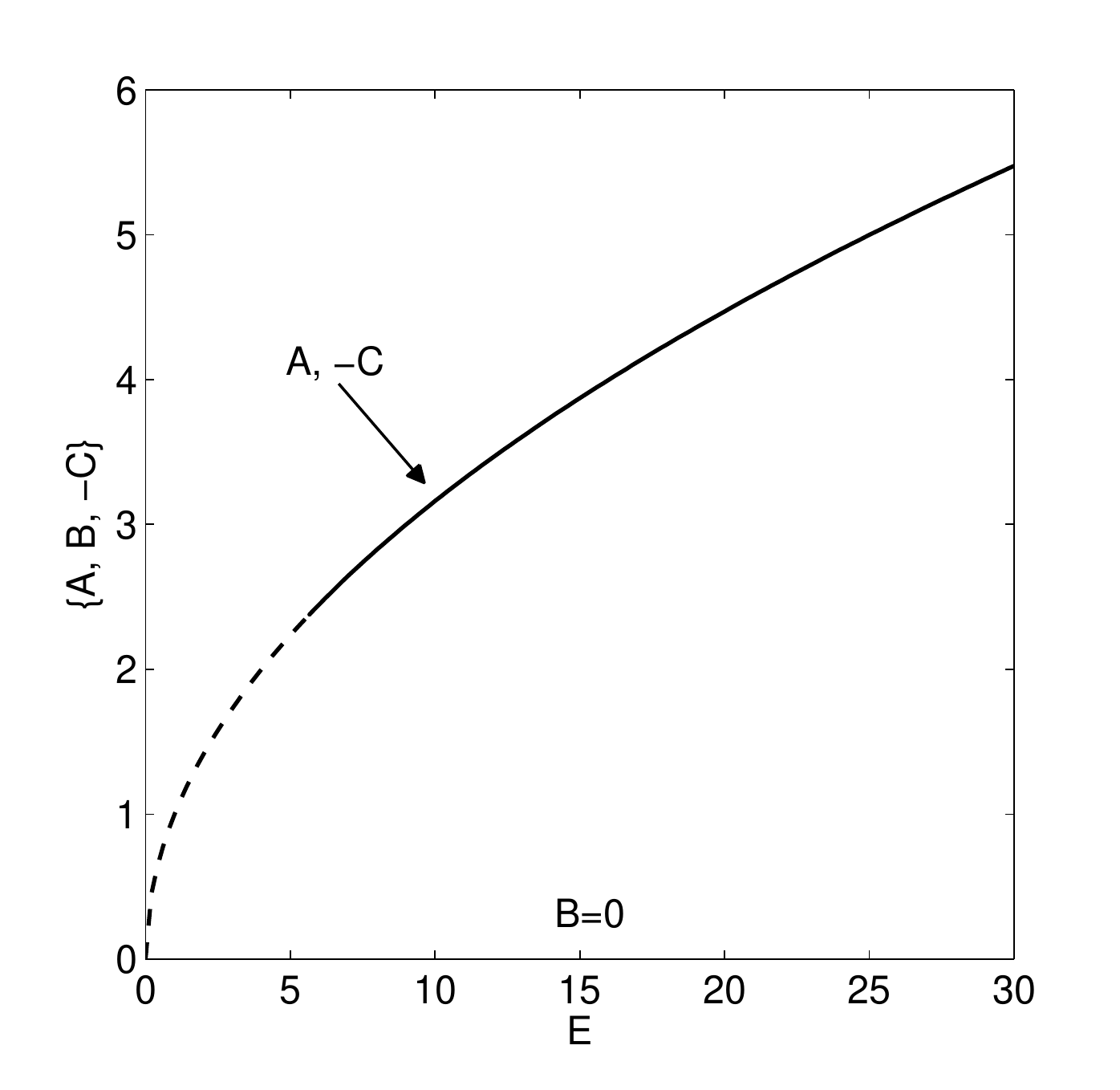}
\label{AntiSymmetricCW1}}
\caption{Amplitudes {$A=-C$ and $B$} of the antisymmetric CW as functions of
$E$.}
\label{AntiSymmetricCW}
\end{figure}

\begin{figure}[th]
\centering%
\subfigure[]{\includegraphics[scale=0.50]{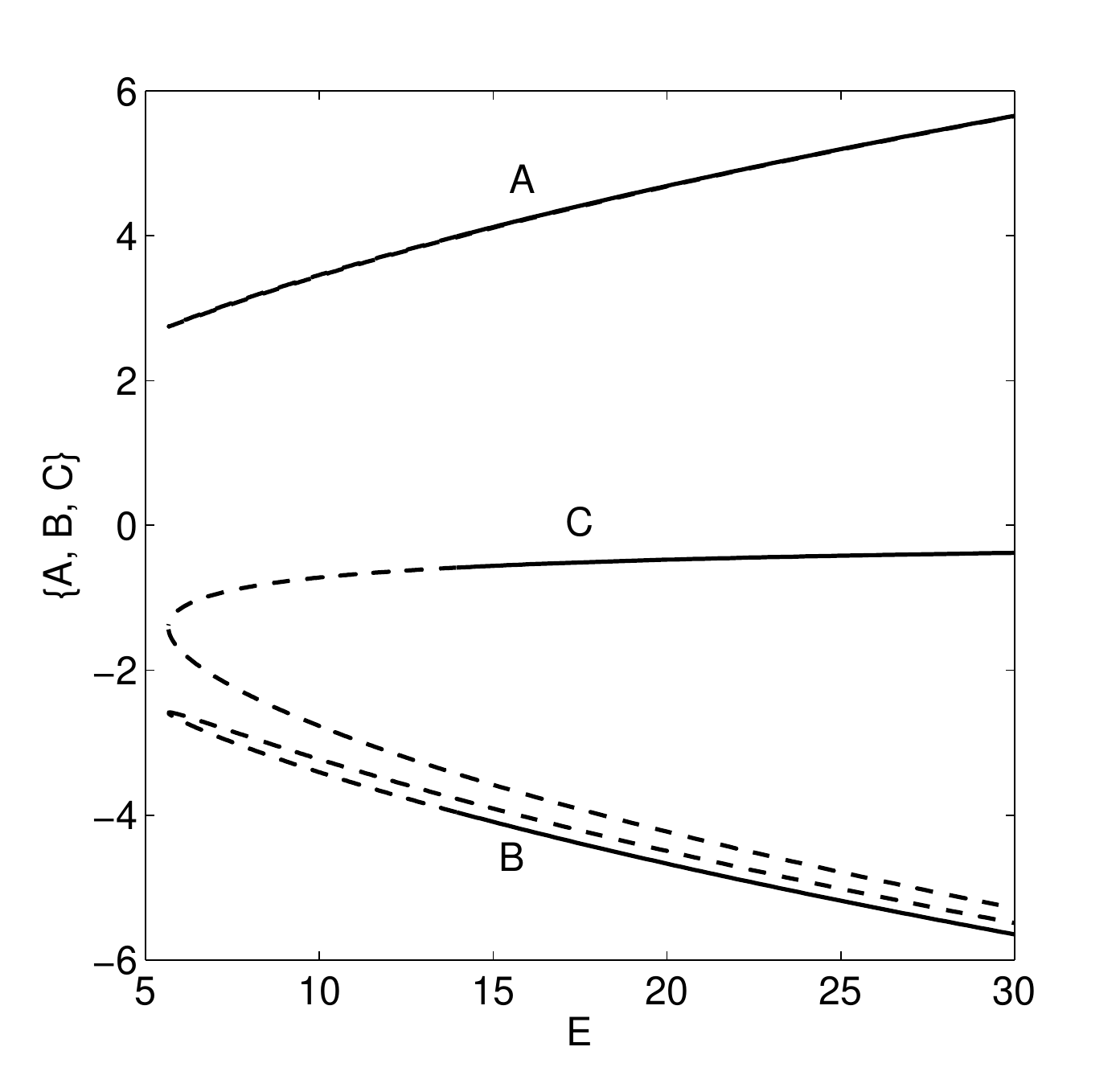}
\label{AntiSymmetricAsymCW1}} \subfigure[]{%
\includegraphics[scale=0.50]{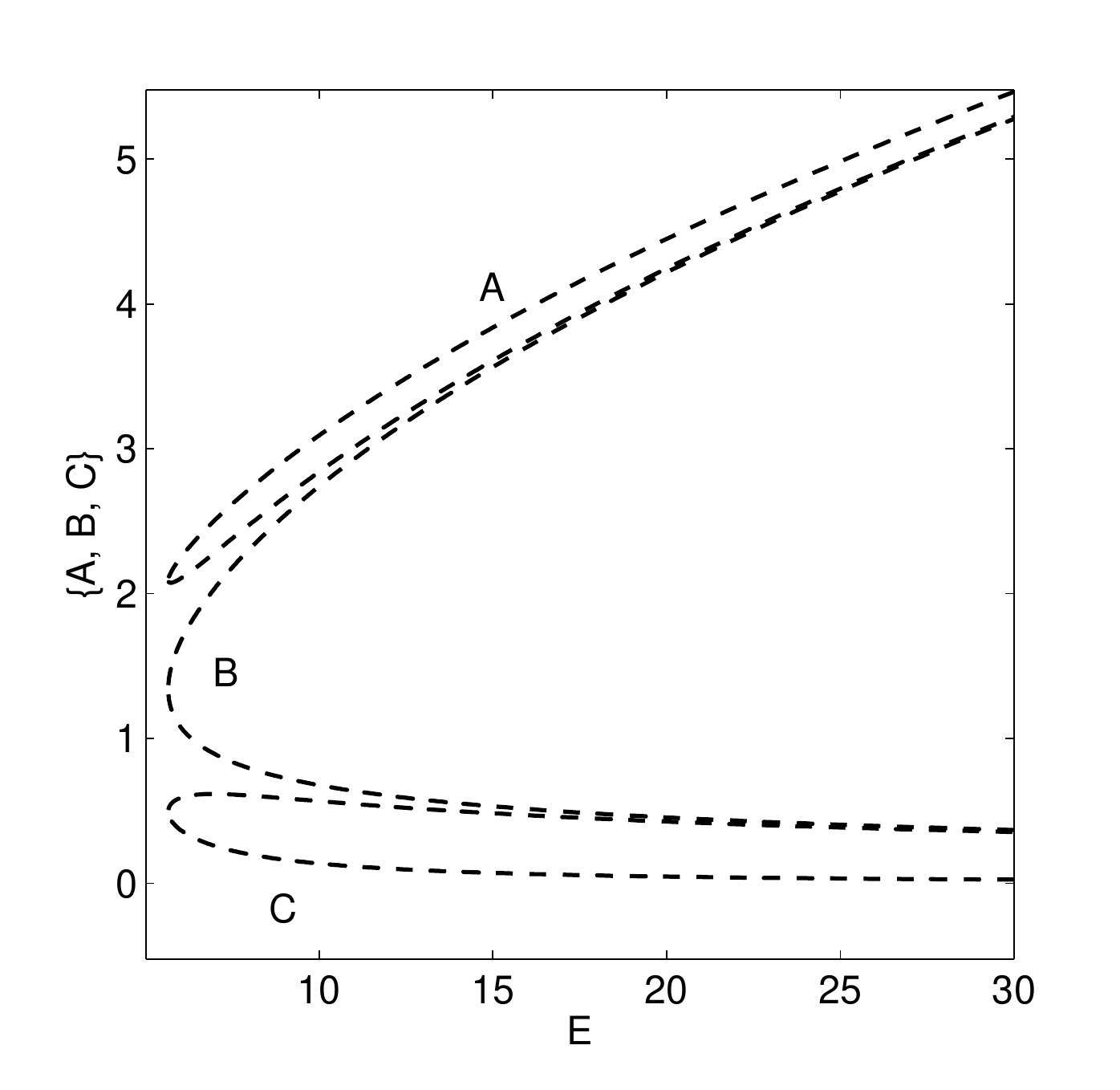}\label{SymmetricAsymCW2}}
\caption{Amplitudes $A$, $B$ and $C$ as functions of $E$, for two families
of asymmetric CW states. Due to the axes scale in panel (a), amplitude $A$
seems to be represented by a single (formally stable) curve, whereas it is
actually composed of two close curves,originating from a common initial
point at $E=5.66$, with a completely unstable lower branch, and the upper
one being stable at $E>13.95$, $A>3.99$.}
\label{AsymCW}
\end{figure}

First, we have examined the existence and stability
of CW solutions for the system based on Eqs. (\ref{abc1}) and (\ref{A_n})-(%
\ref{C_n}). Diagrams which show amplitudes $A$, $B$ and $C$ as functions of
propagation constant $E$, for symmetric, antisymmetric and asymmetric CW
states, are presented in Figs. \ref{SymmetricCW}, \ref{AntiSymmetricCW} and %
\ref{AsymCW}.

For the symmetric case, with $A=C$, three CW families were identified. The
first one, which exists for all $E>0$, and is \emph{completely stable}
against modulational perturbations, is shown in Fig. \ref{SymmetricCW1}. The
second family, which is present at $E>9.70$ and features two coexisting
branches (Fig. \ref{SymmetricCW2}), is entirely unstable.%
The third family, presented in Fig. \ref{SymmetricCW3}, exists at $E>2.82$
and is totally unstable too (although close to the lower edge -- namely, at $%
2.82<E<2.90$ -- the modulational instability of the CW is weak). Direct
simulations (not shown here in detail) demonstrate that unstable CWs are
transformed into chaotic spatiotemporal states.

Figure \ref{AntiSymmetricCW} introduces antisymmetric CW solutions, with $%
A=-C$ and $B=0$. It is found to be partially stable -- namely, at $E>5.64$,
which corresponds to $A=-C>2.38$. 
In fact, the antisymmetric CW is a limit (delocalized) form of compact
antisymmetric solutions, which are presented below in Section \ref%
{sec:LocalizedSymmetricStates}.

Two families of asymmetric CWs were found too, both exhibiting two distinct
branches, meeting at $E=5.66$. In the case of the asymmetric CW family shown
in Fig. \ref{AntiSymmetricAsymCW1}, a stability region is $E>13.95$, $A>3.99$%
, 
$B<-3.96$, 
and $C>-0.58$.
On the other hand, the family of asymmetric CWs displayed in Fig. \ref%
{SymmetricAsymCW2} is completely unstable.

\subsection{Symmetric lattice solitons}

\label{sec:LocalizedSymmetricStates}

As said above, it is natural to expect that the nonlinear lattice supports,
in addition to the exact CLSs (\ref{CLS}), discrete solitons, which may be
subject, in particular, to the symmetry constraint, $a_{n}=c_{n}$, satisfied
at all $n$, or may be asymmetric. The discrete solitons are not compact
modes, but they feature a strong exponential localization.

\begin{figure}[th]
\centering%
\subfigure[]{\includegraphics[scale=0.40]{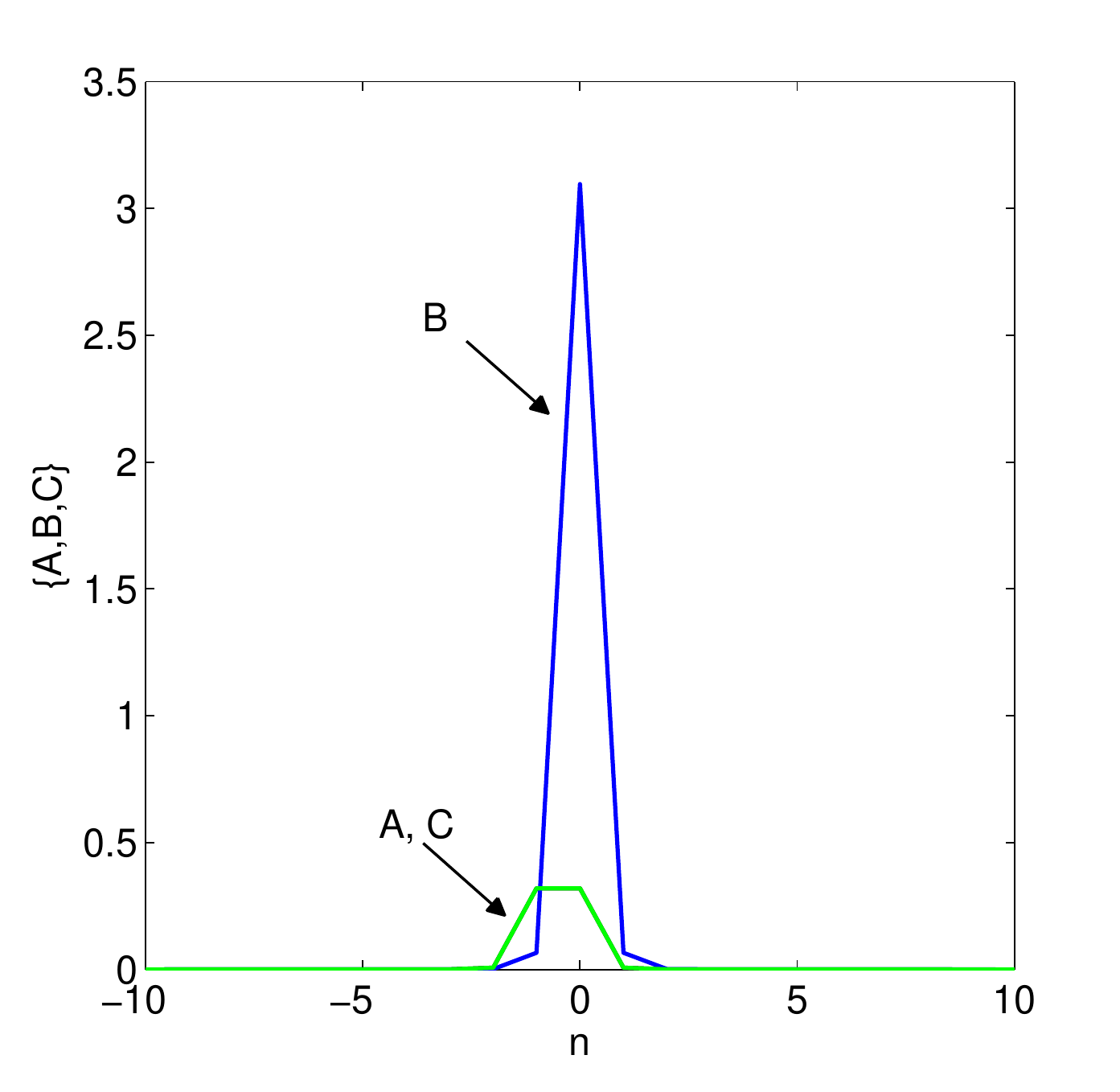}
\label{SymmetricLocalized1_E10}} \subfigure[]{%
\includegraphics[scale=0.40]{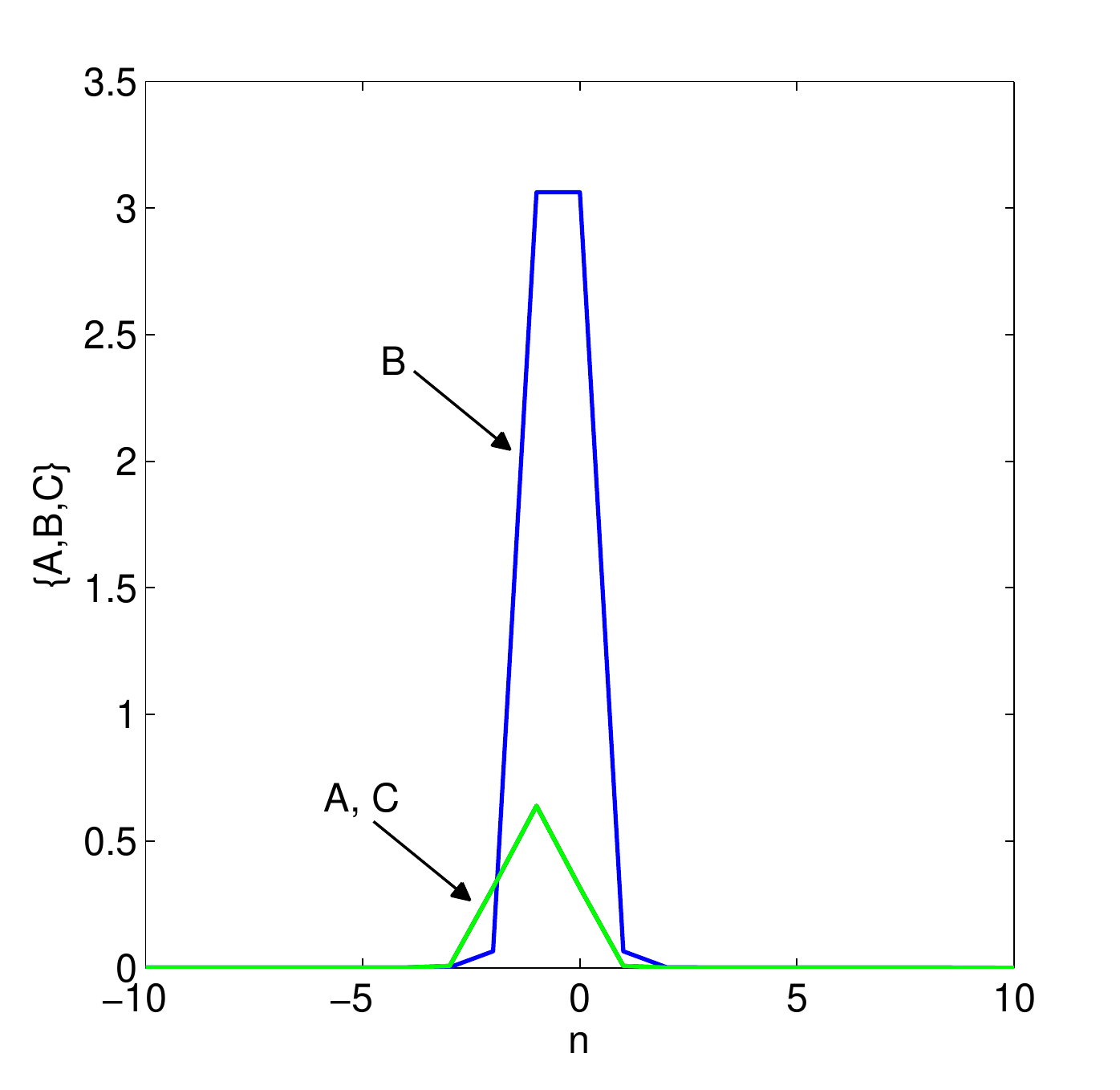}%
\label{SymmetricLocalized2_E10}} \subfigure[]{%
\includegraphics[scale=0.40]{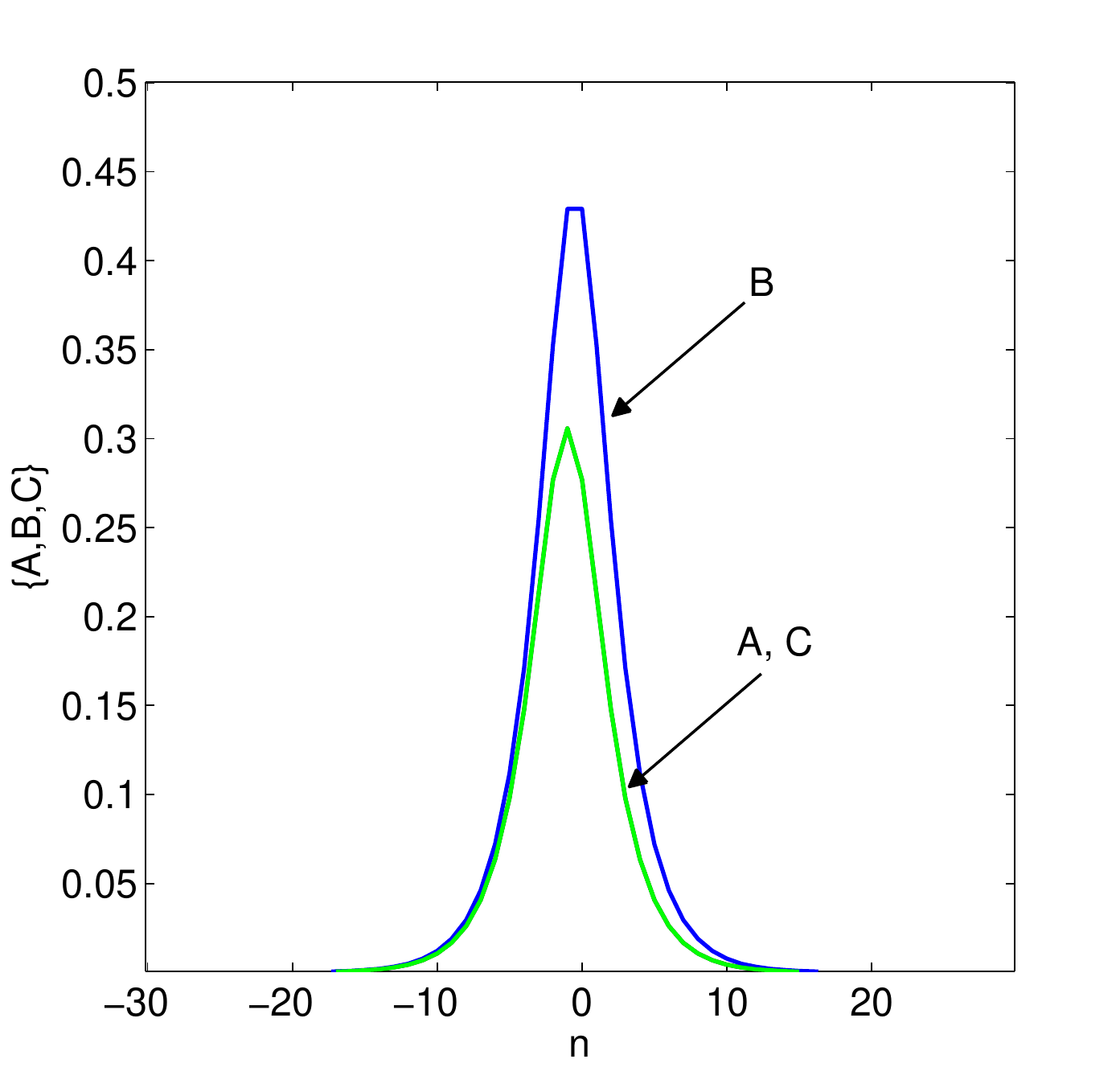}%
\label{SymmetricLocalized_E2p9}}
\caption{(Color online) Examples of the discrete symmetric solitons
corresponding to the marked points in Fig. \protect\ref%
{SymmetricLocalized1and2_NvsE}. Panels (a), (b) and (c) refer to points a ($%
E=10$, the stable lower branch), b ($E=10$, the unstable upper branch), and
c ($E=2.90$, the bifurcation point), respectively. Amplitudes $A_{n}=C_{n}$
and $B_{n}$ are denoted, severally,by green\ and blue lines.}
\label{SymmetricLocalized1and2}
\end{figure}

\begin{figure}[th]
\centering%
\subfigure[]{\includegraphics[scale=0.50]{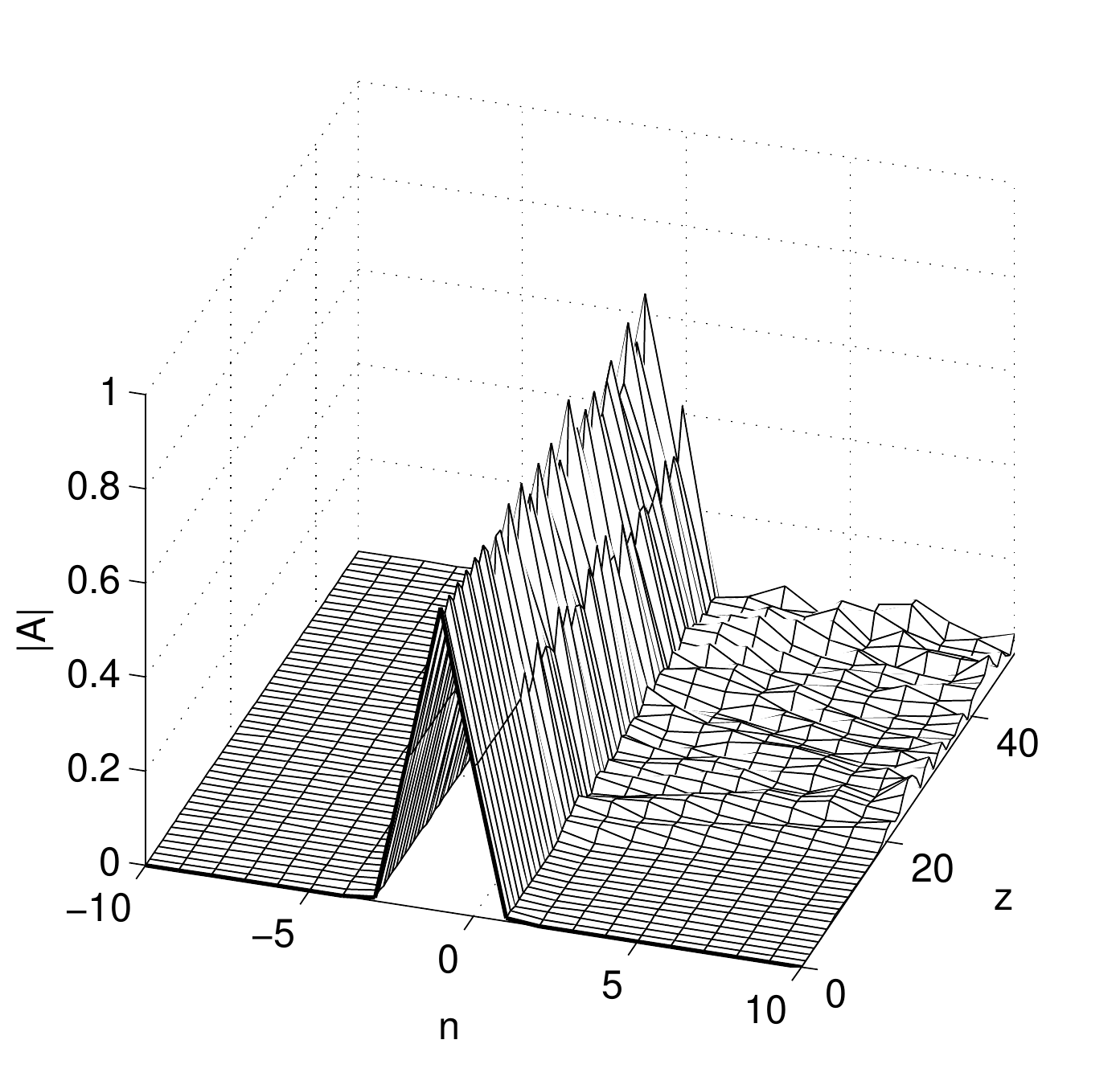}
\label{Symmetric2Evolution_E10_A}} \subfigure[]{%
\includegraphics[scale=0.50]{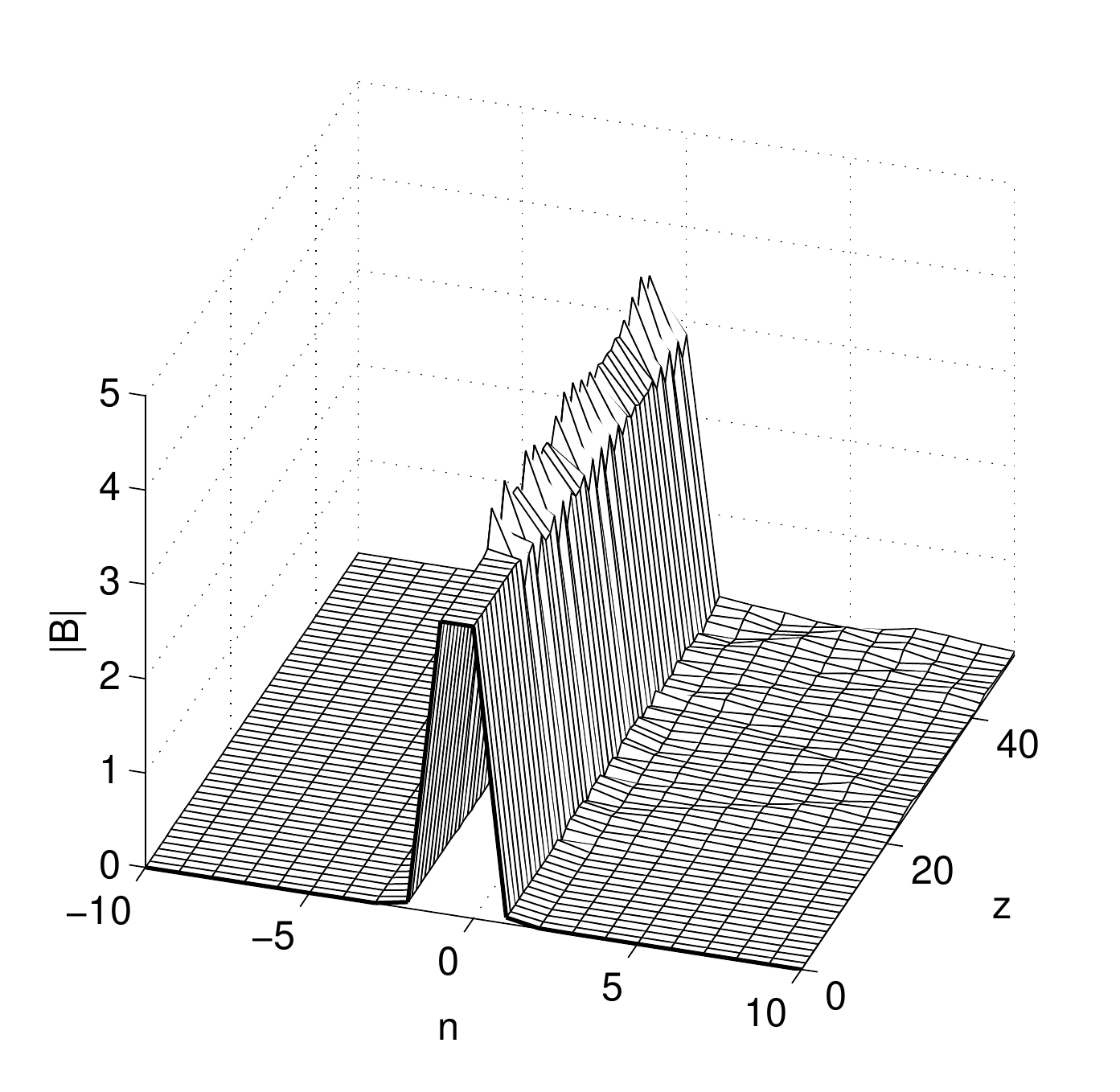}%
\label{Symmetric2Evolution_E10_B}} \subfigure[]{%
\includegraphics[scale=0.50]{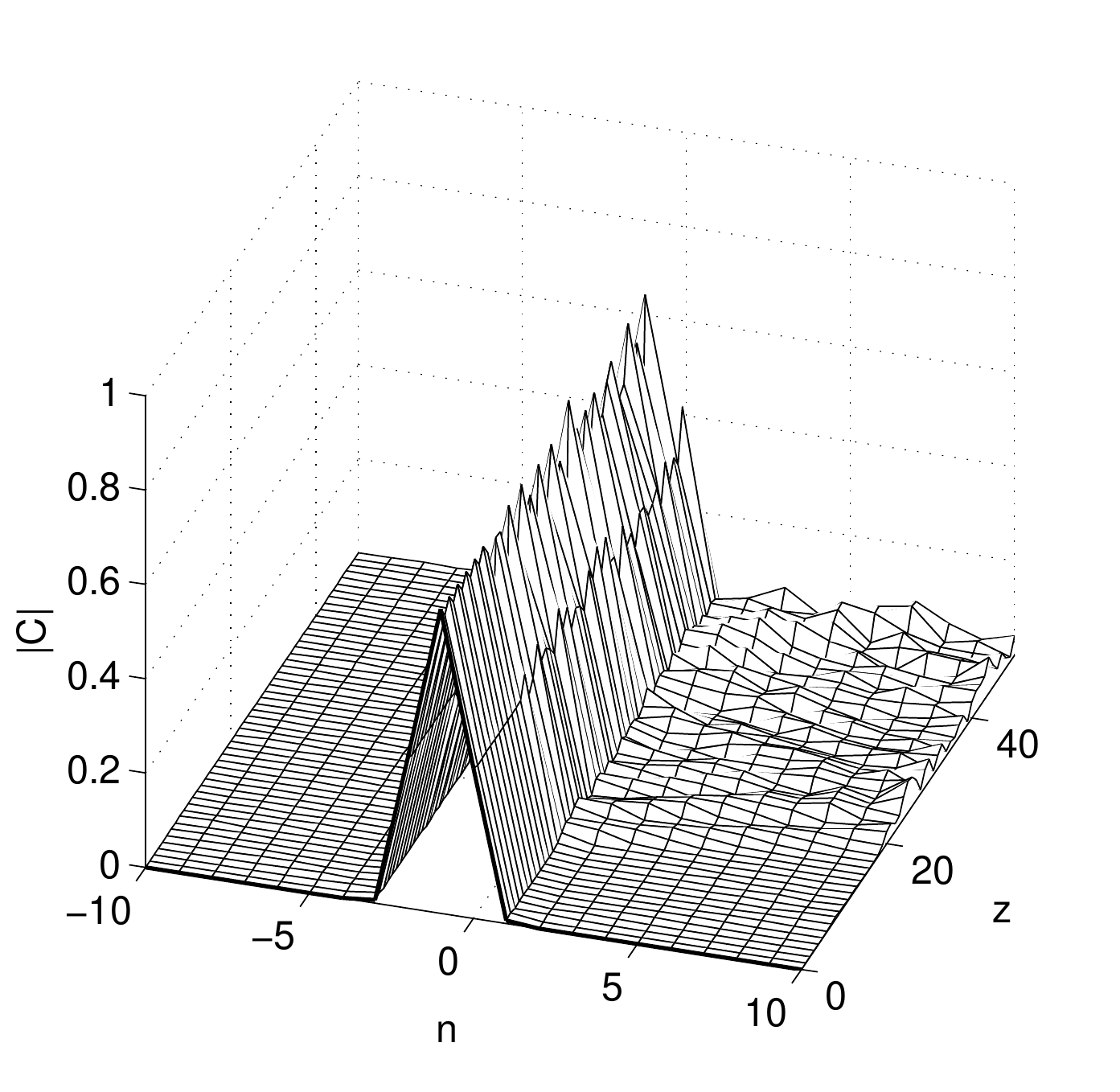}%
\label{Symmetric2Evolution_E10_C}}
\caption{A typical example of the evolution of an unstable symmetric
discrete soliton, belonging to the unstable upper branch in Fig. \protect\ref%
{SymmetricLocalized1and2_NvsE}, with $E=10$. The stationary version of this
soliton is shown in Fig. \protect\ref{SymmetricLocalized2_E10}).}
\label{Symmetric3Evolution_E15}
\end{figure}

With the help of the numerical methods outlined above, several basic
families of symmetric solitons have been detected. The first two families
are illustrated by Fig. \ref{SymmetricLocalized1and2}. While in one case
(Fig. \ref{SymmetricLocalized1_E10}) both amplitudes $A_{n}$ and $C_{n}$
have a maximum with equal values at $n=-1$ and $n=0$, the other solution
(Fig. \ref{SymmetricLocalized2_E10}) has a double maximum in terms of the $B$
amplitude. As concerns the stability, the family demonstrated in Fig. \ref%
{SymmetricLocalized1_E10} was found to be stable, while the one shown in
Fig. \ref{SymmetricLocalized2_E10} is unstable. A systematic numerical
analysis has shown that the family of the solitons with the double maximum
in the $A_{n}=C_{n}$ fields bifurcates from the one characterized by the
double maximum in $B_{n}$, through a typical supercritical bifurcation, see
Fig. \ref{SymmetricLocalized1and2_NvsE}. Before the bifurcation occurs,
i.e., at $E<2.90$, the solution is stable. Past the bifurcation, the upper
branch, represented by the solitons shown in Fig. \ref%
{SymmetricLocalized2_E10}, destabilizes, while the lower one (see an example
in Fig. \ref{SymmetricLocalized1_E10}) remains stable. An example for the
evolution of unstable solitons from the upper branch is presented in Fig. %
\ref{Symmetric3Evolution_E15}. It is seen that the unstable soliton actually
remains a localized mode, which features randomized intrinsic dynamics and
emission of weak phonon waves. Figure \ref{Symmetric3Evolution_E15}
demonstrates that the strongly unstable solution is oscillating around a
stable solution belonging to the lower branch, shown in Fig. \ref%
{SymmetricLocalized1and2_NvsE}. The oscillating unstable soliton emits
spatially asymmetric radiation, due to asymmetric interference between the
unstable and stable modes.

As $E$ decreases, the shape of the discrete solitons becomes Gaussian-like,
i.e., quasi-continuous. This peculiarity can be seen in Fig. \ref%
{SymmetricLocalized_E2p9}, for $E=2.90$, taken at the bifurcation point.

\begin{figure}[th]
\centering%
\subfigure[]{\includegraphics[scale=0.50]{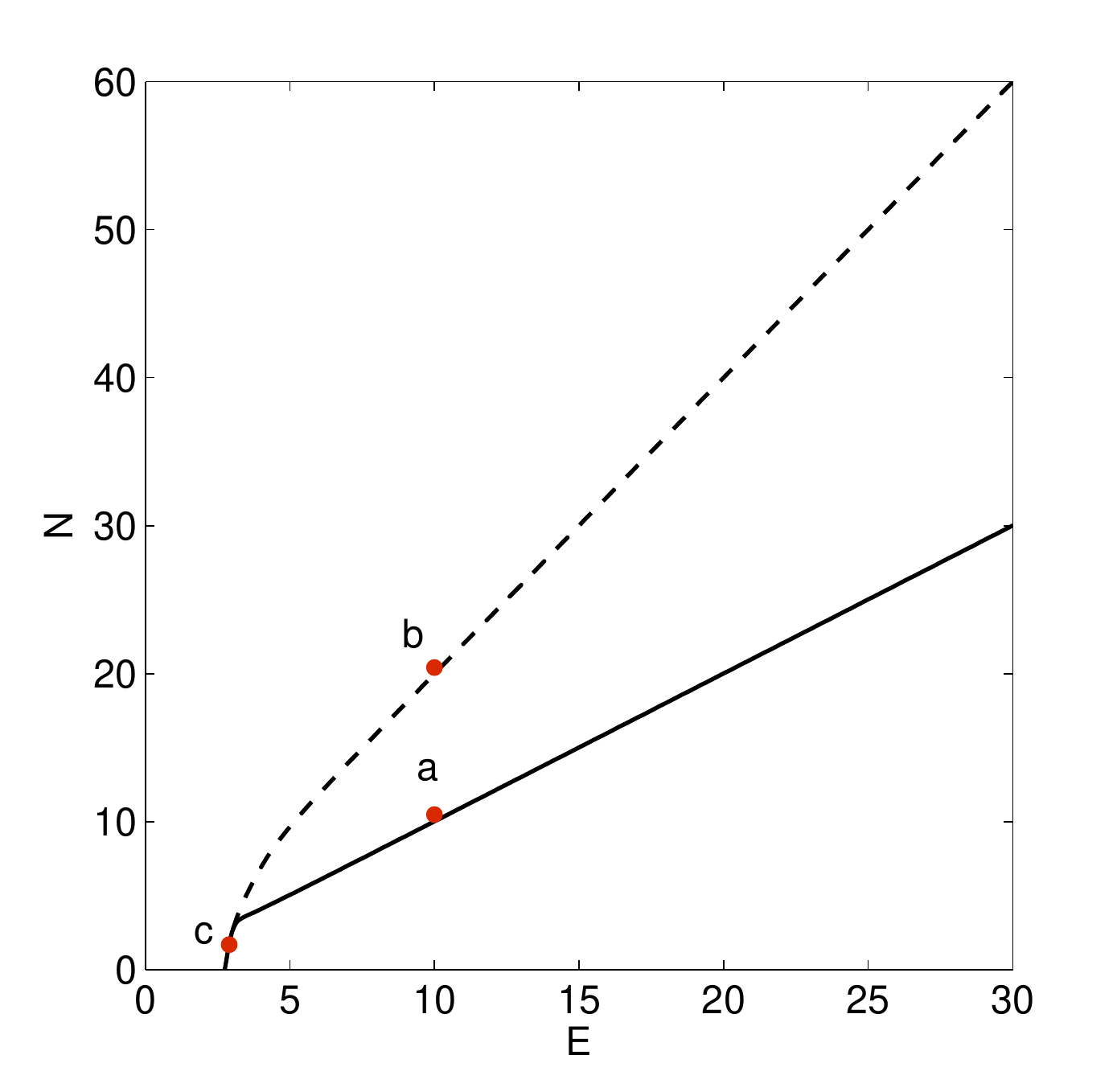}
\label{SymmetricLocalized1and2_NvsE}}
\caption{The stability of the first two families of symmetric discrete
solitons, shown by means of the respective $N(E)$ curves. The marked points
correspond to the solitons displayed in Fig. \protect\ref%
{SymmetricLocalized1and2}.}
\label{SymmetricLocalized1and2_NvsE}
\end{figure}

\begin{figure}[th]
\centering%
\subfigure[]{\includegraphics[scale=0.50]{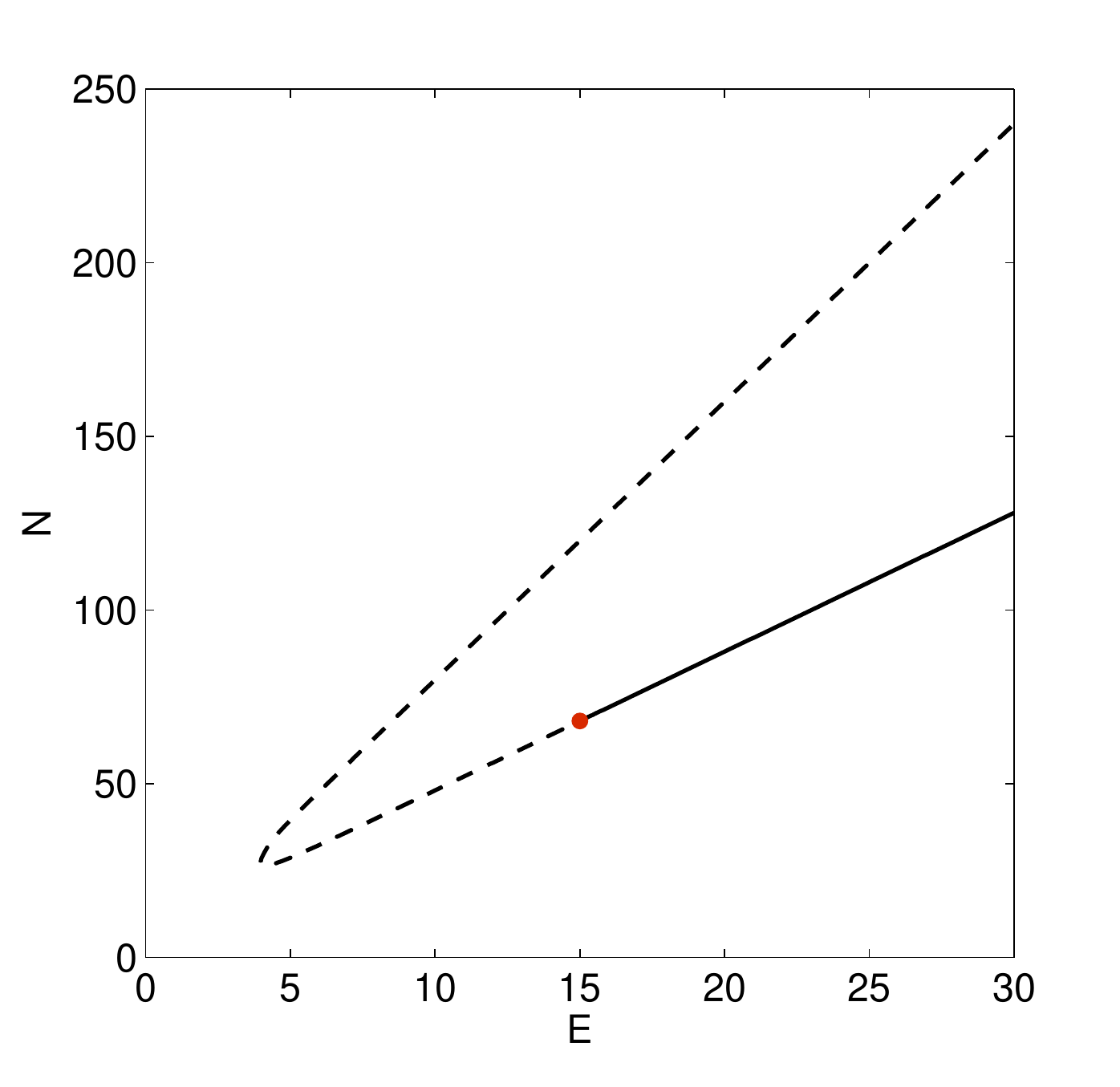}
\label{SymmetricLocalized3_NvsE}} \subfigure[]{%
\includegraphics[scale=0.50]{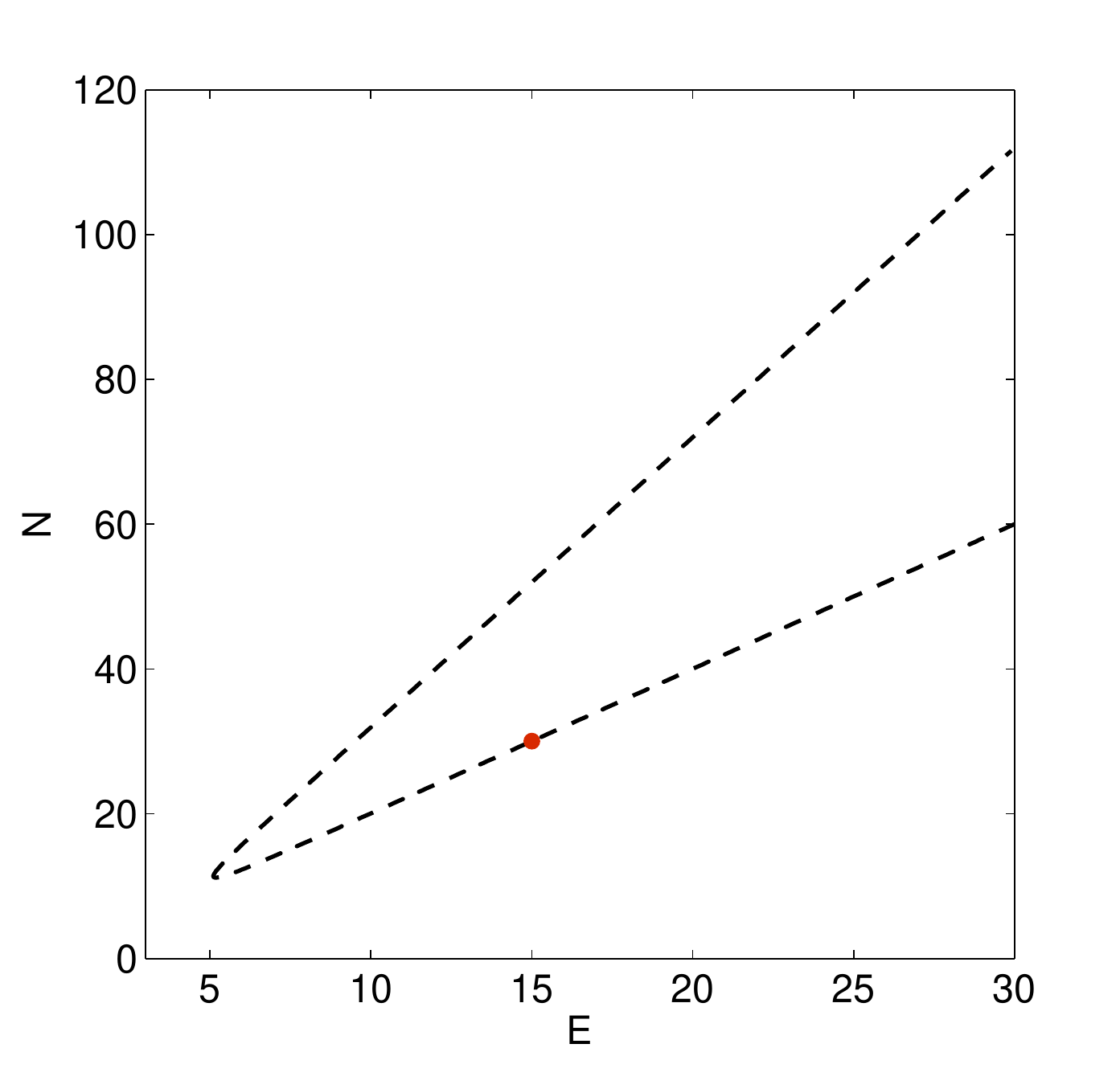}%
\label{SymmetricLocalized4_NvsE}}
\caption{Two additional families of symmetric localized solitons, displayed
by means of $N(E)$ curves.}
\label{SymmetricLocalized3and4_NvsE}
\end{figure}

\begin{figure}[th]
\centering%
\subfigure[]{\includegraphics[scale=0.50]{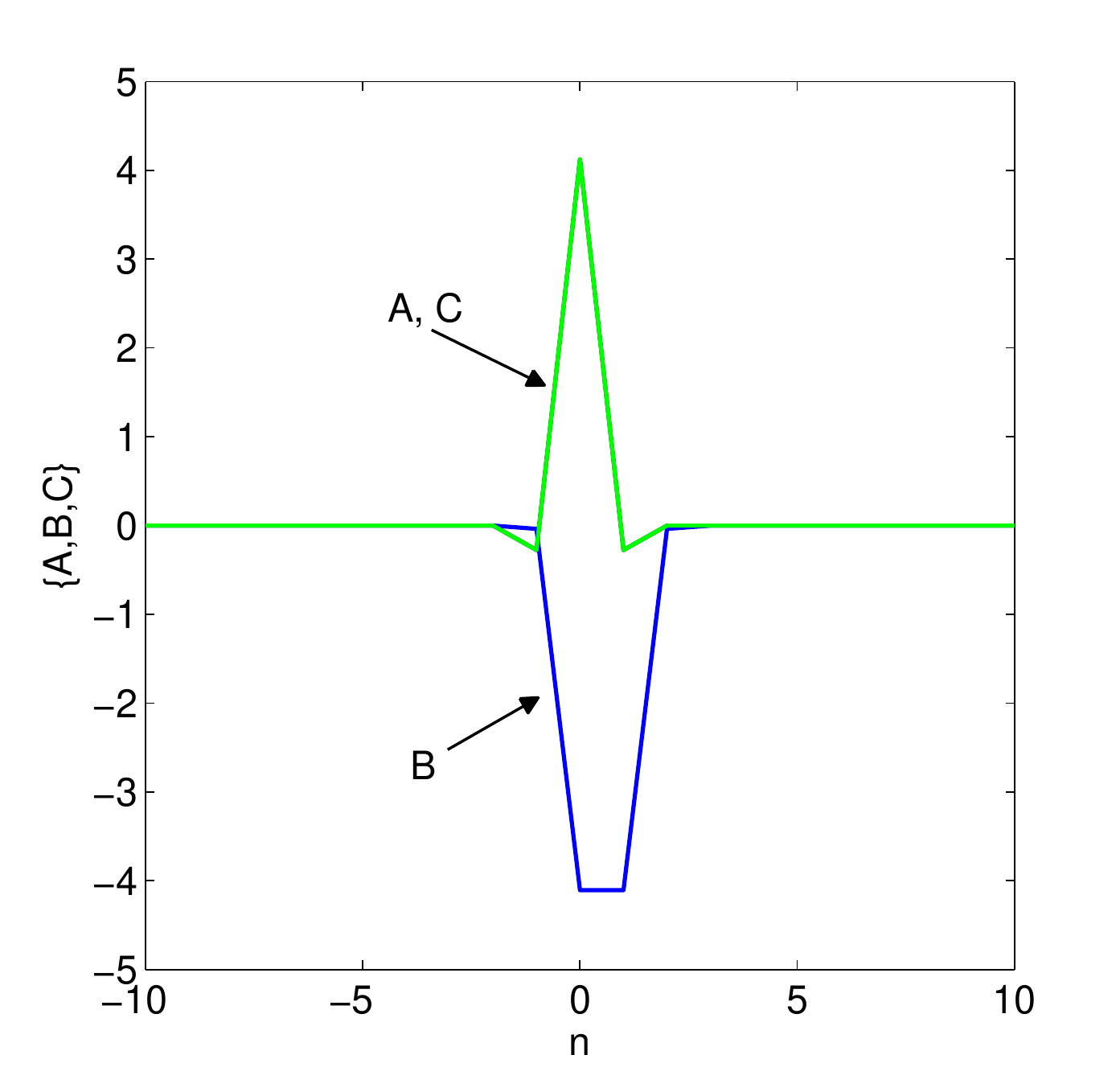}
\label{SymmetricLocalized3_E15}} \subfigure[]{%
\includegraphics[scale=0.50]{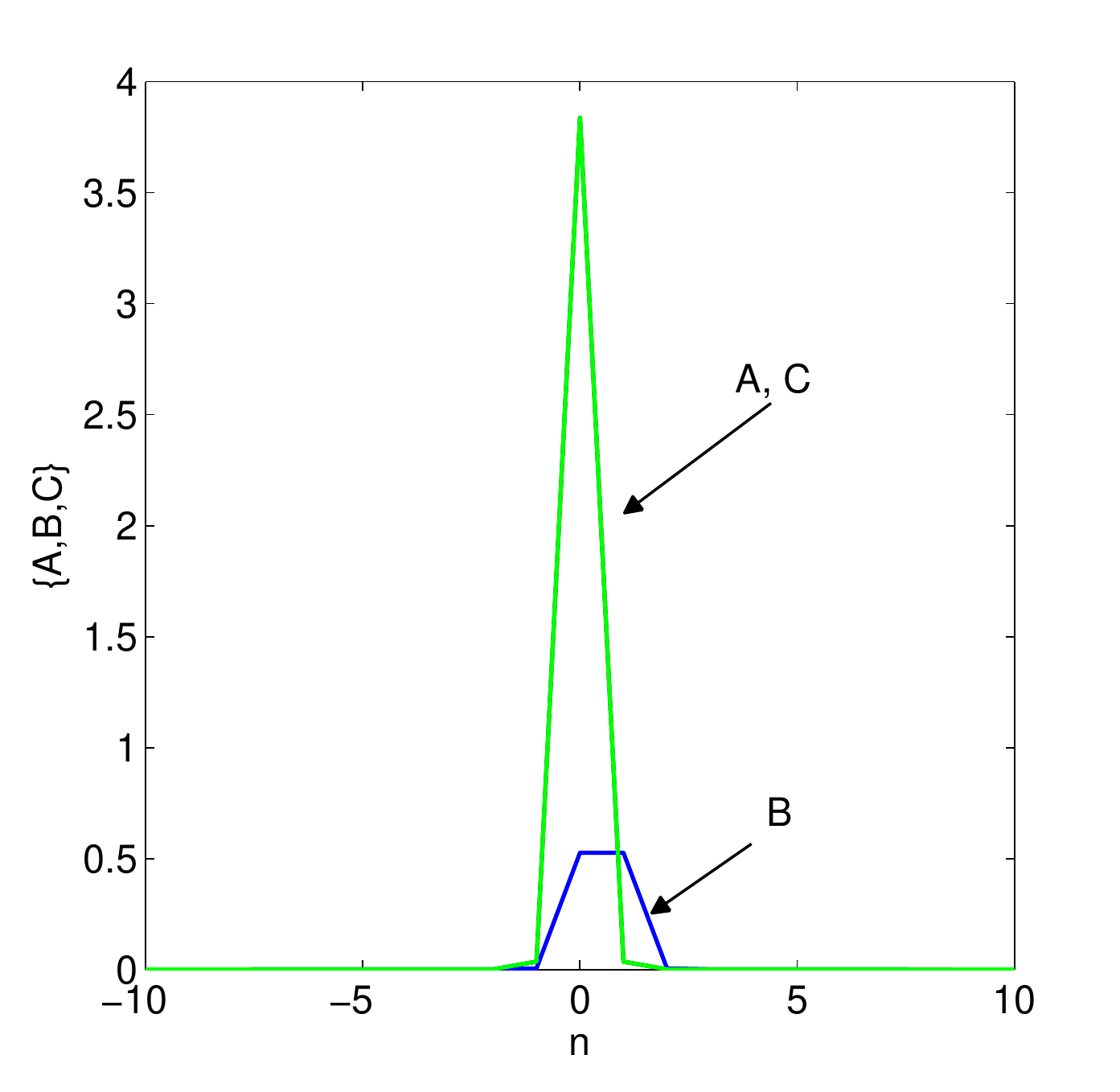}%
\label{SymmetricLocalized4_E15}}
\caption{(Color online) Typical examples of profiles of the symmetric
discrete solitons belonging to the additional families shown in Fig. \protect
\ref{SymmetricLocalized3and4_NvsE}. Panels (a) and (b) correspond to points
marked on the lower branches in Figs. \protect\ref{SymmetricLocalized3_NvsE}
and \protect\ref{SymmetricLocalized4_NvsE}, respectively, both with $E=15$.}
\label{SymmetricLocalized3and4}
\end{figure}

\begin{figure}[th]
\centering%
\subfigure[]{\includegraphics[scale=0.50]{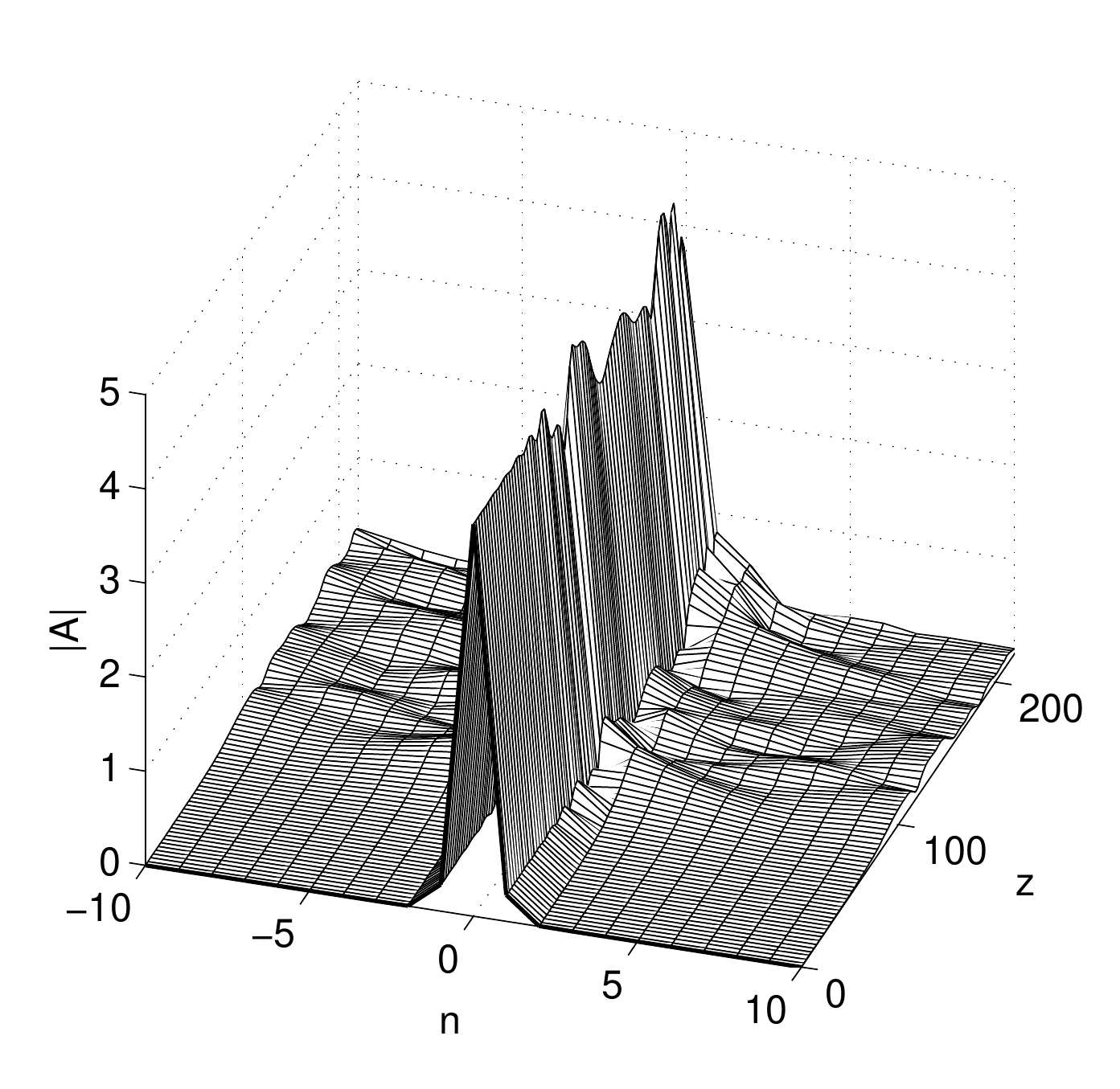}
\label{Symmetric3Evolution_E15_A}} \subfigure[]{%
\includegraphics[scale=0.50]{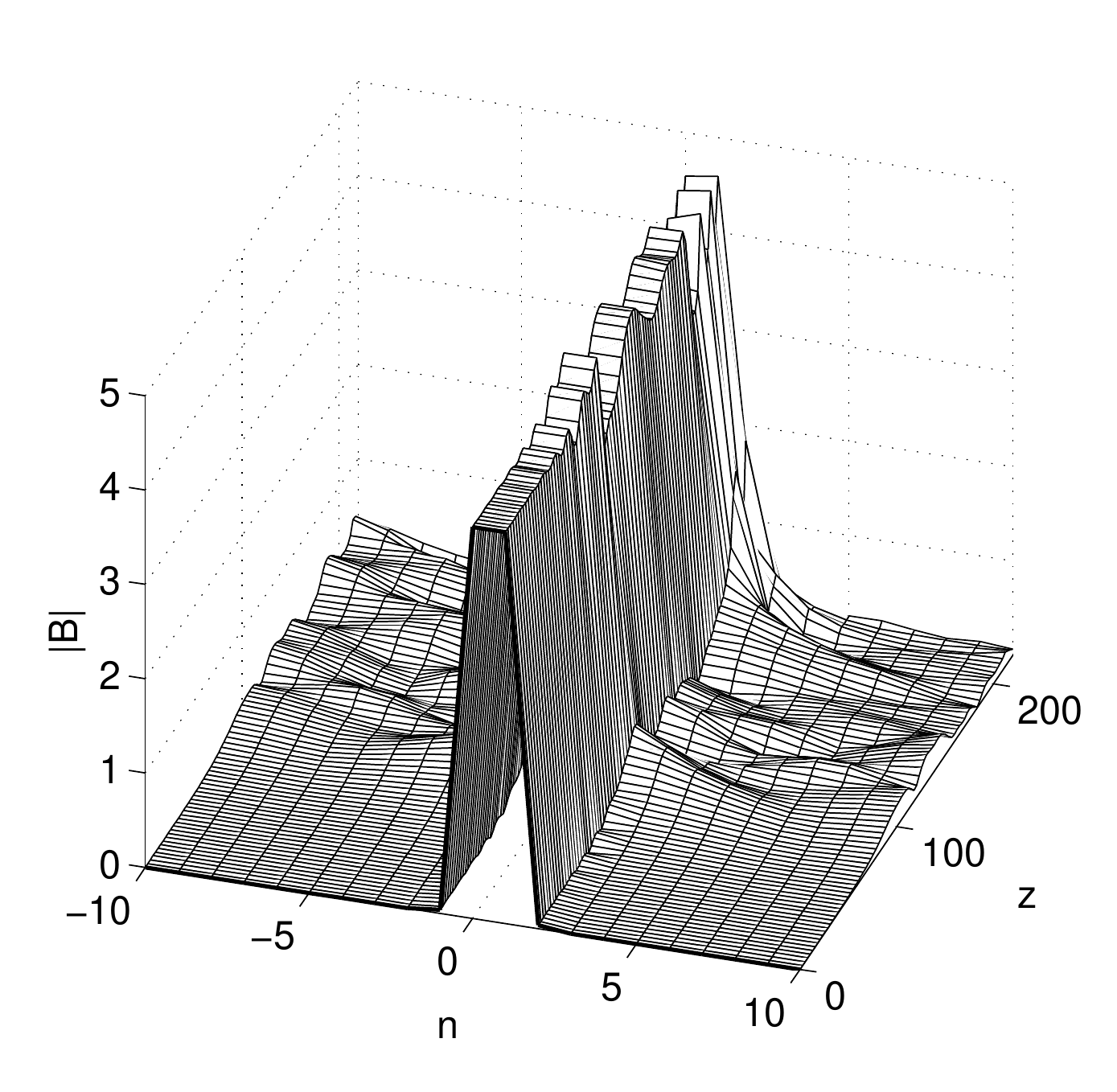}%
\label{Symmetric3Evolution_E15_B}} \subfigure[]{%
\includegraphics[scale=0.50]{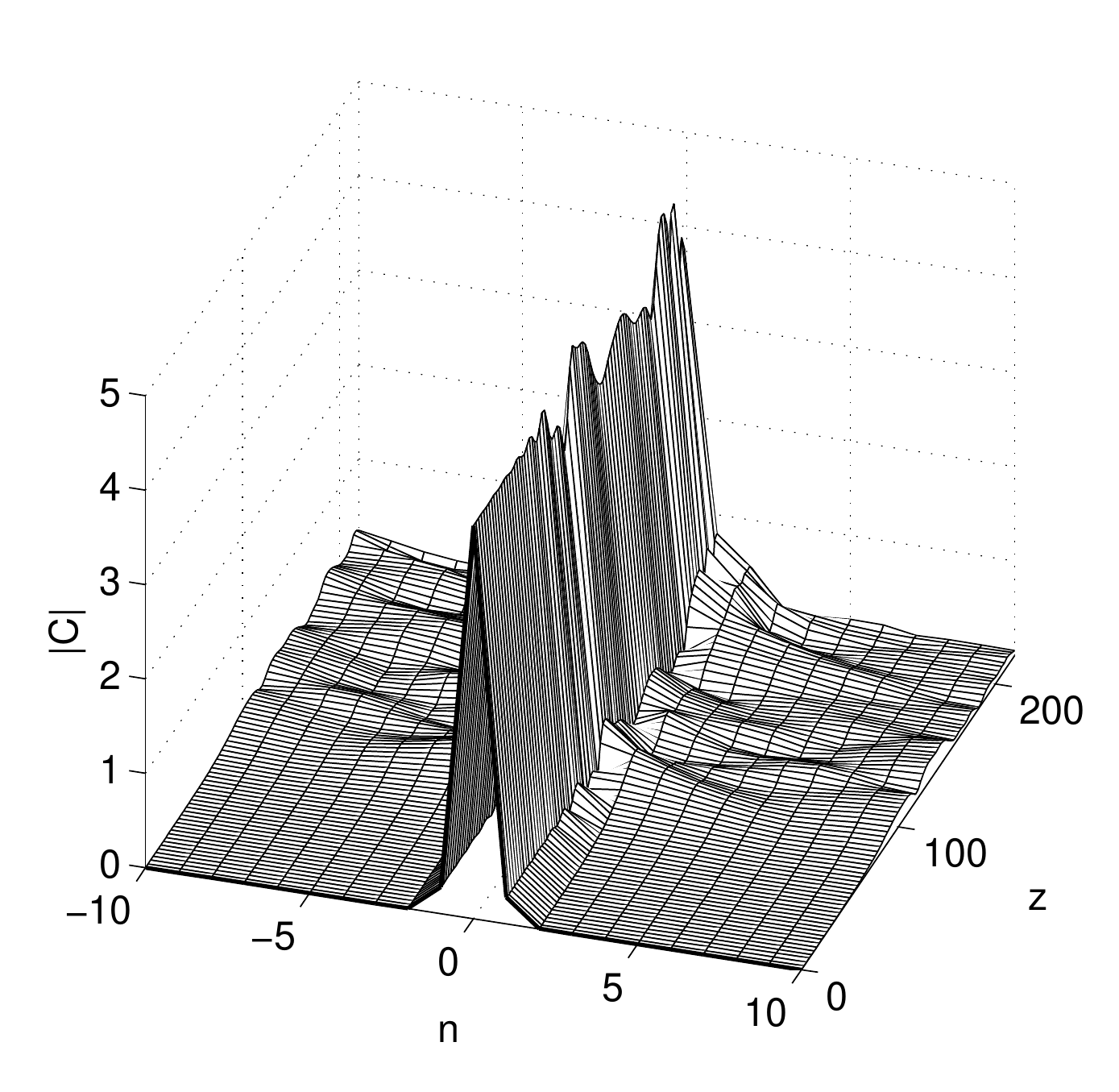}%
\label{Symmetric3Evolution_E15_C}}
\caption{An example for the evolution of an unstable mode that belongs to
the partially stable family in Fig. \protect\ref{SymmetricLocalized3_NvsE},
with $E=15$ (Fig. \protect\ref{SymmetricLocalized3_E15} shows the
unperturbed shape of this discrete soliton).}
\label{Symmetric3Evolution_E15_2}
\end{figure}

Two additional types of symmetric discrete solitons, found in the same
model, are presented in Fig. \ref{SymmetricLocalized3and4}. Both feature
stability diagrams that consist of upper and lower branches merging into
one, as shown in Fig. \ref{SymmetricLocalized3and4_NvsE}. The family
exhibited in Figs. \ref{SymmetricLocalized3_NvsE} and \ref%
{SymmetricLocalized3_E15} is stable in the region of $E>15.46$, $N>69.87$
(the lower branch). An example of the evolution of an unstable soliton is
shown in Fig. \ref{Symmetric3Evolution_E15_2}. It can be concluded that in
this case too, the unstable discrete soliton remains an effectively
localized mode with chaotic intrinsic dynamics, emitting small-amplitude
phonon waves into the lattice. On the other hand, the second additional
family, presented in Figs. \ref{SymmetricLocalized4_NvsE}, \ref%
{SymmetricLocalized4_E15}, is completely unstable.

\subsection{Extended compact antisymmetric states}

\label{sec:LocalizedAntisymmetricStates}

Extended, but nevertheless compact, antisymmetric states, defined by
condition $A_{n}=-C_{n}$, can be constructed as juxtapositions of the
elementary CLS solution (\ref{CLS}), with nonvanishing amplitudes in a
finite set of lattice cells, where it has

\begin{equation}
A_{n}=\sqrt{E}=-C_{n},  \label{obvious antisymmetric states}
\end{equation}%
and zero in all others, with all $B_{n}=0$. The total norm (\ref{N=1}) of
this extended state is given by an obvious expression,

\begin{equation}
N=2nE,  \label{antisymmetric norm}
\end{equation}%
where $n$ denotes a number of cells with nonvanishing amplitudes $A_{n}$ and
$C_{n}$. Examples for $n=1$ and $n=3$ are shown in Fig. \ref{antisymmetric
states}. As mentioned in Section \ref{sec:CWSolutions}, the antisymmetric CW
solution may be considered as a limit case of this family, in the case when $%
n$ comprises the entire domain.

\begin{figure}[th]
\centering%
\subfigure[]{\includegraphics[scale=0.5]{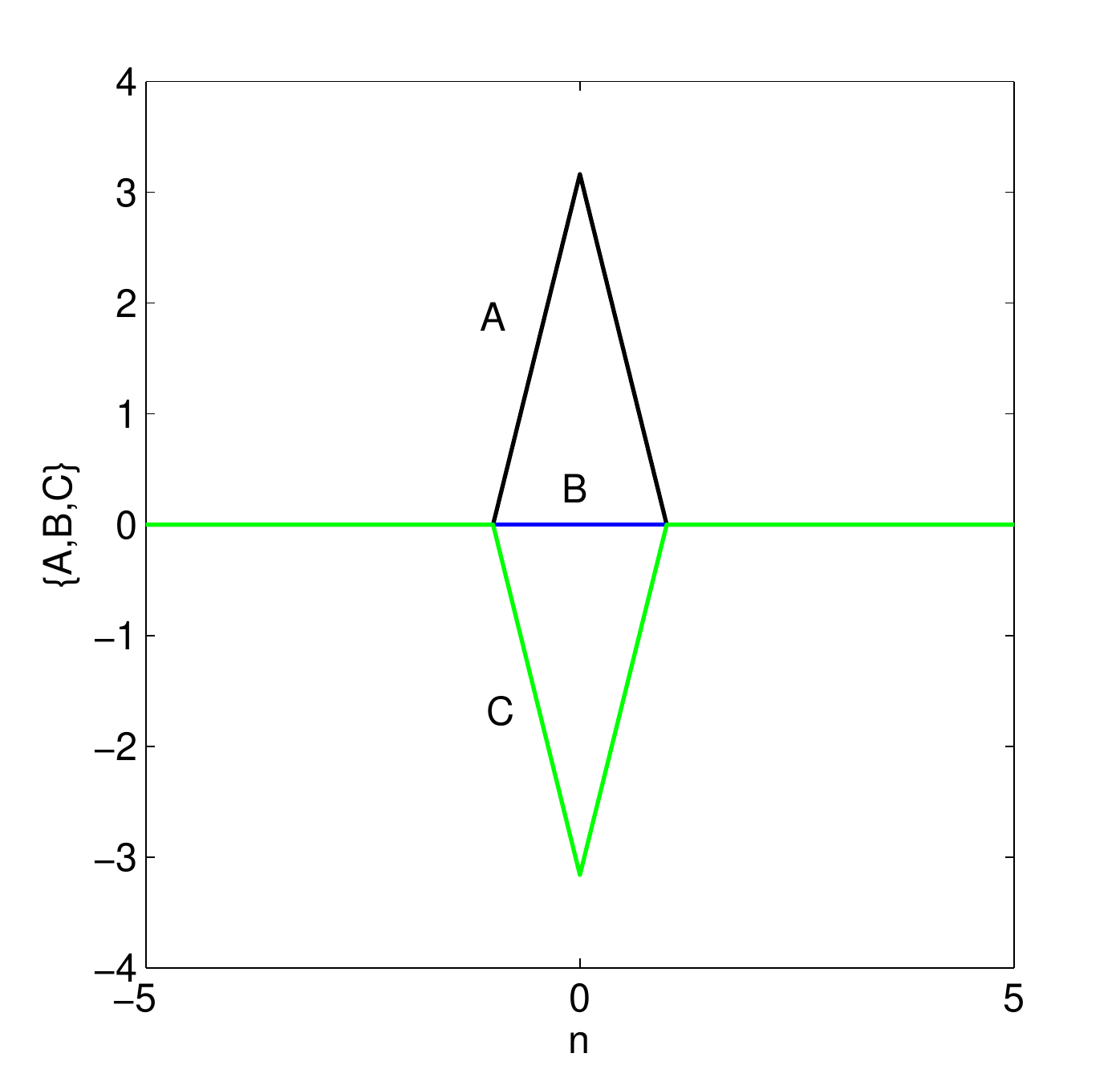}
\label{AntiSymmetric1Site_E10}} \subfigure[]{%
\includegraphics[scale=0.5]{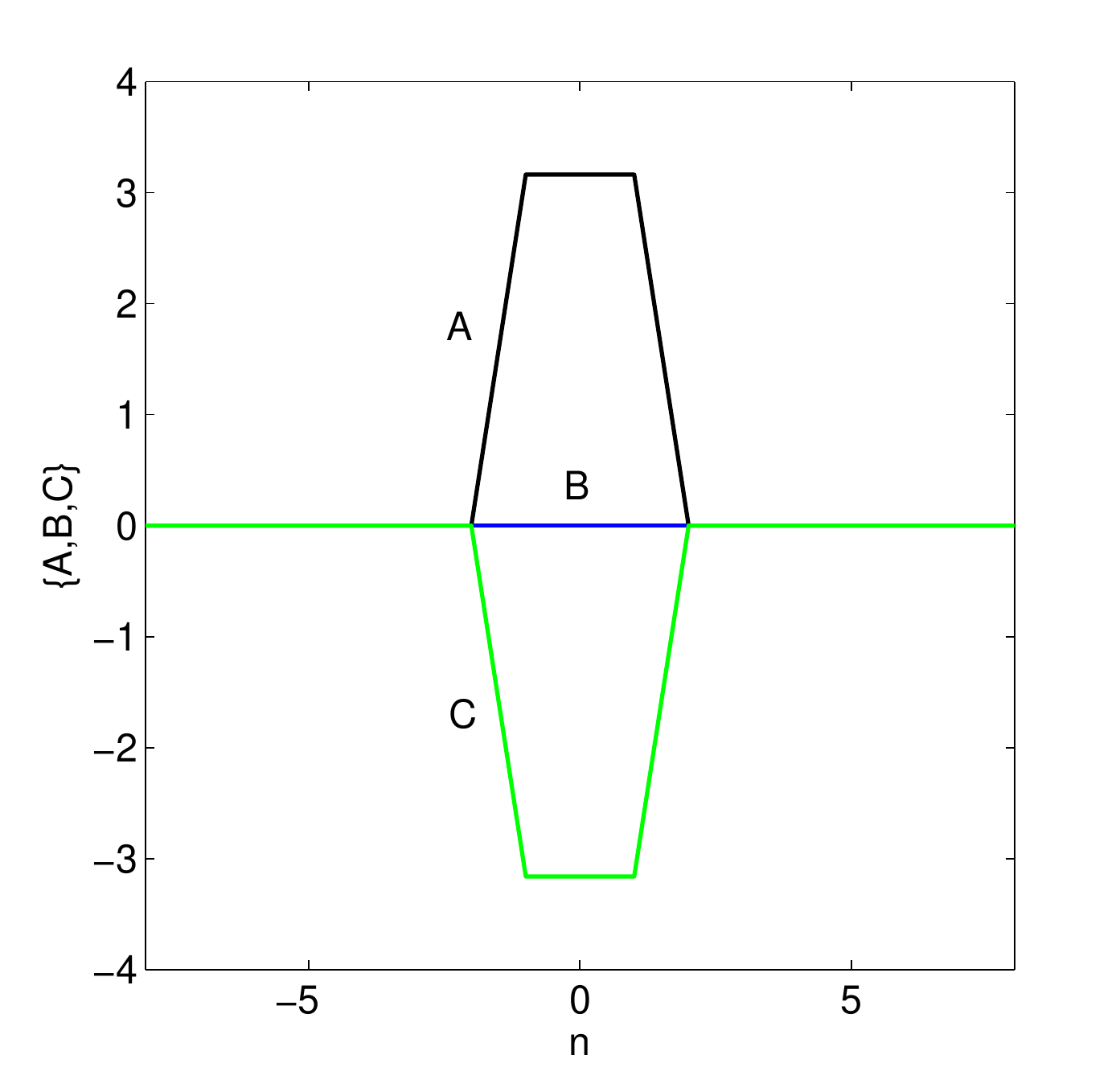}%
\label{AntiSymmetric3Site_E10}}
\caption{(Color online) Compact antisymmetric states extending to one cell
(a) and three cells (b). Amplitudes $A$, $B$, and $C$ are denoted by black,
blue and green lines, respectively.}
\label{antisymmetric states}
\end{figure}

\begin{figure}[th]
\centering%
\subfigure[]{\includegraphics[scale=0.50]{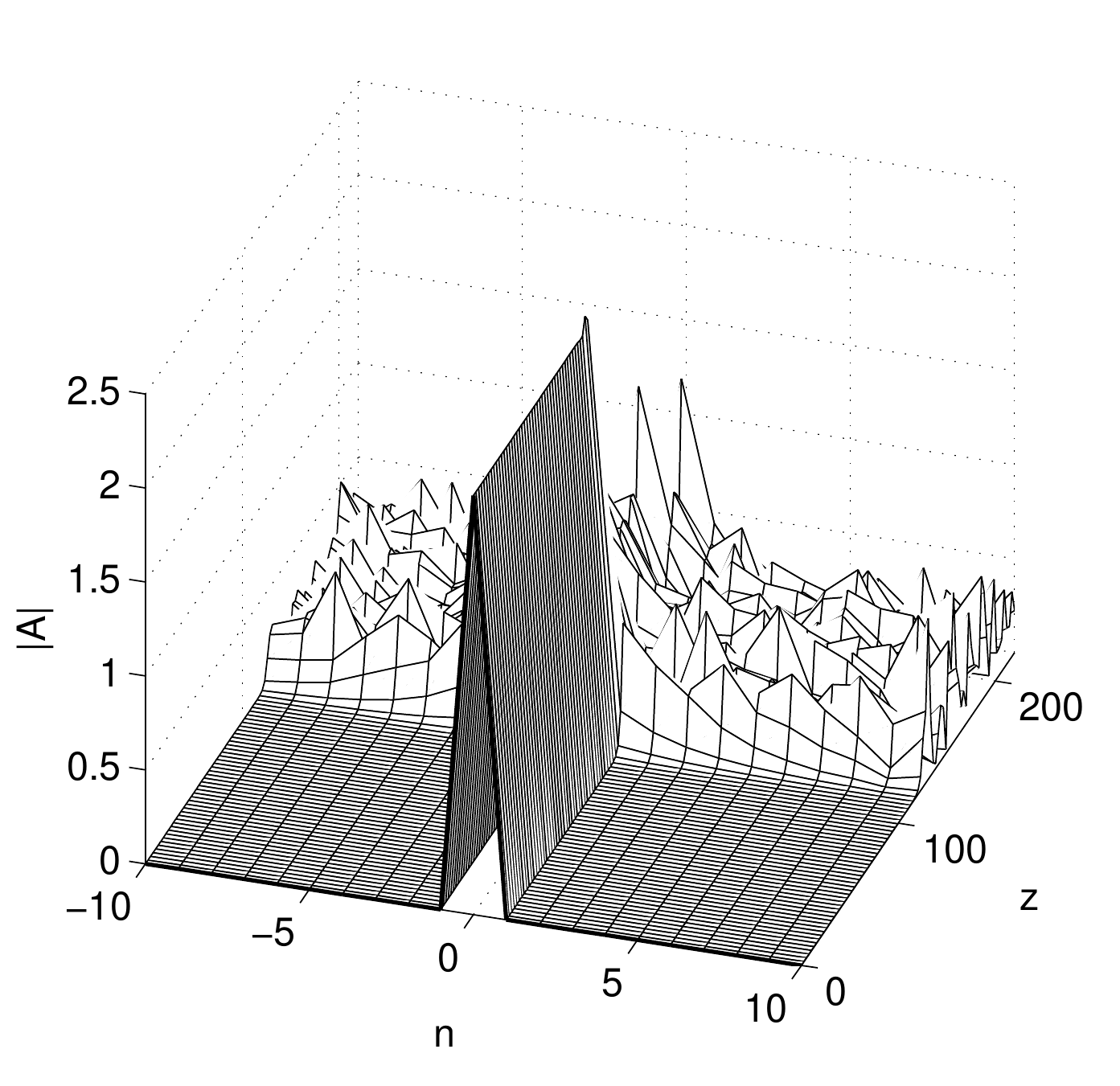}
\label{AntisymmetricEvolution_E5_A}} \subfigure[]{%
\includegraphics[scale=0.50]{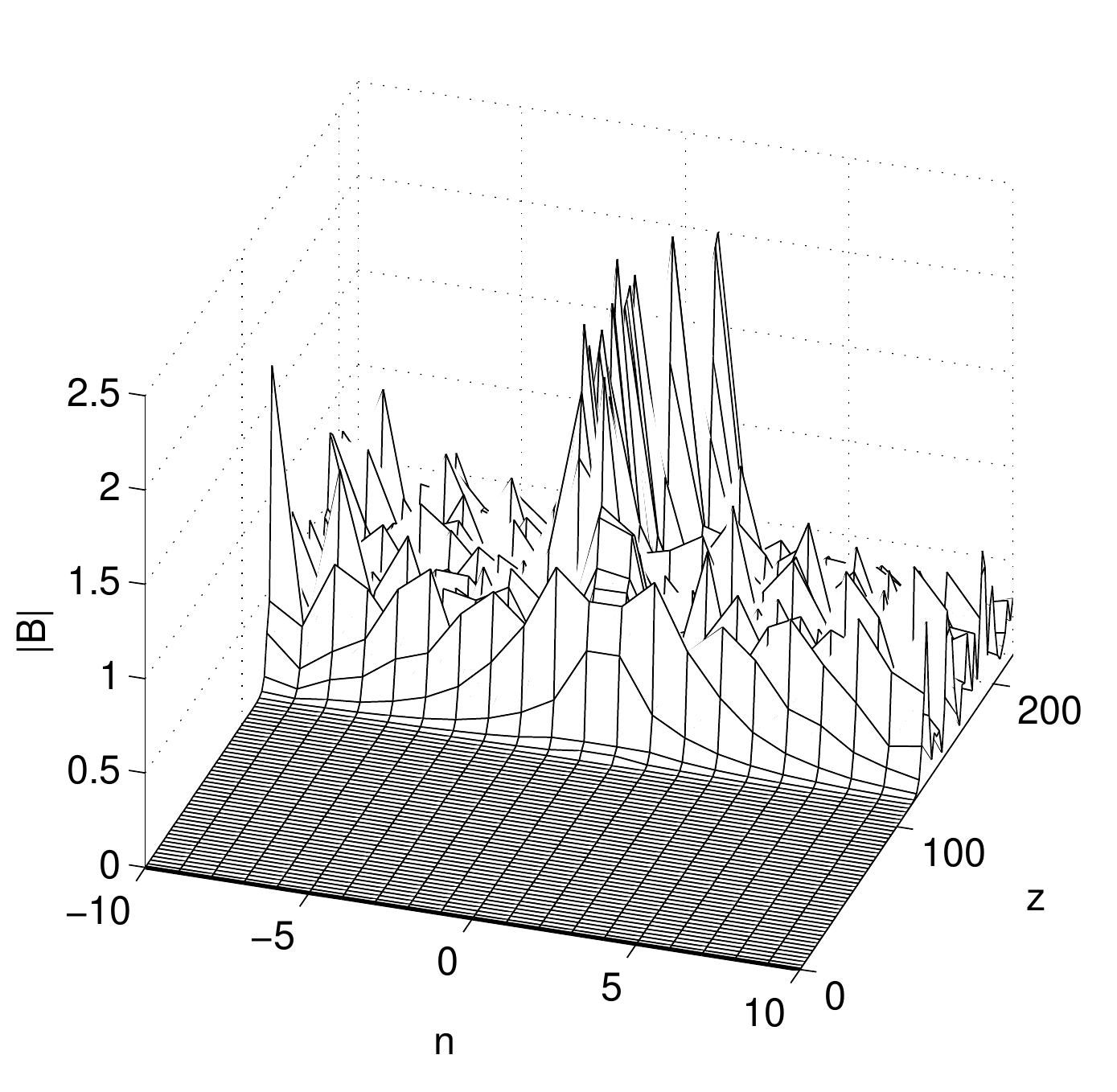}%
\label{AntisymmetricEvolution_E5_B}} \subfigure[]{%
\includegraphics[scale=0.50]{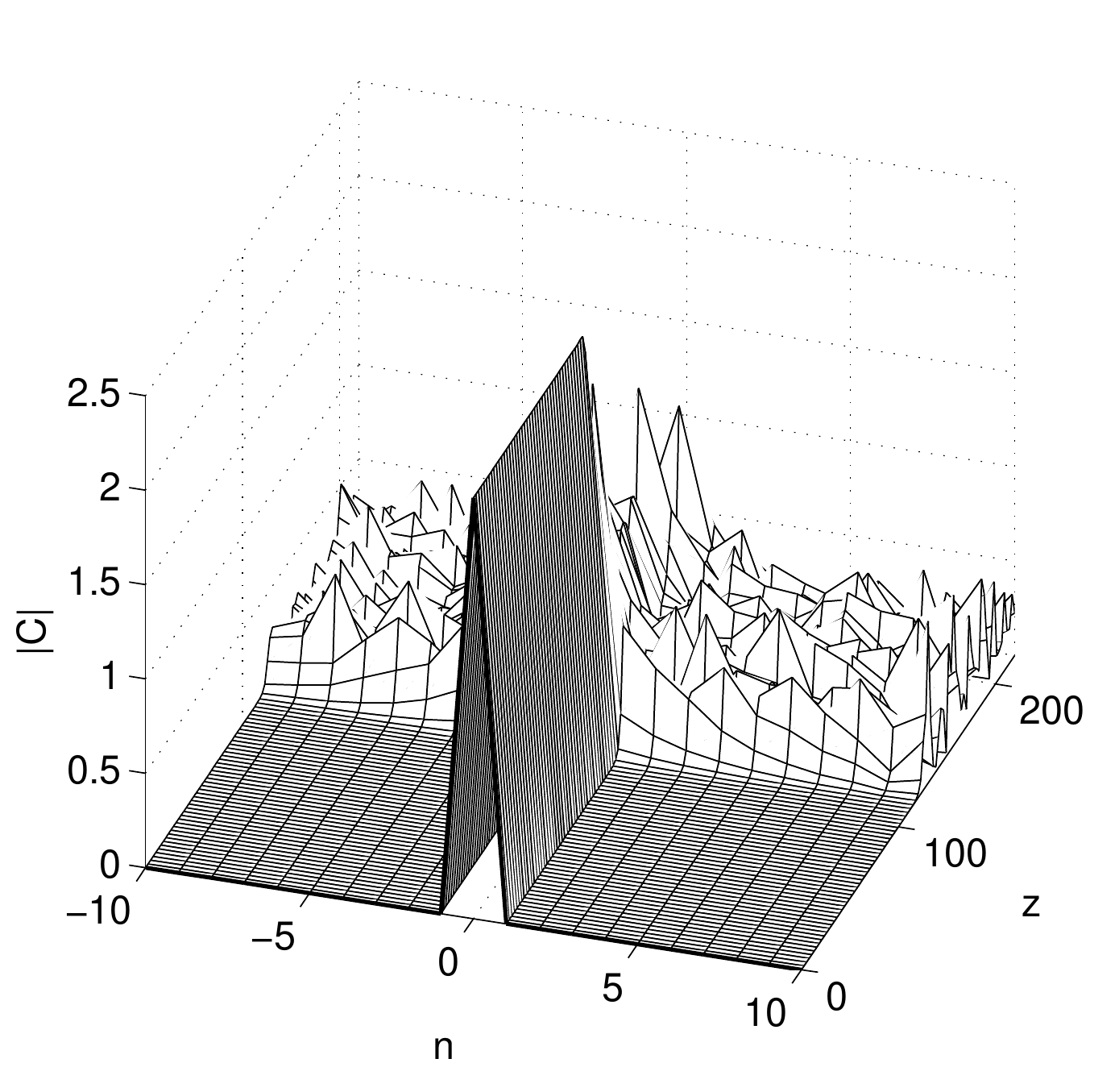}%
\label{AntisymmetricEvolution_E5_C}}
\caption{The evolution of an unstable compact antisymmetric discrete
solution, whose stationary form is given by Eq. (\protect\ref{obvious
antisymmetric states}), with a single cell ($n=1$) and $E=5$.}
\label{AntisymmetricEvolution_E5}
\end{figure}

These compact antisymmetric states are only partially stable. Systematic
numerical analysis has shown that, for smaller values of propagation
constant $E$ (and $N$), all solutions of this type are unstable.
Specifically, for $n=1$ the solutions are stable at
\begin{equation}
E>6.18,N>12.36,  \label{n=1}
\end{equation}
and for $n=3$ the stability region is
\begin{equation}
E>6.74,N>13.48.  \label{n=3}
\end{equation}
In fact, Eqs. (\ref{n=1}) and (\ref{n=3}) define a nontrivial stability
border for the compact modes in the nonlinear chain. A representative
example of the dynamics of an unstable compact mode with $n=1$ and $E=5$ is
displayed in Fig. \ref{AntisymmetricEvolution_E5}, which shows that the
instability destroys it.

\subsection{Asymmetric lattice solitons}

\label{sec:LocalizedAsymmetricStates}

\begin{figure}[th]
\centering%
\subfigure[]{\includegraphics[scale=0.50]{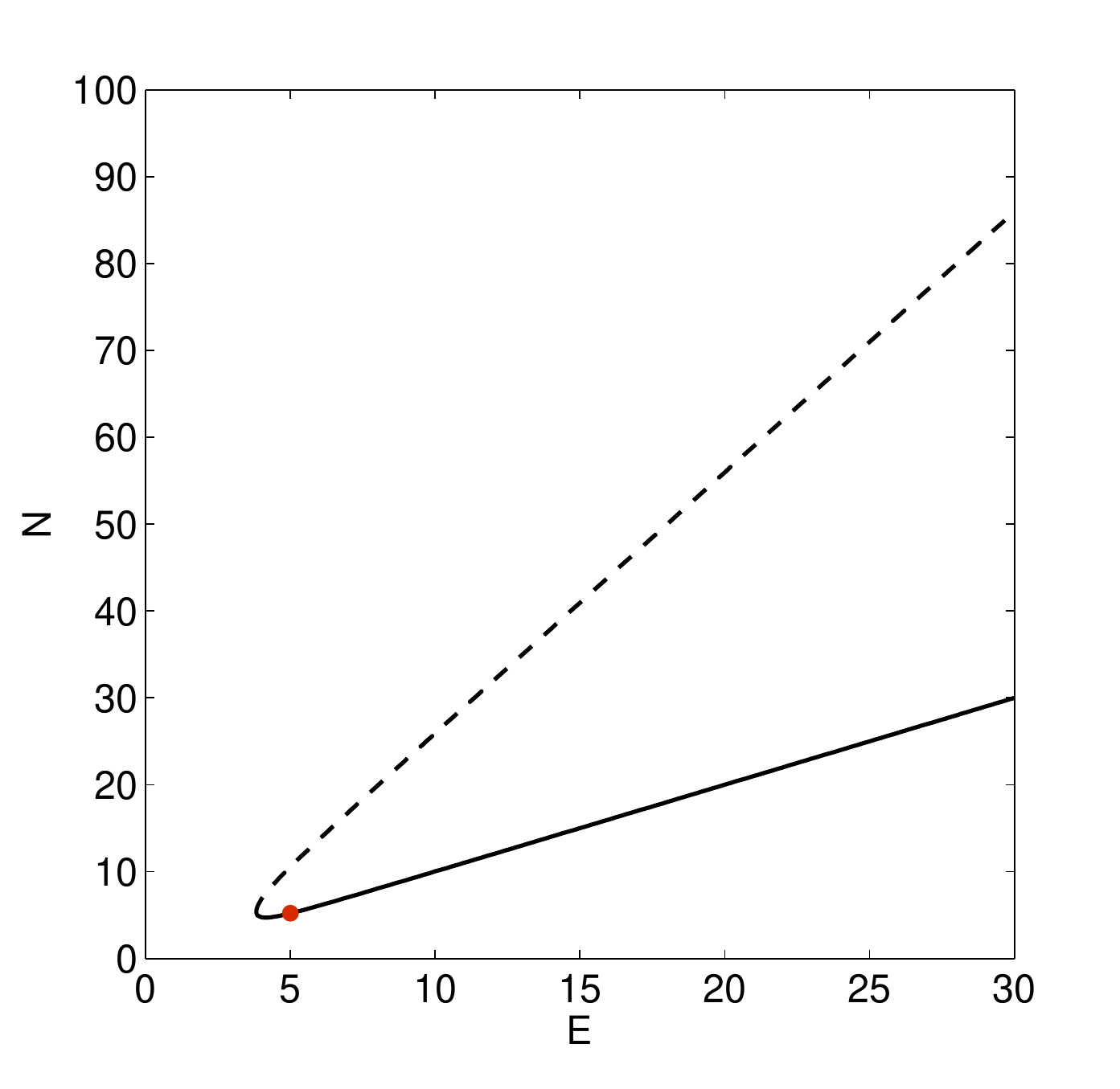}
\label{AsymmetricSymmetric1_NvsE}} \subfigure[]{%
\includegraphics[scale=0.50]{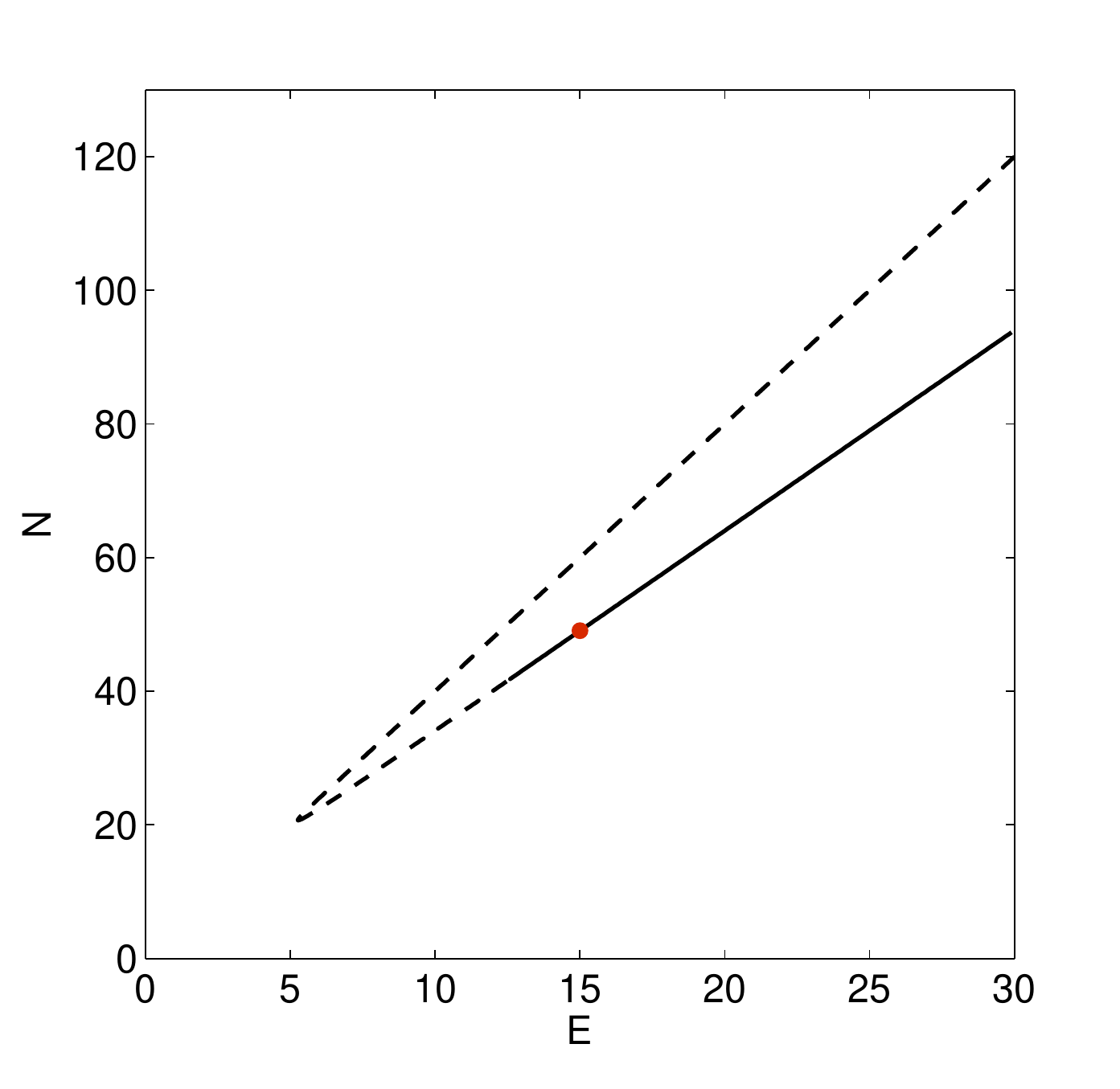}%
\label{AsymmetricAntisymmetric1_NvsE}} \subfigure[]{%
\includegraphics[scale=0.50]{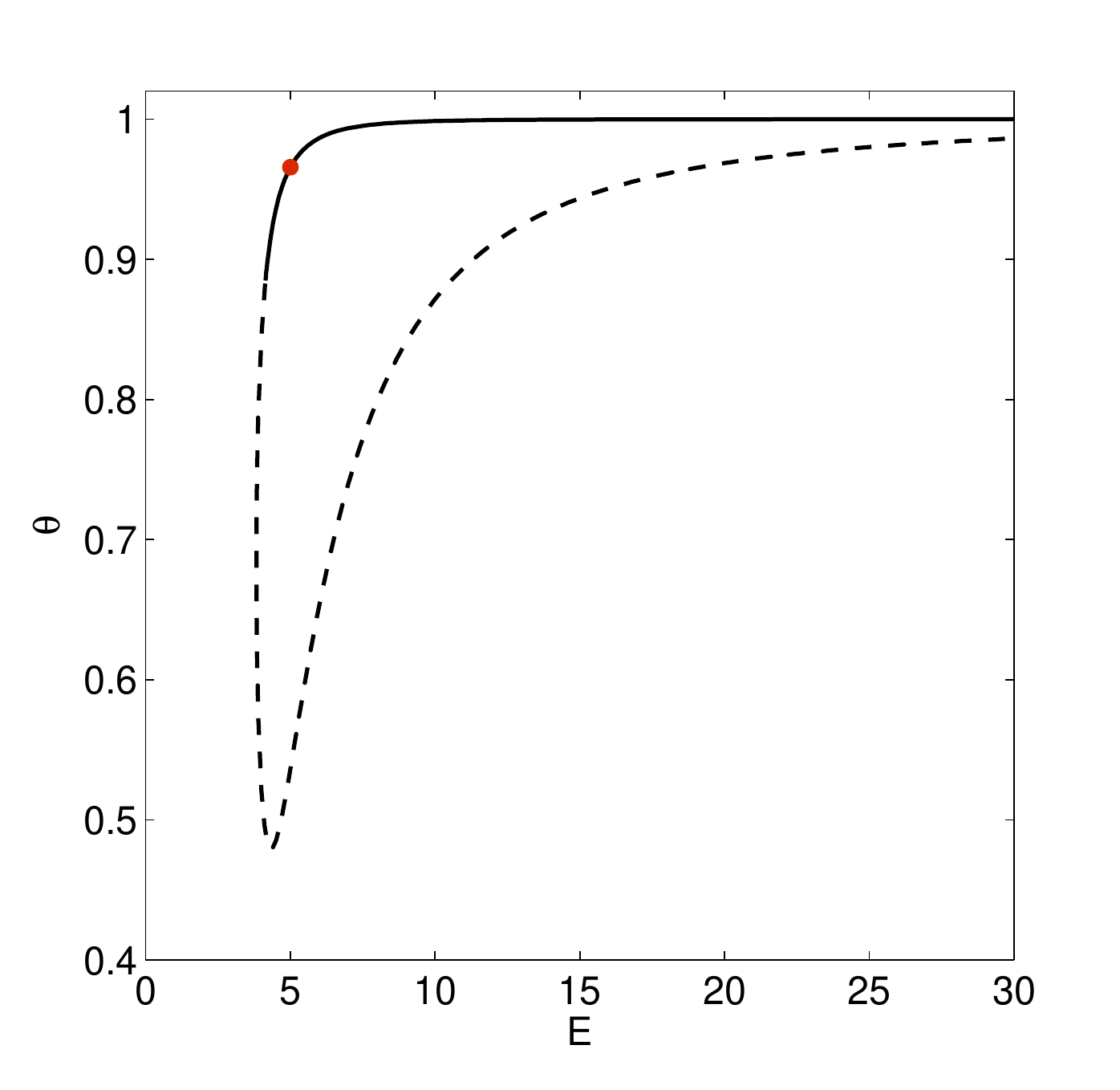}%
\label{AsymmetricSymmetric1_ThetavsE}} \subfigure[]{%
\includegraphics[scale=0.50]{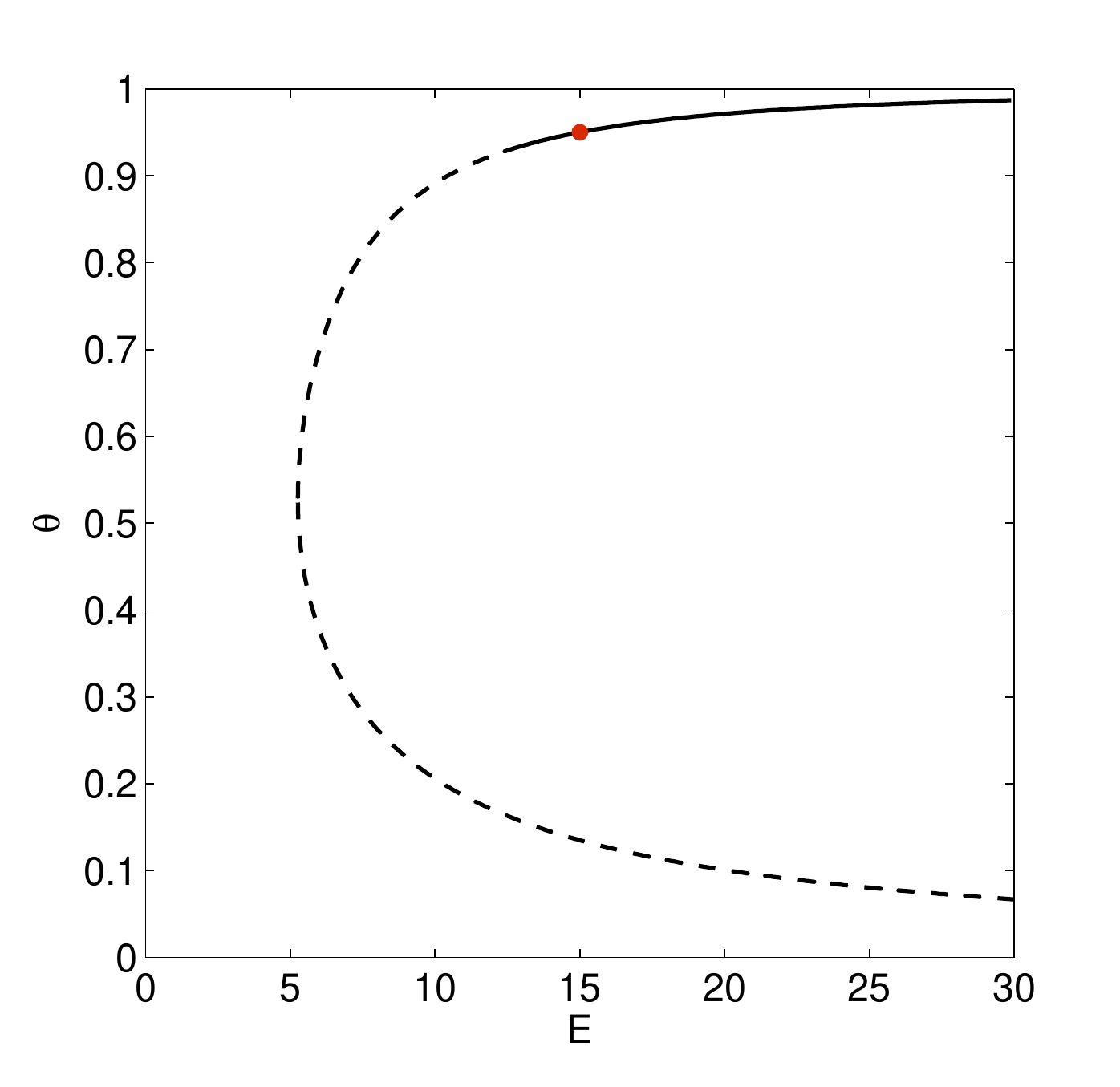}%
\label{AsymmetricAntisymmetric1_ThetavsE}}
\caption{The total norm versus the propagation constant, $E$, for the
families of asymmetric-symmetric lattice solitons (a) and
asymmetric-antisymmetric ones (b) (see definitions in the text). The marked
points correspond to the examples displayed in Fig. \protect\ref%
{AsymmetricProfiles1and2}. (c,d): The same families as in (a,b), but shown
by means of the $\protect\theta (E)$ curves, where the asymmetry ratio, $%
\protect\theta $, is defined in Eq. \protect\ref{theta}.}
\label{AsymmetricDiagrams}
\end{figure}

\begin{figure}[th]
\centering%
\subfigure[]{\includegraphics[scale=0.5]{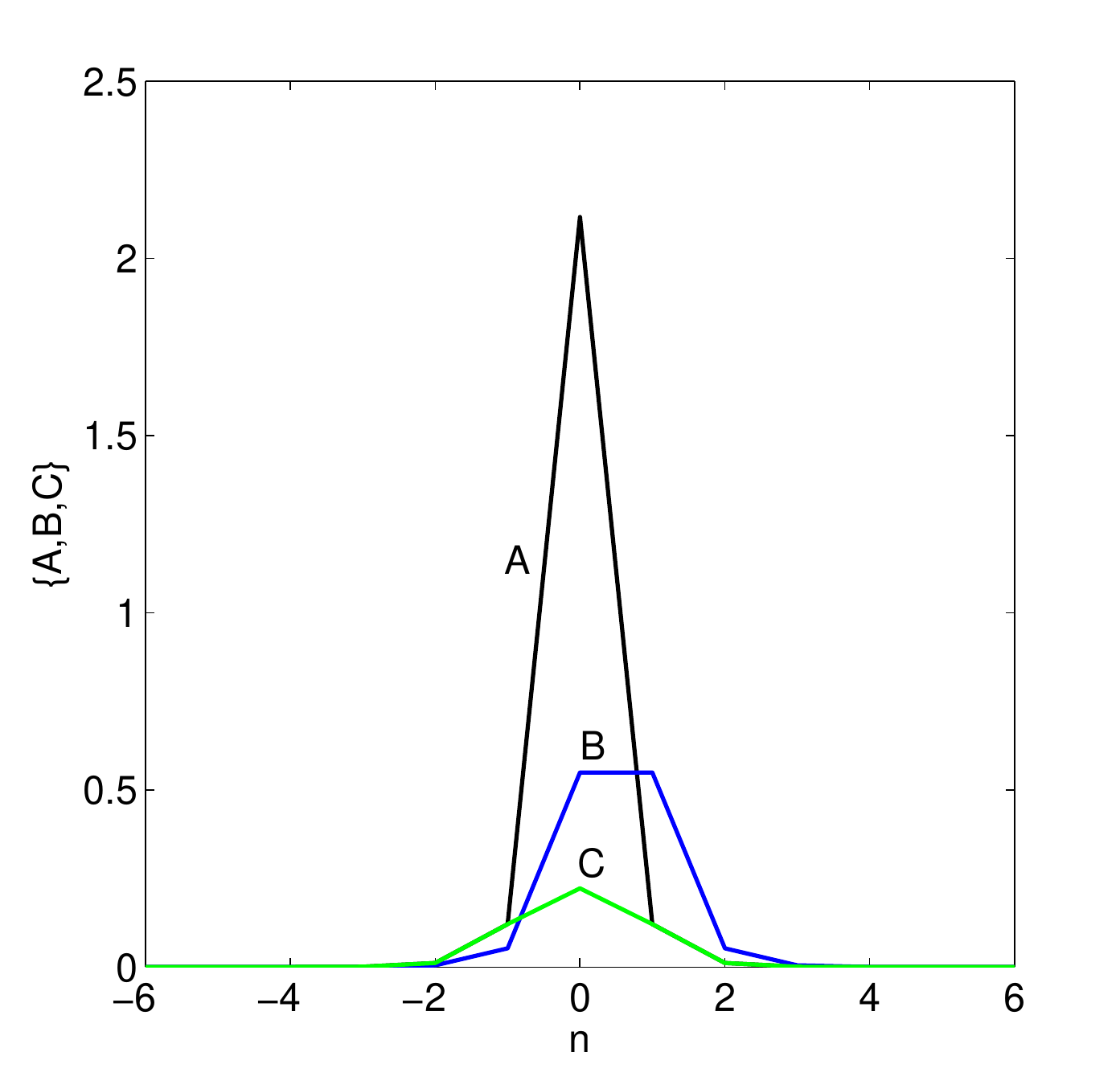}
\label{AsymmetricSymmetric1_E5}} \subfigure[]{%
\includegraphics[scale=0.5]{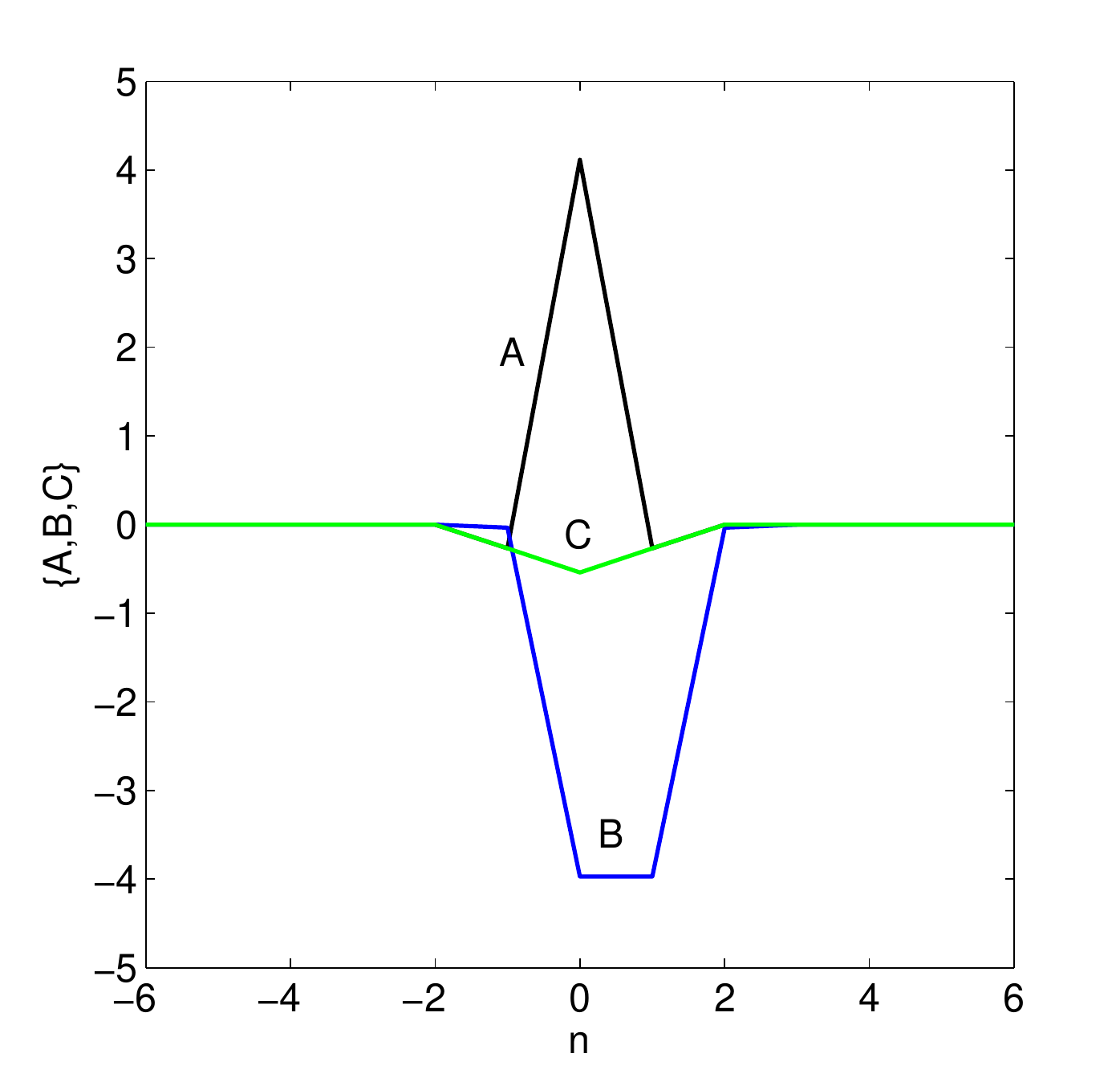}%
\label{AsymmetricAntisymmetric1_E15}}
\caption{(Color online) Examples of profiles of (a) asymmetric-symmetric
lattice solitons, and (b) asymmetric-antisymmetric ones. Panels (a) and (b)
correspond to the marked points in the lower branches of the curves
displayed in Figs. \protect\ref{AsymmetricSymmetric1_NvsE} ($E=5,N=5.2$) and
\protect\ref{AsymmetricAntisymmetric1_NvsE} ($E=15,N=49.04$), respectively.}
\label{AsymmetricProfiles1and2}
\end{figure}

Numerous families of discrete asymmetric lattice states, with $\left\vert
A_{n}\right\vert \neq \left\vert C_{n}\right\vert $, were discovered in the
course of the numerical investigation. Two such fundamental families, which
were found to be partially stable, are presented in Figs. \ref%
{AsymmetricSymmetric1_NvsE}-\ref{AsymmetricAntisymmetric1_NvsE} (the full
stability diagrams) and in Fig. \ref{AsymmetricProfiles1and2} (shape
examples). Additionally, Figs. \ref{AsymmetricSymmetric1_ThetavsE} and \ref%
{AsymmetricAntisymmetric1_ThetavsE} present the asymmetry ratio, $\theta $,
defined here as:

\begin{equation}
\theta =\frac{\sum_{-\infty }^{+\infty }\left( A_{n}^{2}-C_{n}^{2}\right) }{%
\sum_{-\infty }^{+\infty }\left( A_{n}^{2}+C_{n}^{2}\right) },  \label{theta}
\end{equation}%
as a function of $E$, cf. the above definition for the single cell, given by
Eq. (\ref{Theta}).

In particular, the solutions displayed in Fig. \ref{AsymmetricSymmetric1_E5}%
, which may be referred to as \textit{asymmetric-symmetric modes},\textit{\ }%
as they feature $\mathrm{sign}(A_{n})=\mathrm{sign}(C_{n})$, are stable in
the region $E>4.14$, $N>4.7$ [the lower $N(E)\ $branch]. The second family,
which we name the \textit{asymmetric-antisymmetric} \textit{solutions}, with
$\mathrm{sign}(A_{n})=-\mathrm{sign}(C_{n})$, shown in Fig. \ref%
{AsymmetricAntisymmetric1_E15}, is stable at $E>12.58$, $N>41.8$ (the lower
branch).

\begin{figure}[th]
\centering%
\subfigure[]{\includegraphics[scale=0.5]{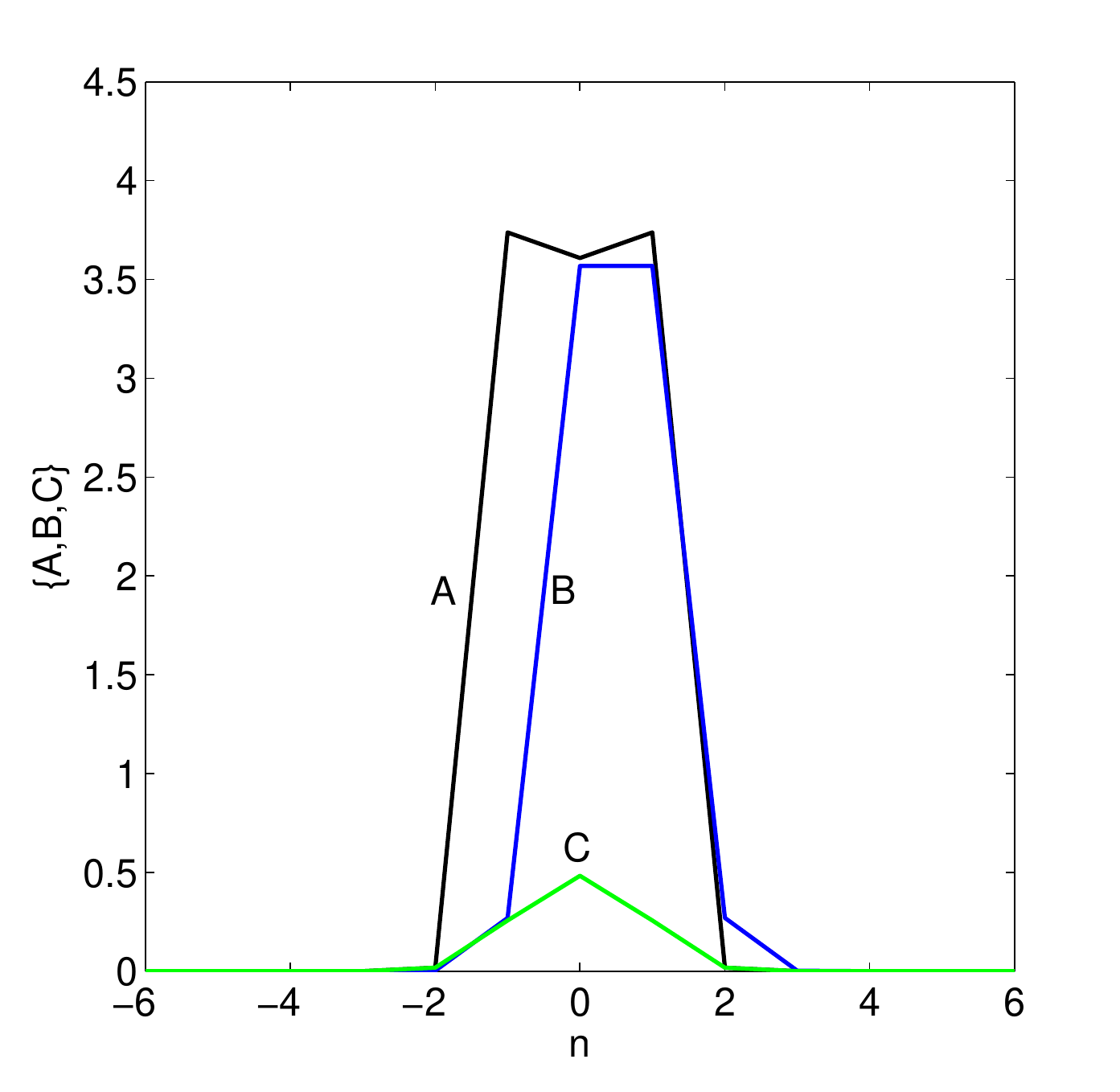}
\label{AsymmetricSymmetric2_E15}} \subfigure[]{%
\includegraphics[scale=0.5]{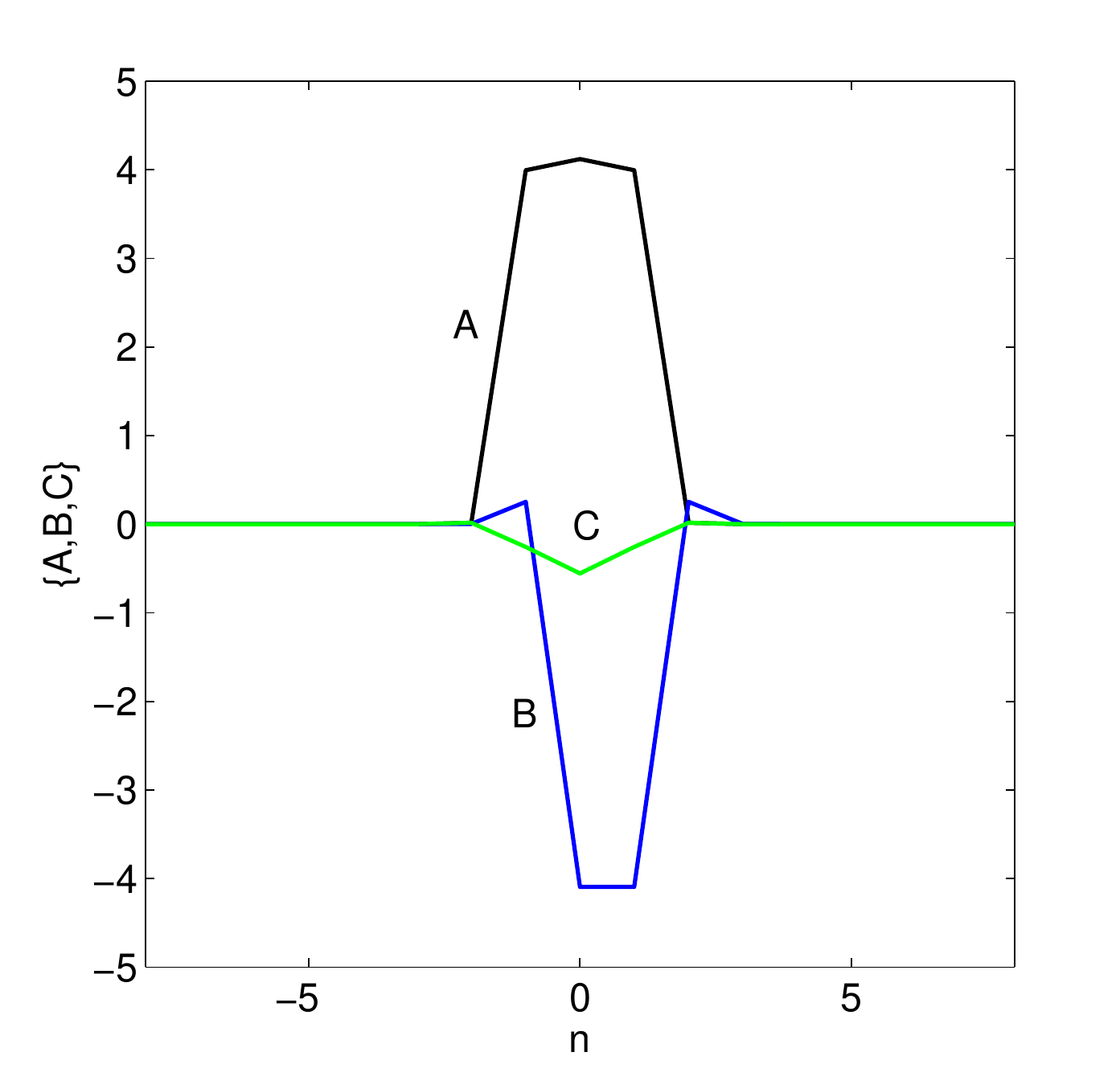}%
\label{AsymmetricAntisymmetric2_E15}}
\caption{(Color online) Typical profiles of two asymmetric lattice solitons
belonging to the two additional families, both with $E=15$.}
\label{AsymmetricProfiles3and4}
\end{figure}

Other asymmetric lattice solitons were also found, two of which are
presented in Fig. \ref{AsymmetricProfiles3and4}. While the family
demonstrated in Fig. \ref{AsymmetricSymmetric2_E15} is entirely unstable,
the one displayed in Fig. \ref{AsymmetricAntisymmetric2_E15} does have a
stability region (not shown here explicitly).

\begin{figure}[th]
\centering%
\subfigure[]{\includegraphics[scale=0.5]{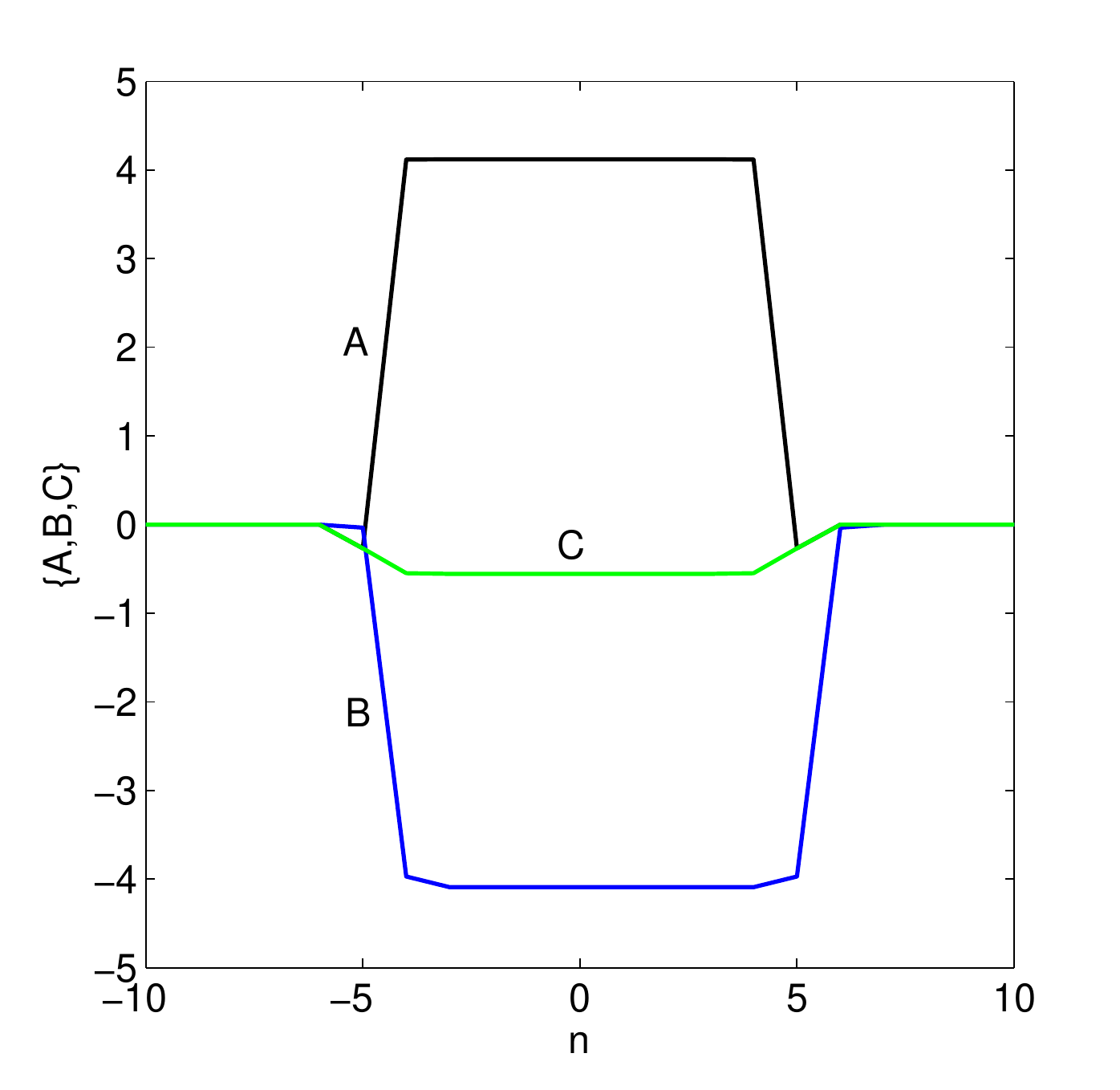}
\label{AsymmetricFT_E15}} \subfigure[]{%
\includegraphics[scale=0.5]{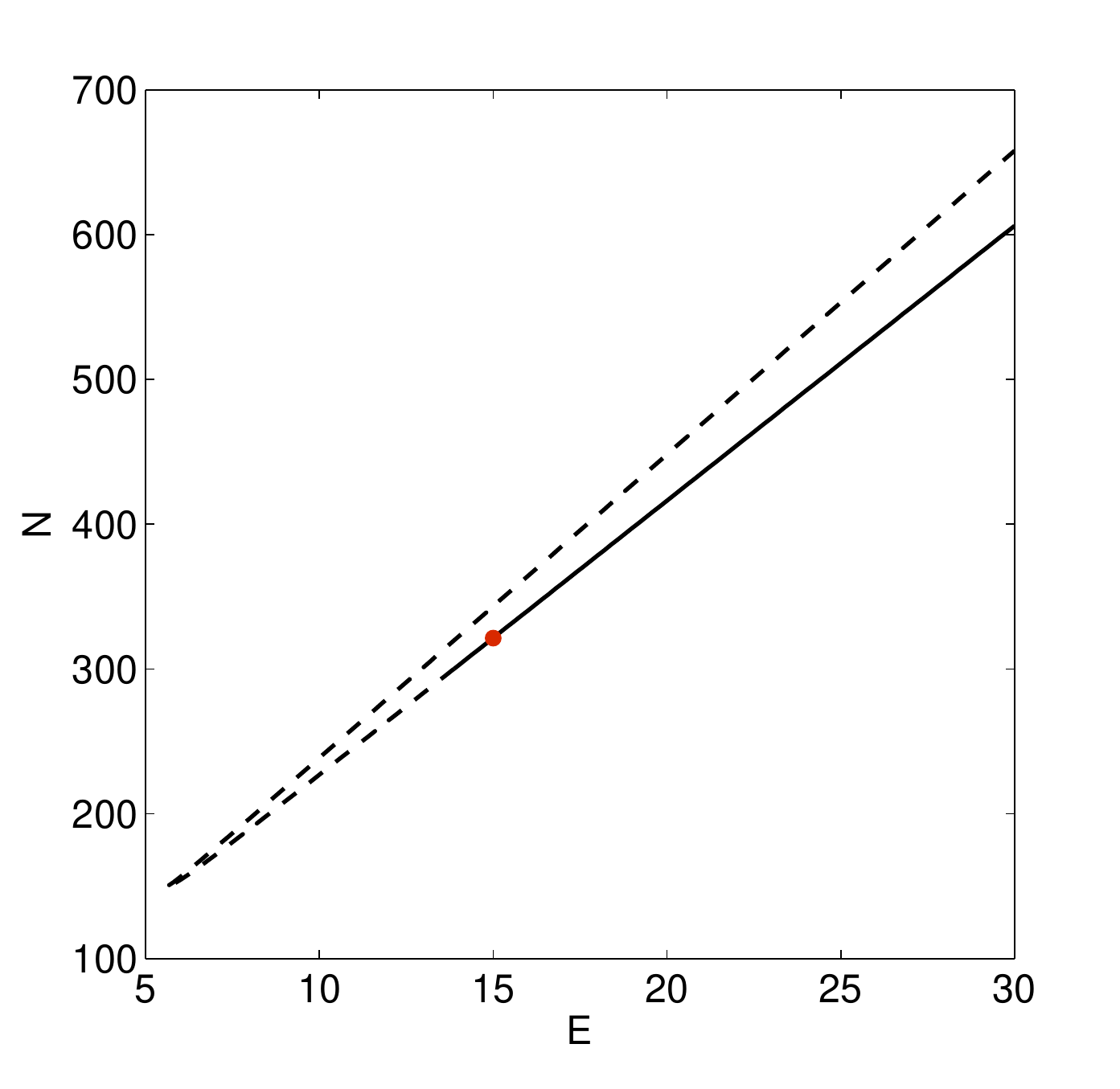}\label{AsymmetricFT_NvsE}}
\caption{(color online) (a) An example an extended confined asymmetric
state, for $E=15$. This mode is a combination of the localized asymmetric
one from Fig. \protect\ref{AsymmetricAntisymmetric1_E15} and an 8-cells
segment of the asymmetric CW from Fig. \protect\ref{AntiSymmetricAsymCW1}.
(b) The stability diagram for the family of these extended confined
asymmetric states. The marked point corresponds to the asymmetric state
shown in panel (a).}
\label{AsymmetricFT}
\end{figure}

Additional varieties of both symmetric and asymmetric localized states can
be built by combining the lattice solitons found above (in particular,
taking any of the four types of the asymmetric solitons mentioned here) and,
accordingly, symmetric or asymmetric CW segments, taken from the CW states
obtained in Section \ref{sec:CWSolutions}, with an arbitrary number of
lattice cells. An example of a so built extended confined asymmetric lattice
mode is given in Fig. \ref{AsymmetricFT}. This solution is a combination of
the asymmetric-antisymmetric one, presented above in Figs. \ref%
{AsymmetricAntisymmetric1_NvsE}, \ref{AsymmetricAntisymmetric1_ThetavsE}, %
\ref{AsymmetricAntisymmetric1_E15}, and an eight-cell section of the
asymmetric CW state, taken from Fig. \ref{AntiSymmetricAsymCW1}. It can be
checked that, for the stability of this combined solution, both its building
blocks must be stable. In particular, for the combined mode shown here, the
stability region is $E>13.88$, $N>300.10$, as its ingredients are stable in
the same interval.

\section{The variational approximation for lattice solitons in the single
infinite chain}

\subsection{The formulation}

A possibility to produce results for solitons in an analytical form, even if
it is an approximate one, is obviously relevant. In this section we aim to
develop a variational approximation (VA) for lattice solitons, and compare
the so produced approximate analytical results to their numerical
counterparts reported in the previous section. To this end, we note that the
Lagrangian corresponding to Hamiltonian (\ref{H}) is\newline
\begin{eqnarray}
L &=&\frac{i}{2}\sum_{n}\left( a_{n}^{\ast }\frac{da_{n}}{dz}-a_{n}\frac{%
da_{n}^{\ast }}{dz}\right)  \notag \\
&+&\frac{i}{2}\sum_{n}\left( b_{n}^{\ast }\frac{db_{n}}{dz}-b_{n}\frac{%
db_{n}^{\ast }}{dz}\right)  \notag \\
&+&\frac{i}{2}\sum_{n}\left( c_{n}^{\ast }\frac{dc_{n}}{dz}-c_{n}\frac{%
dc_{n}^{\ast }}{dz}\right) -H  \notag \\
&\equiv &-EN-H,  \label{Lagra1}
\end{eqnarray}%
from which underlying equations (\ref{abc1}) can be derived as standard
Euler-Lagrange equations, with $E$ playing the role of the Lagrangian
multiplier.

We apply the VA to stationary lattice solitons, following the general lines
of Ref. \cite{MIW}, where the VA was developed for solitons in the discrete
nonlinear Schr\"{o}dinger equation. The stability of stationary solutions
can be tested by checking if they realize a local minimum of the
Hamiltonian. The VA is based on the following\ two simplest \textit{ans\"{a}%
tze} applicable to discrete solitons, which differ by assuming the double
maximum in components $b_{n}$ or $\left( a_{n},c_{n}\right) $:
\begin{equation}
\mathrm{I}:\left\{
\begin{array}{c}
a_{n}(z)=Ae^{-\eta \left\vert n-n_{0}\right\vert }e^{iEz}, \\
b_{n}(z)=Be^{-\eta \left\vert n-n_{0}-\frac{1}{2}\right\vert }e^{iEz}, \\
c_{n}(z)=Ce^{-\eta \left\vert n-n_{0}\right\vert }e^{iEz};%
\end{array}%
\right.  \label{firsttype}
\end{equation}%
\begin{equation}
\mathrm{II}:\left\{
\begin{array}{c}
a_{n}(z)=Ae^{-\eta \left\vert n-n_{0}+\frac{1}{2}\right\vert }e^{iEz}, \\
b_{n}(z)=Be^{-\eta \left\vert n-n_{0}\right\vert }e^{iEz}, \\
c_{n}(z)=Ce^{-\eta \left\vert n-n_{0}+\frac{1}{2}\right\vert }e^{iEz},%
\end{array}%
\right.  \label{secondtype}
\end{equation}%
where $n_{0}$ is an arbitrary integer coordinate of the soliton's center.
The \textit{ans\"{a}tze }contain four variational parameters: three
amplitudes, $A$, $B$, $C$, and inverse width $\eta $.

As mentioned above, ansatz (\ref{firsttype}) can be used to describe
stationary localized states with sharp ${A_{n},C_{n}}$ and flat-top $B_{n}$
profiles (see Figs. \ref{SymmetricLocalized2_E10}, \ref%
{SymmetricLocalized_E2p9} and \ref{AsymmetricSymmetric1_E5}), while ansatz (%
\ref{secondtype}) pertains to sharp $B_{n}$ and flat-top $\left\{ {A}_{n}{,C}%
_{n}\right\} $ profiles, see Fig. \ref{SymmetricLocalized1_E10}). We
substitute these \textit{ans\"{a}tze }into the Hamiltonian (\ref{H}) and
Lagrangian (\ref{Lagra1}) and perform the summation analytically, which
yields
\begin{equation}
H_{\mathrm{I}}=-\frac{2B\left( A+C\right) }{\sinh \left( \eta /2\right) }-%
\frac{B^{4}+\left( A^{4}+C^{4}\right) \cosh \left( 2\eta \right) }{2\sinh
\left( 2\eta \right) },  \label{HamilItype}
\end{equation}%
\begin{equation}
L_{\mathrm{I}}=-\frac{E\left( B^{2}+\left( A^{2}+C^{2}\right) \cosh \left(
\eta \right) \right) }{\sinh \left( \eta \right) }-H_{\mathrm{I}}~;
\label{LagItype}
\end{equation}%
\begin{equation}
H_{\mathrm{II}}=-\frac{2B\left( A+C\right) }{\sinh \left( \eta /2\right) }-%
\frac{A^{4}+C^{4}+B^{4}\cosh \left( 2\eta \right) }{2\sinh \left( 2\eta
\right) },  \label{HamilIItype}
\end{equation}%
\begin{equation}
L_{\mathrm{II}}=-\frac{E\left( A^{2}+C^{2}+B^{2}\cosh \left( \eta \right)
\right) }{\sinh \left( \eta \right) }-H_{\mathrm{II}}.  \label{LagIItype}
\end{equation}%
The stationary states are predicted by numerically solving the corresponding
Euler-Lagrange equations,
\begin{equation}
\frac{\partial L_{\mathrm{I,II}}}{\partial A}=\frac{\partial L_{\mathrm{I,II}%
}}{\partial B}=\frac{\partial L_{\mathrm{I,II}}}{\partial C}=\frac{\partial
L_{\mathrm{I,II}}}{\partial \eta }=0.
\end{equation}

\subsection{Comparison between variational and numerical results for the
infinite chain}

\begin{figure}[tbp]
\begin{center}
\includegraphics[width=11cm]{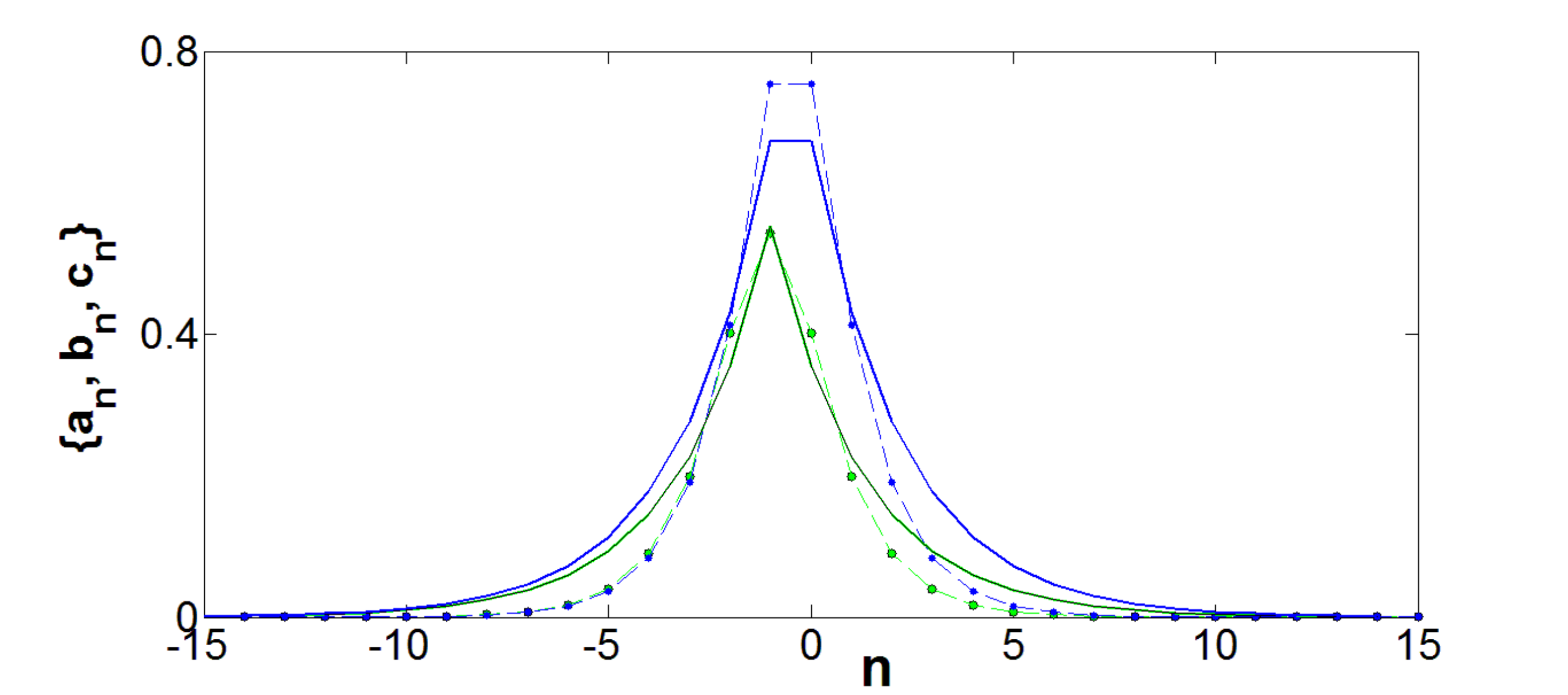}
\end{center}
\caption{(Color online) Comparison between predictions of the VA based on
ansatz (I) with $n_{0}=1$ (solid curves) and numerical results (dashed
curves with small cicles). Here $N=3$, and the results are presented for
symmetric lattice solitons. The respective values of the parameters are $E_{%
\mathrm{VA}}=3.014$, $E_{\mathrm{num}}=3.072$, and $H_{\mathrm{VA}}=8.665$, $%
H_{\mathrm{num}}=8.715$. The blue curves represent $B_{n}$, while the black
and green ones correspond to $A_{n}$ and $C_{n}$, respectively.}
\label{Kstates1}
\end{figure}

\begin{figure}[tbp]
\begin{center}
\includegraphics[width=11cm]{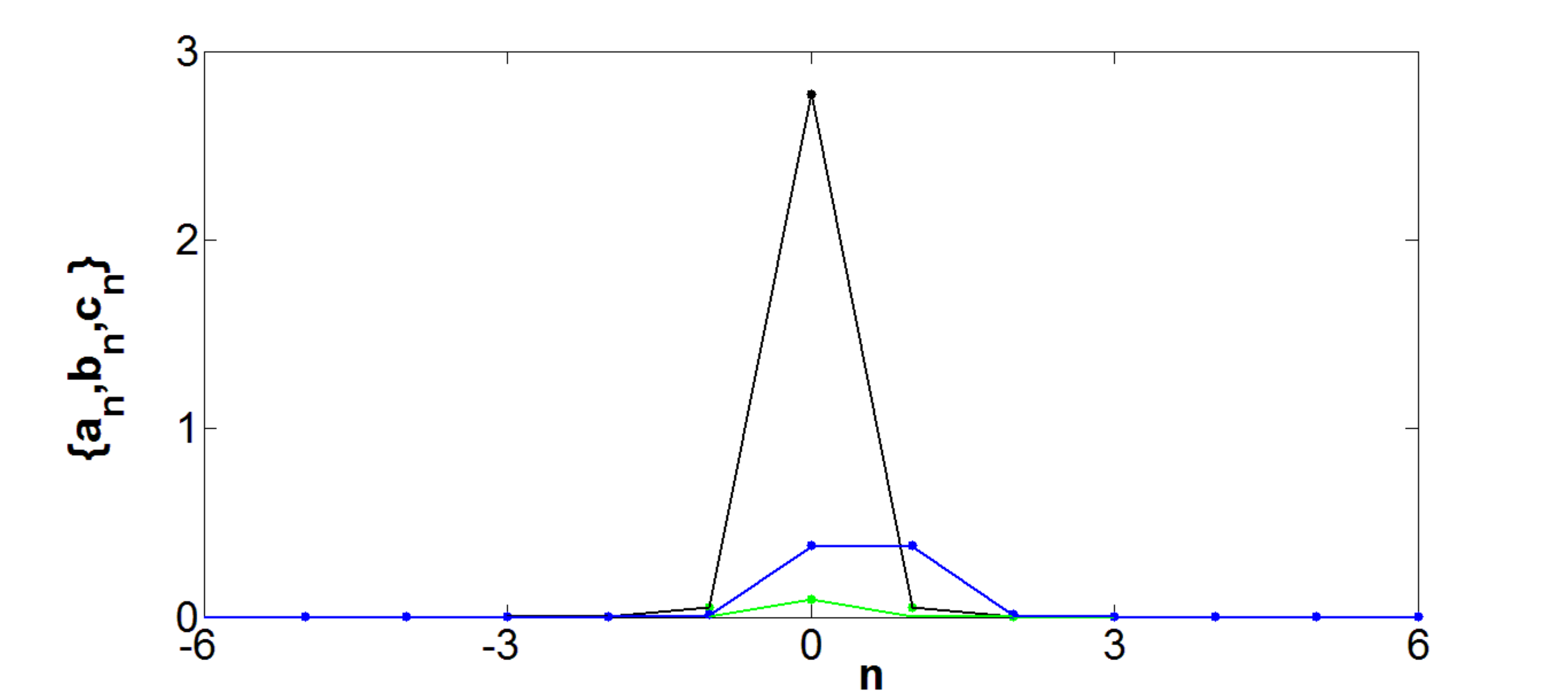}
\end{center}
\caption{(Color online) The same as in Fig. \protect\ref{Kstates1}, but for
an asymmetric state with $N=8$ and $E_{\mathrm{VA}}=7.976$, $H_{\mathrm{VA}%
}=34.09$, $E_{\mathrm{num}}=7.966$, $H_{\mathrm{num}}=34.12$.}
\label{Kstates2}
\end{figure}

In Figs. \ref{Kstates1} and \ref{Kstates2} we compare the VA predictions and
their numerical counterparts for stationary modes with sharp $%
\{A_{n},C_{n}\} $ and flat-top $B_{n}$ shapes, which are approximated by
ansatz I [Eq. (\ref{firsttype})] with norms $N=3$ and $N=8$, respectively.
The VA predicts both symmetric and asymmetric solutions. One can conclude
that the agreement is better for larger $N$. This conclusion is natural, as
larger $N$ correspond to more self-compressed solitons, for which the simple
exponential \textit{ans\"{a}tze }are more appropriate. \newline

The second type of stationary solutions has flat $\{A_{n},C_{n}\}$ and sharp
$B_{n}$ profiles, which can be approximated by ansatz II. In Figs. \ref%
{Hstates1} and \ref{Hstates2} we present the comparisons between the
corresponding VA predictions and numerical results. We again conclude that a
larger norm provides better agreement.\newline

\begin{figure}[th]
\caption{(Color online) (a) The dependences $N(E)$ for lattice solitons, as
predicted by the VA. The same dependences are compared to their numerically
found counterparts in (b). Dashed pink curves in (a) and (b) represent
unstable asymmetric solutions based on Ansatz I. Solutions of the same type
but found numerically are represented by small pink stars in panel (b).
Dotted black curves in (a) and (b) show unstable symmetric solutions based
on Ansatz I. The corresponding numerical solutions are shown by black
triangles in (b). Thin solid red curves in (a) and (b) indicate stable
symmetric solutions produced by Ansatz II, while their numerical
counterparts are shown by red squares in (b). Thick solid blue curves
correspond to stable asymmetric solutions, produced by Ansatz I, whose
numerical counterparts are displayed are bold blue dots in (b). }
\label{muinormaz}\centering%
\subfigure[]{\includegraphics[scale=0.28]{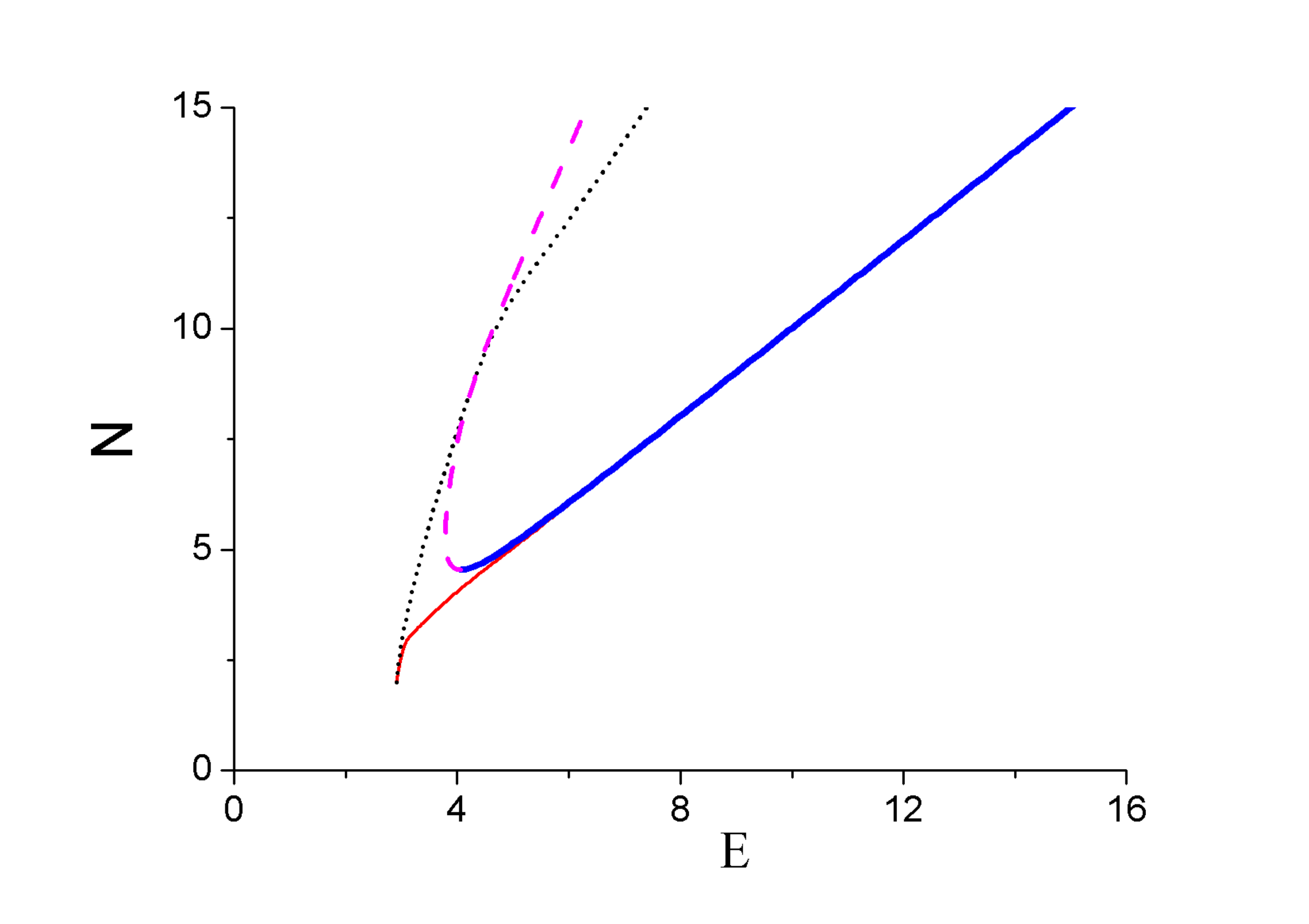}
\label{miunorm1111}}
\subfigure[]{\includegraphics[scale=0.28]{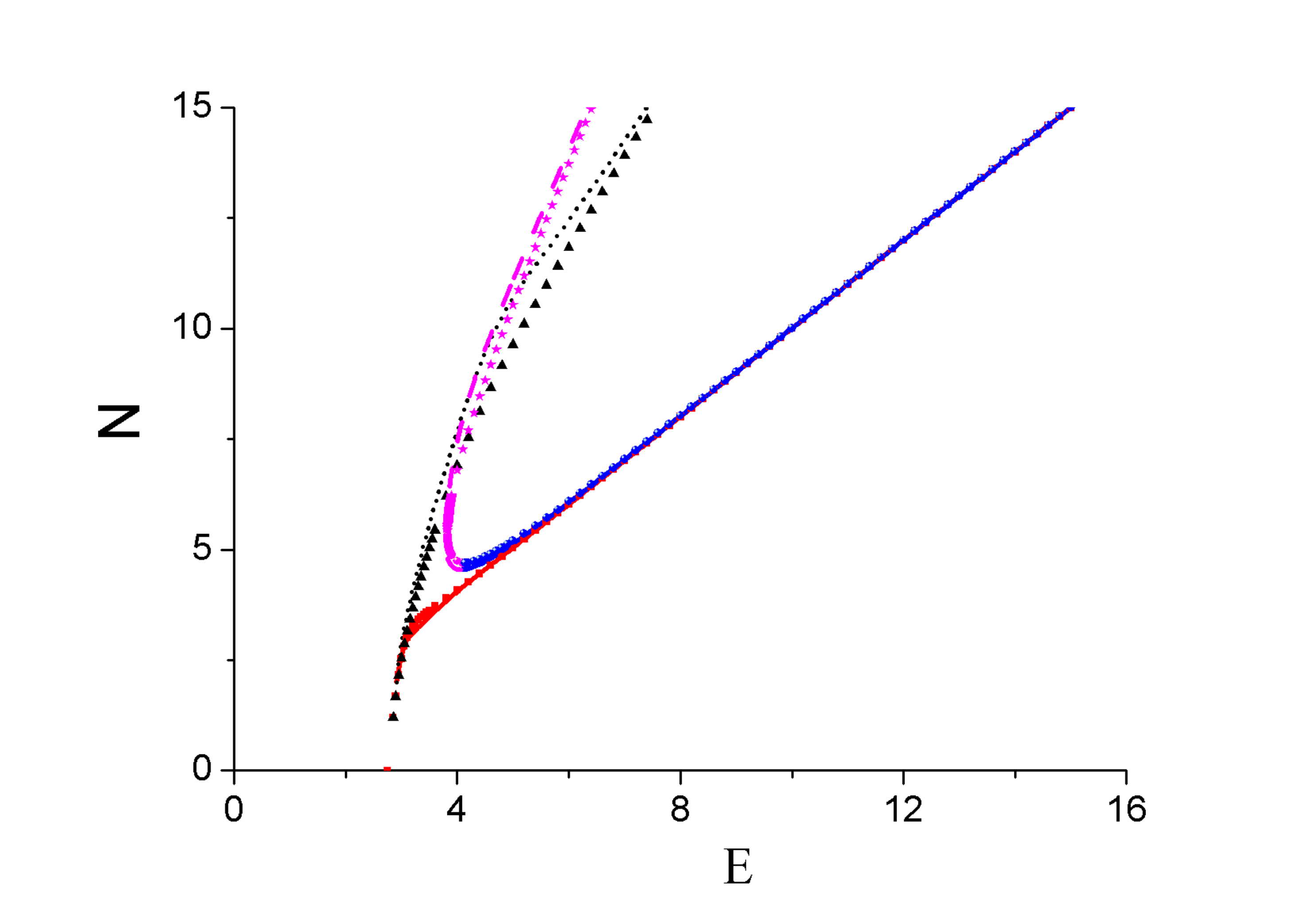}
\label{muinormcomp1111}}
\end{figure}

\begin{figure}[tbp]
\caption{(Color online) The asymmetry parameters, $\protect\theta (E)$, of
lattice solitons. The solid blue curve is the branch of stable asymmetric
solutions predicted by the VA, while the dashed pink curve corresponds to
unstable ones. The corresponding numerical solutions are represented by dots
and stars of the same colors. }
\label{muithetaz}
\begin{center}
\includegraphics[scale=0.28]{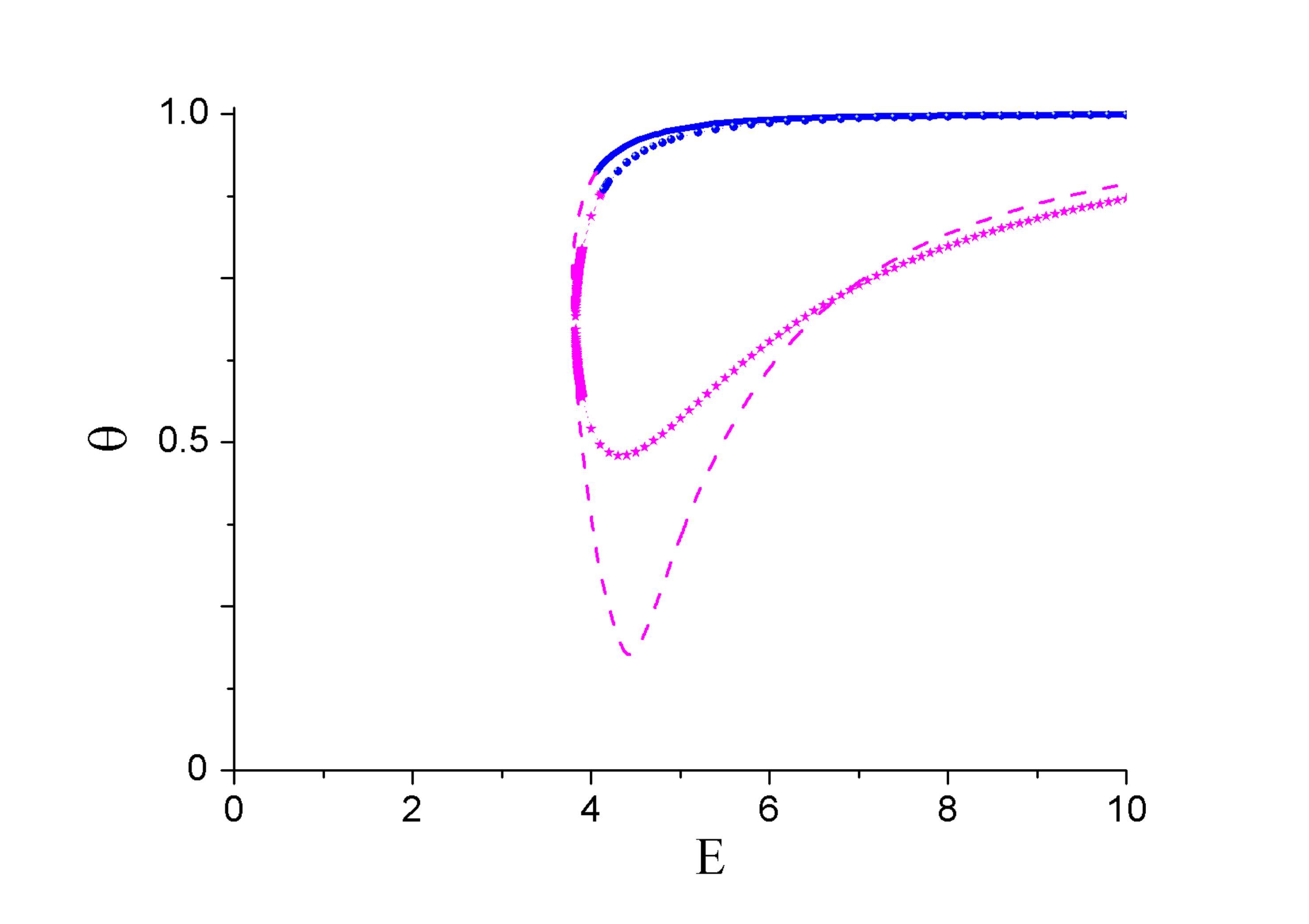}
\end{center}
\end{figure}

Results of the systematic comparison of the variational and numerical
results are summarized in Figs. \ref{muinormaz}-\ref{muithetaz}. In
particular, the VA-predicted $N(E)$ curves are shown in Fig. \ref%
{miunorm1111} for families of symmetric and asymmetric lattice solitons. The
solid and dashed curves again correspond to stable and unstable ones.
Further, these curves are compared to their numerically found counterparts
in Fig. \ref{muinormcomp1111}. One can see that the VA\ predictions agree
well with the numerical data taken from Figs. \ref%
{SymmetricLocalized1and2_NvsE} and \ref{AsymmetricSymmetric1_NvsE}. As
before, a larger propagation constant $E$ provides for better agreement.

%

Finally, in Fig. \ref{muithetaz} we compare dependences $\theta (E)$ for the
asymmetry parameter of the lattice solitons, defined as per Eq. (\ref{theta}%
). The respective numerical data are taken from Fig. \ref%
{AsymmetricSymmetric1_ThetavsE}.

\begin{figure}[tbp]
\begin{center}
\includegraphics[width=11cm]{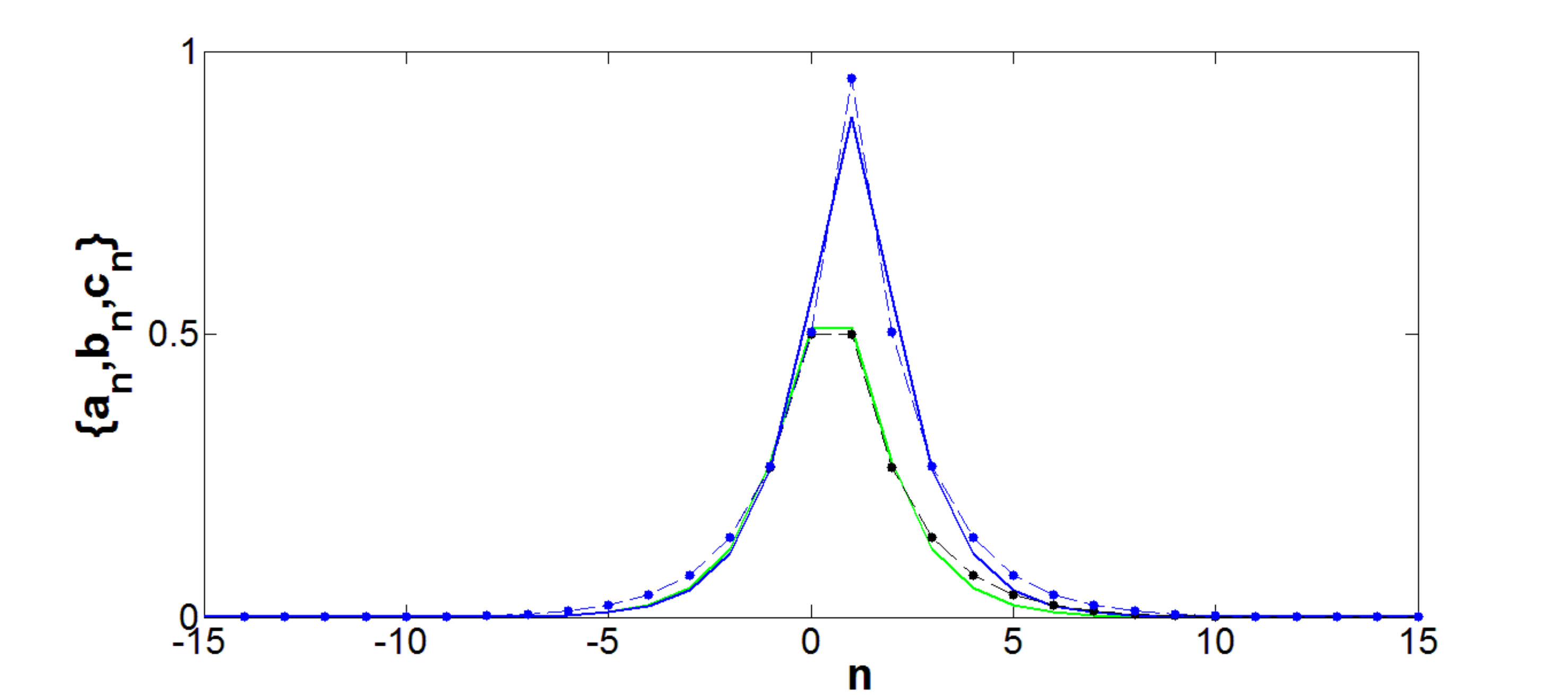}
\end{center}
\caption{(Color online) The same as in Fig. \protect\ref{Kstates1}, but for
the variational predictions based on ansatz II with $n_{0}=-1$ and $N=3$,
for symmetric lattice solitons. The parameters are $E_{\mathrm{VA}}=3.095$, $%
E_{\mathrm{num}}=3.107$, $H_{\mathrm{VA}}=8.703$, and $H_{\mathrm{num}%
}=8.726 $.}
\label{Hstates1}
\end{figure}

\begin{figure}[tbp]
\begin{center}
\includegraphics[width=11cm]{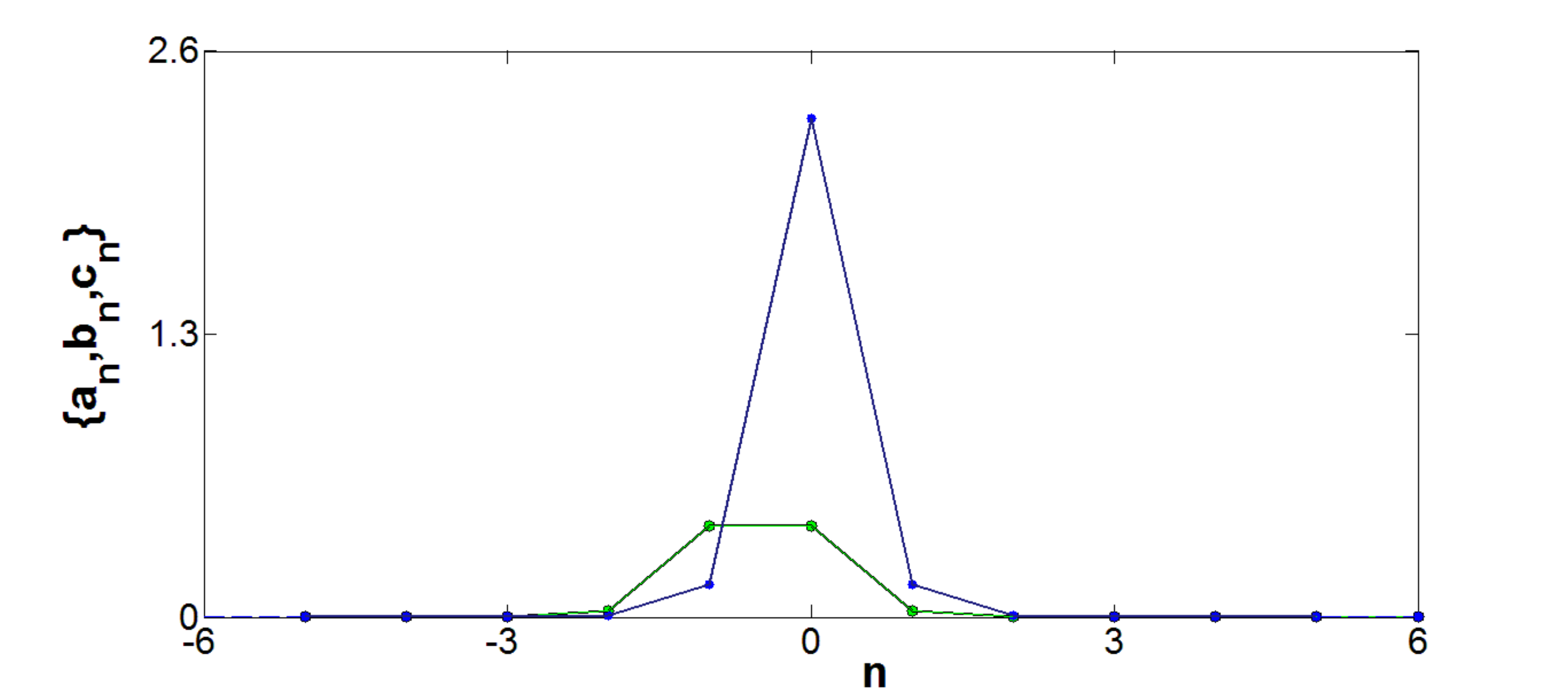}
\end{center}
\caption{(Color online) The same as in Fig. \protect\ref{Hstates1}, but with
$n_{0}=0$, $N=6$ and $E_{\mathrm{VA}}=5.979$ and $H_{\mathrm{VA}}=22.059$, $%
E_{\mathrm{num}}=5.979$, $H_{\mathrm{num}}=22.059$.}
\label{Hstates2}
\end{figure}

\section{The linear double-chain model}

A natural generalization of the model based on Eq. (\ref{abc1}) is the
double system, with each on-site amplitude $a_{n},b_{n},c_{n}$ replaced by
the double set, $\left\{ a_{n}^{(1)},a_{n}^{(2)}\right\} ,\left\{
b_{n}^{(1)},b_{n}^{(2)}\right\} ,\left\{ c_{n}^{(1)},c_{n}^{(2)}\right\} $,
and linear mixing (Rabi coupling) applied at each site. Thus, Eqs. (\ref{a}%
)-(\ref{c}) are replaced by the double system,%
\begin{align}
& i\frac{da_{n}^{(1,2)}}{dz}+\left( b_{n}^{(1,2)}+b_{n+1}^{(1,2)}\right)
+\beta \left[ |a_{n}^{(1,2)}|^{2}+\gamma |a_{n}^{(2,1)}|^{2}\right]
a_{n}^{(1,2)}+\kappa _{ac}a_{n}^{(2,1)}=0,  \label{a12} \\
& i\frac{db_{n}^{(1,2)}}{dz}+\left(
a_{n}^{(1,2)}+a_{n-1}^{(1,2)}+c_{n}^{(1,2)}+c_{n-1}^{(1,2)}\right) +\beta %
\left[ |b_{n}^{(1,2)}|^{2}+\gamma |b_{n}^{(2,1)}|^{2}\right]
b_{n}^{(1,2)}+\kappa _{b}b_{n}^{(2,1)}=0,  \label{b12} \\
& i\frac{dc_{n}^{(1,2)}}{dz}+\left( b_{n}^{(1,2)}+b_{n+1}^{(1,2)}\right)
+\beta \left[ |c_{n}^{(1,2)}|^{2}+\gamma |c_{n}^{(2,1)}|^{2}\right]
c_{n}^{(1,2)}+\kappa _{ac}c_{n}^{(2,1)}=0,  \label{c12}
\end{align}%
where $\gamma \geq 0$ is the relative strength of the on-site XPM
interaction.

In the BEC\ realization of the model, the on-site linear mixing between two
hyperfine atomic states may be induced by a resonant GHz wave, hence in that
case $\kappa _{ac}=\kappa _{b}\equiv \kappa $. In the BEC model, $\gamma =1$
is the most relevant value.\ In the optical realization, the two on-site
modes may represent two different polarizations of light in the same
waveguide. In the case of the linear polarizations, the linear mixing is
imposed by the waveguide's twist, the most relevant respective value of the
XPM coefficient being $\gamma =2/3$; in the case of two circular
polarizations, the mixing is imposed by an elliptic deformation of the
waveguide's cross section, $\gamma =2$ being the most relevant XPM\ value
\cite{Agrawal}. Alternatively, this system may correspond to the system of
optical dual-core waveguides \cite{Christo}, in which case $\gamma =0$. In
the optics model, it is quite natural to have two different linear-mixing
constants, $\kappa _{ac}\neq \kappa _{b}$.

For the double-chain model, the dispersion relation for modes $\sim \exp
\left( iEz+ikx\right) $ takes the following form:

\begin{equation}
\begin{vmatrix}
\ -E & \left( 1+e^{ik}\right) & 0 & \kappa _{ac} & 0 & 0 \\
\left( 1+e^{-ik}\right) & -E & \left( 1+e^{-ik}\right) & 0 & \kappa _{b} & 0
\\
0 & \left( 1+e^{ik}\right) & -E & 0 & 0 & \kappa _{ac} \\
\kappa _{ac} & 0 & 0 & -E & \left( 1+e^{ik}\right) & 0 \\
0 & \kappa _{b} & 0 & \left( 1+e^{-ik}\right) & -E & \left( 1+e^{-ik}\right)
\\
0 & 0 & \kappa _{ac} & 0 & \left( 1+e^{ik}\right) & -E%
\end{vmatrix}%
=0  \label{determinant}
\end{equation}%
In the special case of $\kappa _{ac}=\kappa _{b}=\kappa $, Eq. (\ref%
{determinant}) can be easily factorized:

\begin{equation}
\left( E^{2}-\kappa ^{2}\right) \left[ E^{4}-4\left( 1+e^{ik}\right) \left(
1+e^{-ik}\right) \left( E^{2}+\kappa ^{2}\right) -2\kappa ^{2}E^{2}+4\left(
1+e^{ik}\right) ^{2}\left( 1+e^{-ik}\right) ^{2}+\kappa ^{4}\right] =0.
\label{factorized equation}
\end{equation}%
Then, two eigenvalues represent FBs, $E=\pm \kappa $, with the top and
bottom signs corresponding to the symmetric and antisymmetric modes:
\begin{equation}
\left( a_{n}^{(1)},b_{n}^{(1)},c_{n}^{(1)}\right) =\pm \left(
a_{n}^{(2)},b_{n}^{(2)},c_{n}^{(2)}\right) ,  \label{symm}
\end{equation}%
while other eigenvalues produce dispersive branches, $E=\pm \kappa \pm 4\cos
\left( k/2\right) $.

In the general case, with $\kappa _{ac}\neq \kappa _{b}$, the analysis of
Eq. (\ref{determinant}) is facilitated by considering symmetric and
antisymmetric eigenstates, defined as per Eq. (\ref{symm}), which makes it
possible to reduce the $6\times 6$ determinant in Eq. (\ref{determinant}) to
$3\times 3$ ones. In this general case, two eigenvalues corresponding to the
split FBs are found in an \emph{exact form}:

\begin{equation}
E=\pm \kappa _{ac},  \label{flat band eigenvalues}
\end{equation}%
and four dispersive branches are found exactly too:

\begin{equation}
E=\pm \frac{1}{2}\left( \kappa _{ac}+\kappa _{b}\right) \pm \frac{1}{2}\sqrt{%
\left( \kappa _{ac}-\kappa _{b}\right) ^{2}+32\cos ^{2}\left( k/2\right) }.
\label{dispersive states eigenvalues}
\end{equation}%
In both cases, the $\pm $ sign before the first term identifies the
symmetric and antisymmetric eigenmodes, the $\pm $ sign in front of the
second term being an independent one.

For the flatband states corresponding to eigenvalues (\ref{flat band
eigenvalues}), the eigenvectors can also be obtained in an exact form:

\begin{equation*}
\psi _{n}^{\left( 1\right) }=\pm \psi _{n}^{(2)}=\left( 1,0,-1\right) f_{n},
\end{equation*}%
where $f_{n}$ is an arbitrary discrete function. In such a case, the CLS,
i.e., the single-cell excitation, corresponds to the following eigenvectors:

\begin{equation*}
\psi _{n}^{\left( 1\right) }=\frac{1}{\sqrt{2}}\left( 1,0,-1\right) \delta
_{n,n_{0}}=\pm \psi _{n}^{\left( 2\right) },
\end{equation*}%
cf. Eq. (\ref{branches}).

All the bands corresponding to Eqs. (\ref{flat band eigenvalues}) and (\ref%
{dispersive states eigenvalues}) are plotted in Fig. \ref{plot of energy
levels}.

\begin{figure}[t]
\centering\subfigure[]{\includegraphics[scale=0.35]{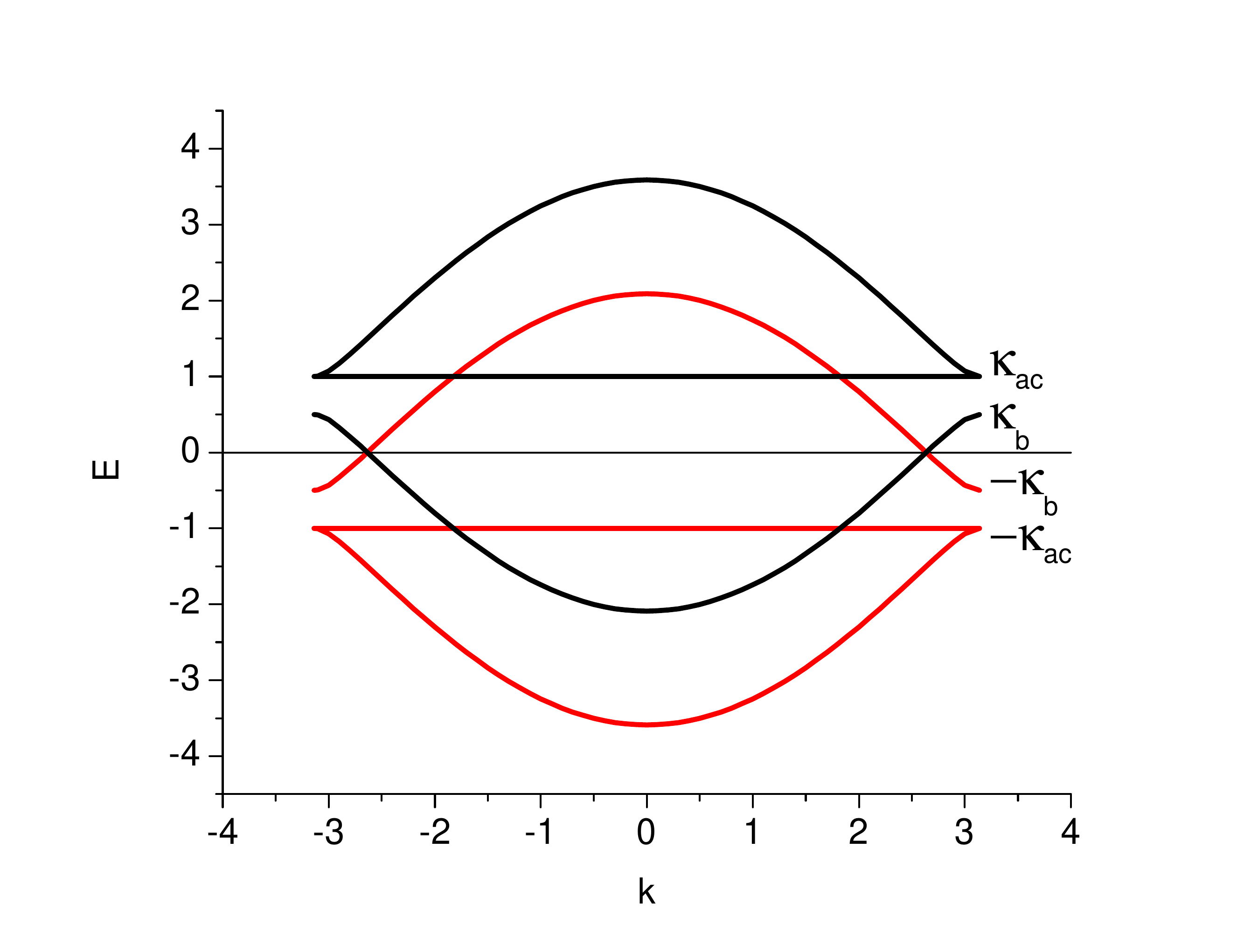}%
\label{bigger_kappa_ac}}\subfigure[]{%
\includegraphics[scale=0.35]{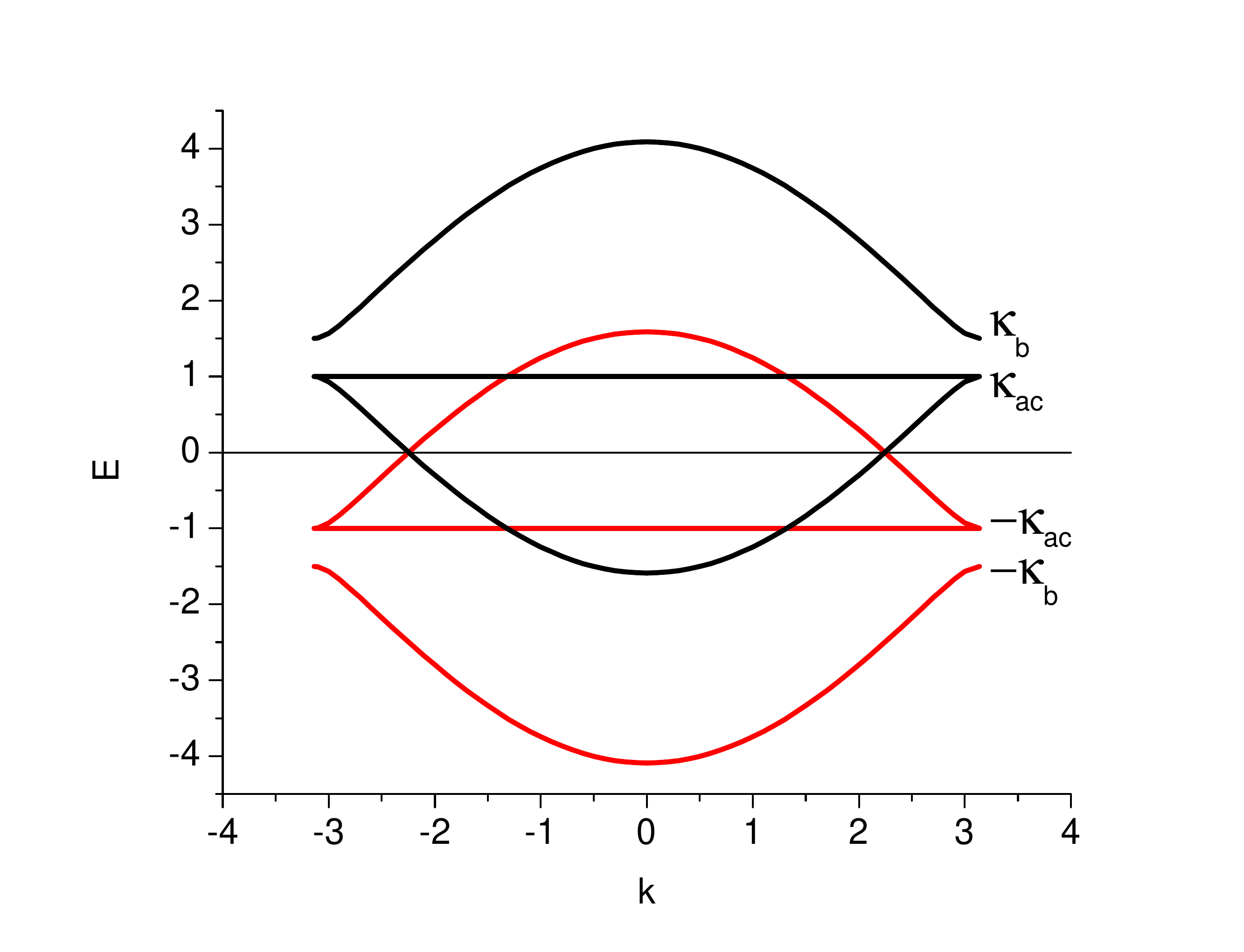}\label{smaller_kappa_ac}}
\caption{(Color online) Exactly found dispersion branches corresponding to
the symmetric modes (black lines) and antisymmetric ones (red lines) in the
double-diamond-chain system, plotted for (a) $\protect\kappa _{ac}>\protect%
\kappa _{b}$ ($\protect\kappa _{ac}=1,\protect\kappa _{b}=0.5$), and (b) $%
\protect\kappa _{ac}<\protect\kappa _{b}$ ($\protect\kappa _{ac}=1,\protect%
\kappa _{b}=1.5$). Flat-band states are represented by the horizontal lines,
$E=\pm \protect\kappa _{ac}$.}
\label{plot of energy levels}
\end{figure}

The linear system with broken symmetry between the top and bottom sites, $a$
and $c$, should also be mentioned. In that case, there are 3 different
coupling constants $\kappa _{a}$, $\kappa _{b}$ and, $\kappa _{c}$ and the
respective dispersion equation takes the form of

\begin{equation}
\begin{vmatrix}
\ -E & \left( 1+e^{ik}\right) & 0 & \kappa _{a} & 0 & 0 \\
\left( 1+e^{-ik}\right) & -E & \left( 1+e^{-ik}\right) & 0 & \kappa _{b} & 0
\\
0 & \left( 1+e^{ik}\right) & -E & 0 & 0 & \kappa _{c} \\
\kappa _{a} & 0 & 0 & -E & \left( 1+e^{ik}\right) & 0 \\
0 & \kappa _{b} & 0 & \left( 1+e^{-ik}\right) & -E & \left( 1+e^{-ik}\right)
\\
0 & 0 & \kappa _{c} & 0 & \left( 1+e^{ik}\right) & -E%
\end{vmatrix}%
=0,  \label{determinant2}
\end{equation}%
cf. Eq. (\ref{determinant}). The reduction based on Eq. (\ref{symm}) again
allows one to reduce the $6\times 6$ determinant (\ref{determinant2}) to
ones of size $3\times 3$, but the respective solutions turn out to be much
more cumbersome than above. The respective eigenvalues have been
analytically computed with the help of \textit{Mathematica}, but are not
included here. Effects of the on-site nonlinearity in the double-chain
system, and the respective nonlinear modes, will be considered elsewhere.

\section{Conclusion}

The objective of this work is to report the development of the known FB
(flatband) system, based on the \textquotedblleft diamond chain", in two
directions: adding the on-site cubic nonlinearity, which is naturally
present in the optical and matter-wave (BEC) implementation of the FB
lattices, and, on the other hand, to introduce the double FB\ system, with
two components coupled by the Rabi mixing at each lattice site. First, we
have produced a full analytical solution for all stationary states
(antisymmetric, symmetric, and asymmetric ones) in the system with three
degrees of freedom, which represents an isolated nonlinear cell of the
lattice. The asymmetric states emerge from their symmetric counterpart via a
spontaneous-symmetry-breaking bifurcation, whose character is weakly
subcritical. In the infinite nonlinear one-component chain, antisymmetric
CLSs (compact localized states) of different lattice sizes, which are a
hallmark of FB systems, have been found in an exact form too. Their
stability was studied partly analytically (to demonstrate that they are not
subject an antisymmetry-breaking bifurcation), and partly numerically,
revealing a nontrivial stability boundaries for the compact modes, as given
by Eqs. (\ref{n=1}) and (\ref{n=3}). These stability boundaries are specific
to the nonlinear system. Along with the CLSs, various types of symmetric,
antisymmetric and asymmetric CW (continuous-wave) states and lattice
solitons (which are exponentially localized, but not compact modes) have
been found too, in the numerical form and by dint of the VA (variational
approximation). The VA for symmetric and asymmetric solitons demonstrates
good accuracy, in comparison with their numerically generated shapes. It is
found that different branches of the CW and soliton families may be
completely or partly stable, some of them being fully unstable. Unstable
lattice solitons typically evolve into confined quasi-soliton states, with
randomized inner evolution, that emit small-amplitude phonon waves. Finally,
an exact solution for eigenmodes of the linear double diamond chain was
produced, with two split FBs present in the spectrum.

\section*{Acknowledgments}

We appreciate stimulating discussions with S. Flach. K.B.Z. and B.A.M.
appreciate hospitality of the Center for Theoretical Physics of Complex
Systems at the Institute for Basic Science (Daejeon, Korea). N.V.H.
acknowledges support provided by Prof. Marek Trippenbach during his stay in
Warsaw for postdoc research. K.B.Z. acknowledges support from the National
Science Center of Poland through Project FUGA No. 2016/20/S/ST2/00366. M.T.
and N.V.H acknowledge also support from the National Science Center of
Poland through Project HARMONIA No. 2012/06/M/ST2/00479.

\end{document}